\documentclass[superscriptaddress,secnumarabic,
amssymb,amsmath,nobibnotes,aps,prd,showpacs,nofootinbib]{revtex4}%
\usepackage{graphicx}
\usepackage{epsf}
\usepackage{bm}
\usepackage{amsmath}
\usepackage{amsfonts}
\usepackage{amssymb}
\usepackage{epstopdf}
\usepackage{natbib}
\usepackage{color}%
\providecommand{\U}[1]{\protect\rule{.1in}{.1in}}
\newcommand{\be}{\begin{equation}}
\newcommand{\ee}{\end{equation}}

\newcommand{\mincir}{\raise
-3.truept\hbox{\rlap{\hbox{$\sim$}}\raise4.truept\hbox{$<$}\ }}
\newcommand{\magcir}{\raise
-3.truept\hbox{\rlap{\hbox{$\sim$}}\raise4.truept\hbox{$>$}\ }}

\usepackage{color}

\begin{document}

\title{Constraints on quintessence scalar field models using cosmological
observations}

\author{Weiqiang Yang}
\email{d11102004@163.com}
\affiliation{Department of Physics, Liaoning Normal University, Dalian, 116029, P. R. China}

\author{M. Shahalam}
\email{shahalam@zjut.edu.cn}
\affiliation{Institute for Theoretical Physics $\&$ Cosmology,
Zhejiang University of Technology, Hangzhou, 310023, P. R. China}

\author{Barun Pal}
\email{terminatorbarun@gmail.com}
\affiliation{Department of Mathematics, Netaji Nagar College for Women, Kolkata 700092, West Bengal, India}

\author{Supriya Pan}
\email{supriya.maths@presiuniv.ac.in}
\affiliation{Department of Mathematics, Presidency University, 86/1 College Street, Kolkata 700073, India}

\author{Anzhong Wang}
\email{Anzhong_Wang@baylor.edu}
\affiliation{Institute for Theoretical Physics $\&$ Cosmology,
Zhejiang University of Technology, Hangzhou, 310023, P. R. China}
\affiliation{GCAP-CASPER, Department of Physics, Baylor University, Waco, TX, 76798-7316, USA}

\begin{abstract}
We consider a varieties of quintessence scalar field models in a homogeneous and isotropic geometry of the universe with zero spatial curvature aiming to provide stringent constraints using a series of cosmological data sets, namely, the cosmic microwave background observations (CMB), baryon acoustic oscillations (BAO), joint light curve analysis (JLA) from supernovae type Ia, redshift space distortions (RSD), and the cosmic chronometers (CC). From the qualitative  evolution of the models, we find all of them are able to execute a fine transition from the past decelerating phase to the presently accelerating expansion where in addition, 
the equation of state of the scalar field (also the effective equation of state) might be close to that of the $\Lambda$CDM cosmology depending on its free parameters. From the observational analyses, we find that the scalar field parameters are unconstrained irrespective of all the observational datasets. In fact, we find that the quintessence scalar field models are pretty much determined by the CMB observations since any of the external datasets such as BAO, JLA, RSD, CC does not add any constraining power to CMB.  Additionally, we observe a strong negative correlation between the parameters $H_0$ (present value of the Hubble parameter), $\Omega_{m0}$  (density parameter for the matter sector, i.e., cold dark matter plus baryons) exists, while no correlation between $H_0$, and $\sigma_8$  (amplitude of the matter fluctuation) are not correlated. We also comment that the present models are unable to reconcile the tension on $H_0$.  
Finally, we conclude our work with the Bayesian analyses which report that the non-interacting $\Lambda$CDM model is preferred over all the quintessence scalar field models. 
\end{abstract}


\maketitle
\section{Introduction}
Since the detection of accelerating universe by measuring the luminosity distances of type Ia Supernovae \cite{Riess:1998cb, Perlmutter:1998np} $-$ a new era of modern cosmology has began. Subsequent investigations by different groups  \cite{Hinshaw:2012aka, Ade:2015xua} conveyed that current acceleration of our universe could be an effect of some hypothetical fluid with large negative pressure  known as dark energy \cite{Copeland:2006wr}, which is completely unknown by its character and origin. Usually there are two distinct approaches to describe such accelerating expansion -- one route is through  the modifications of the matter sector in the context of Einstein gravity \cite{Copeland:2006wr, AT,Visinelli:2018utg}, and the other way around is to modify the Einstein's gravitational theories \cite{DeFelice:2010aj, Sotiriou:2008rp, Capozziello:2011et,Myrzakulov:2015qaa, Cai:2015emx,Sebastiani:2016ras, Nojiri:2017ncd,Dutta:2017fjw,Casalino:2018tcd} which leads to some extra geometrical terms (alternatively known as geometrical dark energy fluids in order to make a difference between the accelerating effects coming from the matter modifications or geometry modifications).
Apart from above two approaches there is another alternative to describe this accelerating universe -- the gravitational particle production mechanism, see \cite{Chakraborty:2014fia,Nunes:2015rea,deHaro:2015hdp,Pan:2016jli,Nunes:2016aup,Nunes:2016eab,Nunes:2016tsf,Salo:2016xlw,Paliathanasis:2016dhu,Pan:2016bug} and references therein.  However, overall, the actual dynamics of these mysterious components are unknown, but thanks to the recent observational evidences, we have an estimate of such dark fluids. According to the observational predictions, such dark energy fluids contribute nearly 68\% of the total energy density of our universe \cite{Ade:2015xua}.  Additionally, another bulk content of the matter sector, about 28\% of the total energy density of our universe, is occupied by some non-luminous dark matter component.  Thus, overall, almost 96\% of the total energy content of our universe has been filled up by these dark fluids, namely, dark energy and dark matter and probing their nature, evolution and the origin is one of the most intriguing facts of modern cosmology. 

In the present work we  confine ourselves to Einstein gravity and thus incorporate dark energy fluid through  matter sector in the field equations. The dark energy fluids may appear in many forms 
$-$ from the simplest cosmological constant to more complicated ones. Although the cosmological constant characterized by the constant equation of state  $w_{\Lambda} = -1$, is consistent with present observations, however, a time varying dark energy equation of state has been found to be efficient in relieving the tensions between several cosmological parameters, see \cite{Zhao:2017cud}. The release of tensions between cosmological parameters in presence of a constant dark energy equation of state is not usually possible until some extra degrees of freedom in terms of the non-gravitational interaction, for instance, is allowed into the picture. Moreover, the problem related to the  magnitude of the cosmological constant (see \cite{Weinberg, Sean}) further  motivated to consider some  dark energy fluids evolving with cosmic time.

The scalar field models, arising in the context of particle physics theory, is a very natural choice for the dynamical dark energy models.  Usually there is a division in the scalar field models, one is the canonical scalar field models and the other one is the non-canonical scalar field models. We consider a specific type of canonical scalar field model, namely the quintessence which is a single minimally coupled scalar field with a canonical kinetic term having positive sign. In case of phantom scalar field models, which is also a canonical scalar field model, the sign of its kinetic term is negative contrary to the quintessence scalar field model. We note that the canonical scalar field models are more informative compared to the non-canonical scalar field models.  The physics of canonical scalar field models, which is mostly contained 
in the potential, $V (\phi (t))$ where $\phi (t)$ is the 
underlying field, have gained a considerable interest in the cosmological community due to explaining various stages of the universe evolution, see \cite{Guth:1980zm,Linde:1981mu,Peebles:1987ek, Ratra:1987rm,Barrow:1993ah,Barrow:1993hn,Liddle:1998xm,Peebles:1998qn,Sahni:1999qe,Rubano:2001xi,Basilakos:2011rx,Tsamparlis:2011cw,Paliathanasis:2014zxa,Paliathanasis:2015gga,Myrzakulov:2015nqa,Myrzakulov:2015kda,Cognola:2016gjy,deHaro:2016hpl,deHaro:2016hsh,deHaro:2016cdm,deHaro:2016ftq,deHaro:2017nui,Haro:2018jtb,Dimakis:2016mip,Paliathanasis:2015cza,Vagnozzi:2017ilo,Akrami:2018ylq,Casalino:2018wnc} 
(also see \cite{Paliathanasis:2014yfa,Dimakis:2017zdu,Giacomini:2017yuk,Papagiannopoulos:2017whb,Barrow:2018zav}). The selection of quintessence scalar field models should not be much arbitrary, according to recently introduced swampland \cite{Obied:2018sgi,Agrawal:2018own} and refined swampland \cite{Ooguri:2018wrx} criteria.  In particular, these criteria imply that the quintessence potential should not be steeper than order unity in Planck units \cite{Kinney:2018nny}. The swapland criteria has recently got massive attention in the cosmological community, see for an incomplete list \cite{Kinney:2018nny,Heisenberg:2018yae,Marsh:2018kub,Kinney:2018kew,Thompson:2018ifr,Tosone:2018qei,Raveri:2018ddi,Palti:2019pca,vandeBruck:2019vzd}.  
However, apart from the canonical scalar field models, non-canonical scalar field models have also been investigated widely in the literature. A class of non-canonical scalar field models include the k-essence,
tachyon, ghost condensates and the dilatonic scalar field 
models, see \cite{Copeland:2006wr,AT} for more details on them. 
However, in the present work we shall consider some specific  potentials for  quintessence scalar field models aiming to impose stringent cosmological constraints using the latest cosmological sources, namely,  the cosmic microwave background temperature and polarization data \cite{Adam:2015rua,Aghanim:2015xee}, baryon acoustic oscillations distance measurements \cite{Beutler:2011hx,Ross:2014qpa,Gil-Marin:2015nqa}, redshift space distortion \cite{Gil-Marin:2016wya}, Supernovae Type Ia \cite{Betoule:2014frx}, and finally the Hubble parameter measurements from the cosmic chronometers \cite{Moresco:2016mzx}. 
The underlying geometry has been chosen to be the spatially flat  Friedmann-Lema\^{i}tre-Robertson-Walker (FLRW) metric where the matter fields of the universe 
is minimally coupled to the gravity $-$ described by the usual Einstein gravity.

The work has been organized as follows. In section \ref{sec2},  we describe the evolution equations for any quintessence scalar field model at the level of background and perturbations in a spatially flat FLRW universe. The section \ref{sec-models} introduces a class of quintessence models and their qualitative behavior that we wish to study in this work.
In section \ref{sec-data}, we introduce the observational data, the statistical technique and the results of the models. In particular, the subsection \ref{sec-observational} contains the observational constraints on the models; the subsection \ref{subsec-om} includes a geometrical probe, namely, the $Om$ diagnostic; the subsection \ref{sec-bayesian} provides with the Bayesian model comparison and in subsection \ref{sec-comparisons} we present an overall comparison between all the quintessence scalar field models. Finally, with section \ref{sec-conclu} we close the present work summarizing the main findings.

\section{Basic Framework}
\label{sec2}

As usual we consider a homogeneous and isotropic space-time of the universe characterized by the Friedmann-Lema\^{i}tre-Robertson-Walker line element 

\begin{eqnarray}
d{\rm s}^2 = -dt^2 + a^2 (t) \left[\frac{dr^2}{1-Kr^2} + r^2 \left( d\theta^2 + \sin^2 \theta \; d \phi^2\right) \right], 
\end{eqnarray}
where $a(t)$ is the expansion scale factor of the universe and $K$ is the curvature scalar which for $0, +1, -1$, respectively represent the spatially flat, closed and open universe. In agreement with the observational evidences \cite{Ade:2015xua} we consider the spatial flatness of this universe, that means, we shall consider $K =0$ throughout the analysis of the present work. 
Let us now consider the action in such a universe where the matter fields minimally coupled to gravity as follows
\begin{eqnarray}
\mathcal{A} = \int \sqrt{-g} d^4 x\, \left[ \frac{R}{2\kappa}  - g^{\mu \nu} \partial_{\mu} \phi \partial_{\nu} \phi - 2 V (\phi)\right] + S_{m}+ S_{r}
\end{eqnarray} 
where $\kappa = 8 \pi G$ is the Einstein's gravitational constant ($G$ is the Newton's gravitational constant); 
$S_{m}$, $S_{r}$ are respectively the actions representing the matter and radiation sectors; $V (\phi)$ is the potential of the scalar field $\phi$.  The corresponding field equations for flat FLRW universe are given by
\begin{eqnarray}
H^2  = \frac{8\pi G}{3} \; \rho_{\rm eff},\label{f1}\\
2\dot{H} + 3 H^2 = - 8 \pi G \; p_{\rm eff},\label{f2} 
\end{eqnarray}
where an overhead dot represents the cosmic time differentiation;
$H \equiv \dot{a}/a$, is the Hubble parameter; $\rho_{\rm eff} = \rho_r+\rho_b+ \rho_c + \rho_{\phi}$ is the total energy density of the universe.  Here, $\rho_i$ ($i = r, b, c, \phi$) represents the energy density of the $i$-th fluid where the symbol $r, b, c, \phi$ corresponds to the radiation, baryons, cold dark matter and the scalar field sector, respectively.~\footnote{We fix the total neutrino mass to $M_{\nu}=0.06\,{\rm eV}$, the minimal mass allowed within the normal ordering, as in the \textit{Planck} baseline analyses. This is justified by the current very tight limits on neutrino masses~\cite{Palanque-Delabrouille:2015pga,Giusarma:2016phn,Vagnozzi:2017ovm,Giusarma:2018jei,Aghanim:2018eyx}, which favour the normal ordering~\cite{Simpson:2017qvj,Schwetz:2017fey} (see also~\cite{Gerbino:2016sgw}).}  Similarly, one can define the total pressure by adding the pressure  of each fluid, i.e., $p_{\rm eff}= p_r +p_b + p_c +p_{\phi}$ in which $p_i$ stands for the pressure term for the $i$-th fluid ($i = r, b, c, \phi$). For the flat FLRW universe, the energy density and the pressure of the scalar field model, respectively are, $\rho_{\phi} = \dot{\phi}^2/2 + V(\phi)$, \  $p_{\phi} = \dot{\phi}^2/2 - V(\phi)$. 
Additionally, we assume that each fluid follows the barotropic equation of state $p_i = w_i \rho_i$, $i = r, b, c, \phi$ ($w_i$ being the barotropic index for the $i$-th fluid) where $w_r = 1/3$, $w_b = 0$, $w_c =0$, and $w_{\phi} = p_{\phi}/\rho_{\phi} = \frac{\dot{\phi}^2-2V (\phi)}{\dot{\phi}^2+2 V(\phi)}$.  
Since there is no such interaction between any two fluids, hence, each fluid satisfies their
own conservation equation, i.e., $\dot{\rho}_i + 3 H (p_i +\rho_i) = 0$. Following this conservation equation, one can see the evolution of the component fluids as follows: $\rho_r \propto a^{-4}$, $\rho_b \propto a^{-3}$, $\rho_c \propto a^{-3}$. 
The evolution of the remaining fluid, i.e., the quintessence scalar field sector is governed by the  following equation
\begin{eqnarray}\label{K-G-eqn}
\ddot{\phi} = - 3 H \dot{\phi} - \frac{dV}{d \phi}
\end{eqnarray}
which can also be obtained from equations (\ref{f1}) and (\ref{f2}).  
However, one can clearly understand that the dynamics of the scalar field can only be determined once the 
potential $V (\phi)$ is prescribed and there is no such specific rule to select the potential. In this work, we shall consider a few  phenomenological forms for the potential and render stringent constraints from the latest observational data coming from different cosmological sources. 

Let us define some important cosmological parameters exhibiting the qualitative evolution of the scalar field model. The effective equation of state for the total fluid  characterized by $\rho_{\rm eff}, p_{\rm eff}$ as 

\begin{eqnarray}\label{eff-eos}
w_{\rm eff} = \frac{p_{\rm eff}}{\rho_{\rm eff}} = -1 - \frac{2\dot{H}}{3H^2}
\end{eqnarray}
and the deceleration parameter $q$ that quantifies the accelerating or decelerating 
phase of the universe, is given by 
\begin{eqnarray}\label{deceleration}
q = - 1 -\frac{\dot{H}}{H^2}.
\end{eqnarray}
One may also recall the usual definitions of  the density parameters for the fluids as follows: $\Omega_i = \rho_i/\rho_c$, where $\rho_c = 3 H_0^2/8\pi G$, is the critical density of the universe. 

Having the above set of equations in hand, in principle one can determine the evolution of the quintessence scalar field model at the background level but not at the level of perturbations. Since the observed structure formation of the universe is related to the evolution at the level of perturbations, thus, we now proceed to the investigations of the perturbation equations. In order to do so, we consider the perturbed FLRW metric in the conformal Newtonian gauge. The perturbed FLRW metric in this gauge takes  the following form 
\begin{eqnarray}\label{perturbed-flrw}
ds^2 = a^2 (\tau) \left[ - (1+ 2 \Psi) d \tau^2 + (1+ 2 \Phi) \delta_{ij} dx^i dx^j \right] 
\end{eqnarray}
where $\tau$ is the conformal time; $\Psi$, $\Phi$ are the perturbation quantities. For the above perturbed FLRW metric (\ref{perturbed-flrw}),  one may write down the Einstein's field equations in the conformal Newtonian gauge as \cite{Chen:2015oga}:

\begin{eqnarray}
\delta^{\prime}_A + 3 a H \left(c_{s, A}^2 - w_A \right) \delta_A = - (1+w_A) (\theta_A + 3 \Phi^{\prime}),\\
\theta_A^\prime + \left(a H (1- 3 w_A) + \frac{w_A^\prime}{1+w_A}\right) \theta_A = k^2 \left(\frac{c_{s,A}^2}{1+w_A} \delta_A + \Psi + \sigma_A \right),\\
k^2 \Phi = 4 \pi G a^2 \rho_A \left( \delta_A + 3 a H (1+w_A) \frac{\theta_A}{k^2}\right),
\end{eqnarray} 
and $\Psi = -\Phi$. Here, we recognize $\delta_A = \delta \rho_A /\rho_A$, as the density perturbation for the fluid $A$; $\theta_A = i \vec{k}. \vec{v}_A$, as the divergence of the peculiar
velocity $v_A$ of the component $A$. The prime attached to any quantity denotes the differentiation with respect to the conformal time $\tau$. The quantity $c_{s,A}^2 = \delta p_A/\delta \rho_A$, is the sound speed of the fluid $A$ which we consider to be non-negative in order to avoid any kind of instabilities. Finally, we note that the above perturbations equations could also be written using the synchronous gauge. So, we have presented the equations in Appendix B for completeness.

\section{Scalar field Models}
\label{sec-models}
The scalar field models play an essential role to describe the different dynamical phases of our universe. It is widely known that both early and late accelerating phases of the universe, are almost satisfactorily described by the scalar field models. However, in this section we shall introduce some scalar field models  belonging to the quintessence class and investigate their qualitative behavior.  The quintessence scalar field models are usually classified into two distinct groups, namely the thawing (slow-roll) and freezing (fast-roll) scalar field models. The models belonging to the thawing class are very sensitive to the initial conditions, while on the other hand, freezing models are independent for a wide range of the initial conditions. Concerning the freezing models, they are again subdivided into two classes based on the nature of the models $-$ whether they have scaling and tracking behaviour. The tracker models could produce late-time cosmic acceleration but for scaling models late time acceleration is not possible, see Refs. \cite{wand,shinji,stein} for more discussions in this direction. In the present work we shall focus on both kind of scalar field models, that means thawing and freezing scalar field models. Since the present literature includes a large number of potentials, therefore, the main aim of this work is to introduce some `generalized' potentials that could recover some well known models that have been already studied.

\subsection{Model 1}
\label{sec-model1}

Let us first consider the following scalar field model characterized by the potential 
\begin{eqnarray} \label{model1}
V(\phi) = V_0  \text{cosh} \left[\beta \left (\frac{\phi}{M_p}\right)^u \right]
\end{eqnarray}
where $V_0$, $u$ and $\beta$ are real constants in which  $\beta$, $u$ are dimensionless while $V_0$ is dimensionful; $M_p$ represents the reduced Planck mass. The parameter $\beta$ in (\ref{model1}) quantifies the deviation of the model from the constant potential $V (\phi) = V_0$.  One can clearly notice that the potential (\ref{model1}) is the sum of two exponential potentials: $$V (\phi) =\frac{V_0}{2} \left(e^{\beta (\phi/M_p)^{u}}+ e^{-\beta (\phi/M_p)^{u}}\right),$$
and it is the generalized version of the potential $V = V_0 \cosh(\beta \phi/M_p)~$ \cite{Lee:2004vm} which is obtained by setting $u = 1$ in (\ref{model1}).   
The original $\cosh$ potential ($u = 1$) has the following asymptotic form
\begin{eqnarray} 
 V(\phi) \simeq \frac{V_0}{2} \exp \left (\frac{\beta\phi}{M_p}\right),~~\mbox{for}~~\frac{\beta |\phi|}{M_p} \gg 1
\label{pot1exp}
\end{eqnarray} 
\begin{eqnarray} 
 V(\phi) \simeq V_0 \left[ 1+ \frac{1}{2}  \left (\frac{\beta \phi}{M_p}\right)^{2}\right],
~~\text{for} ~~ \frac{\beta |\phi|}{M_p} \ll 1 
\label{pot1sq}
\end{eqnarray}

\begin{figure*}[tbp]
\centering
\includegraphics[width=1.9in,height=1.9in,angle=0]{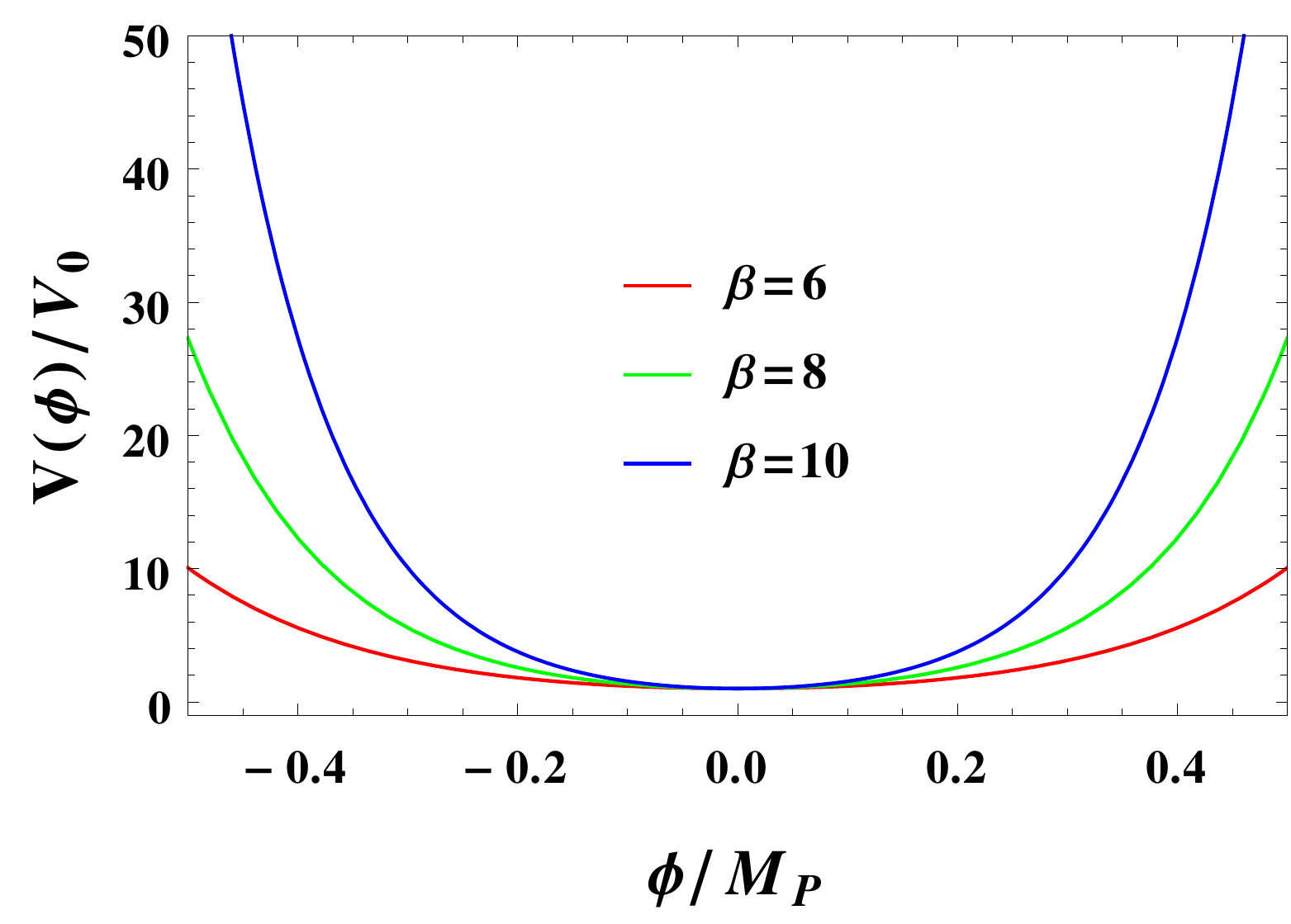}
\includegraphics[width=1.85in,height=1.85in,angle=0]{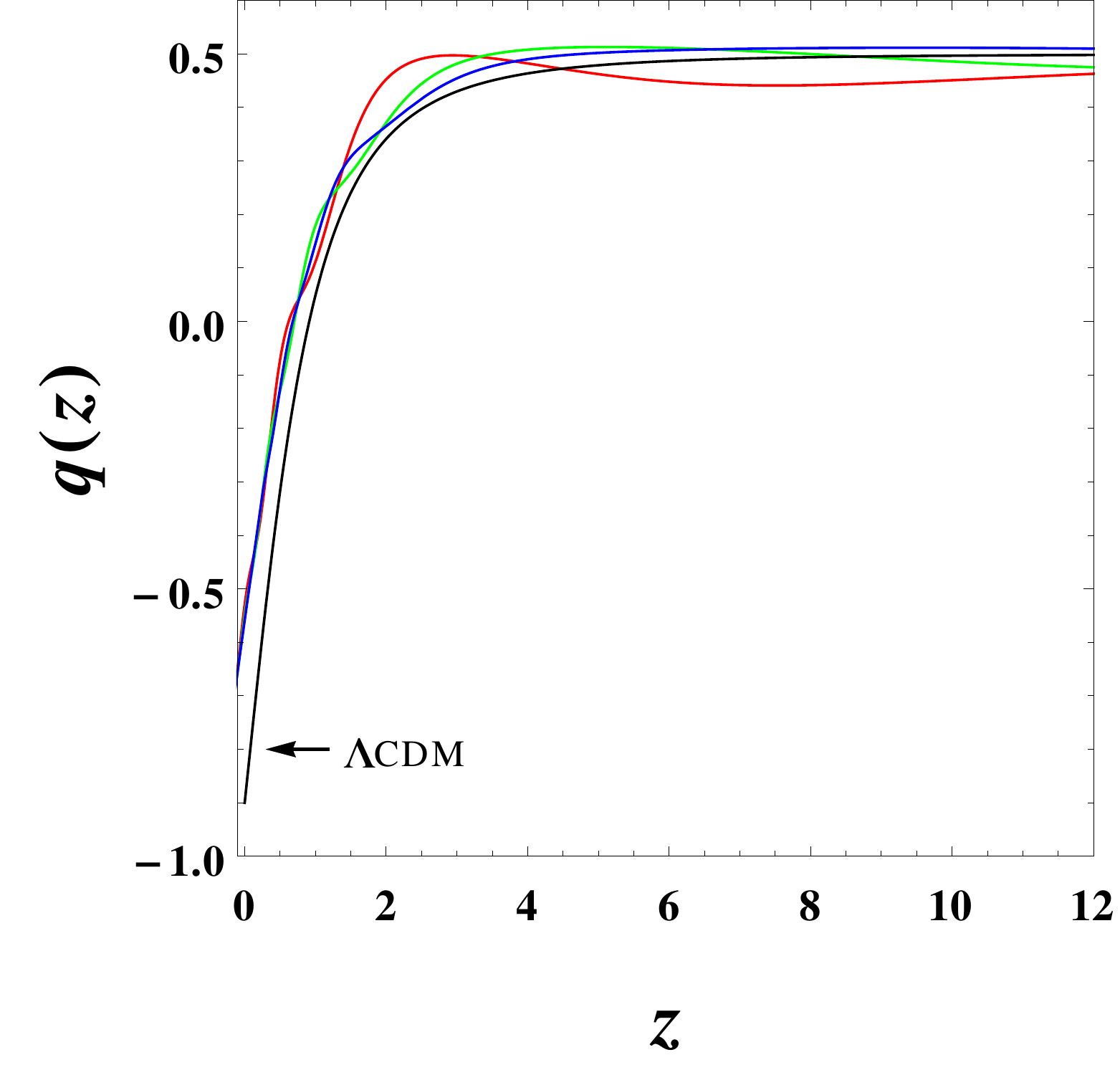}
\includegraphics[width=1.9in,height=1.9in,angle=0]{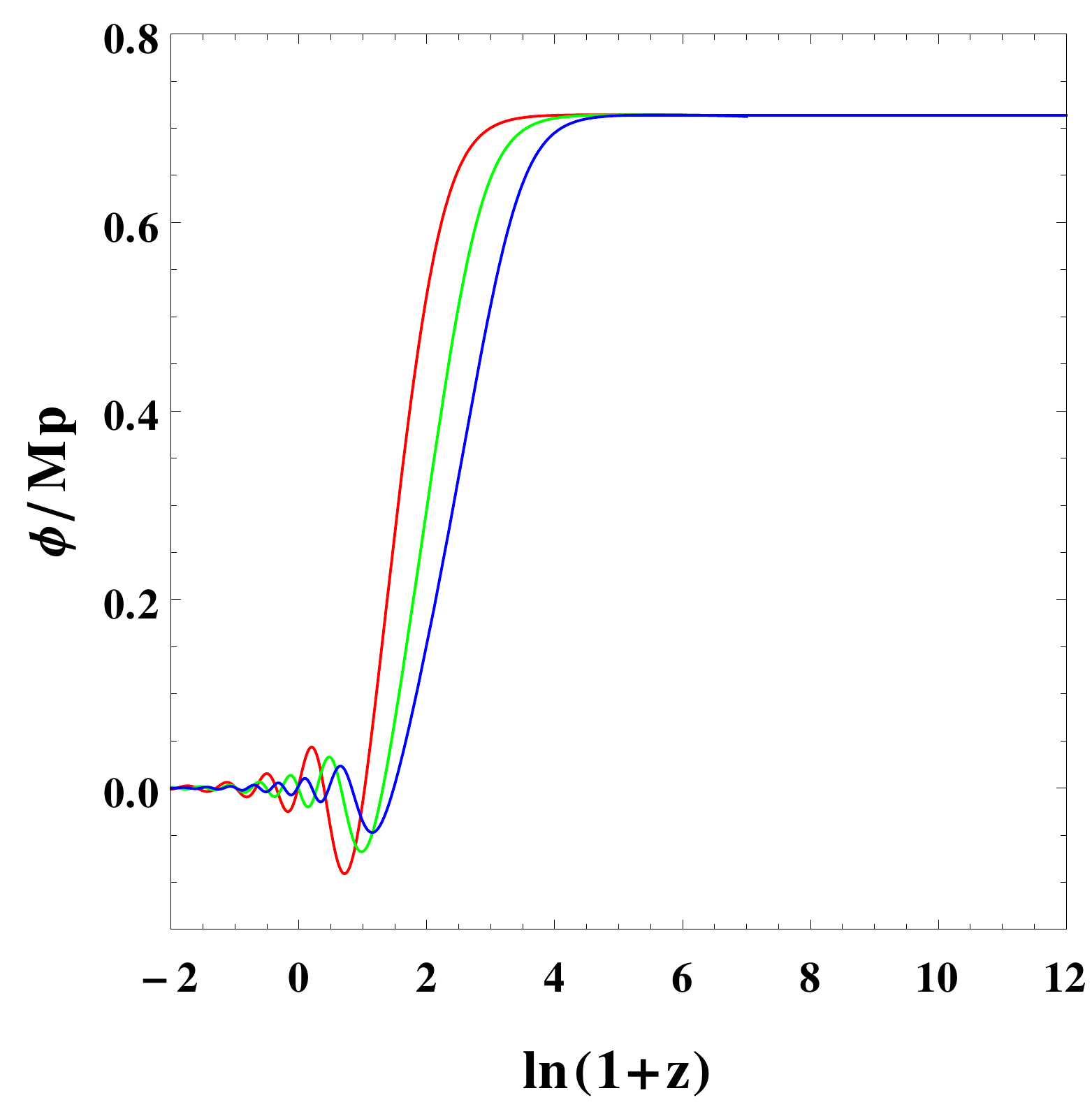}
\includegraphics[width=1.9in,height=1.9in,angle=0]{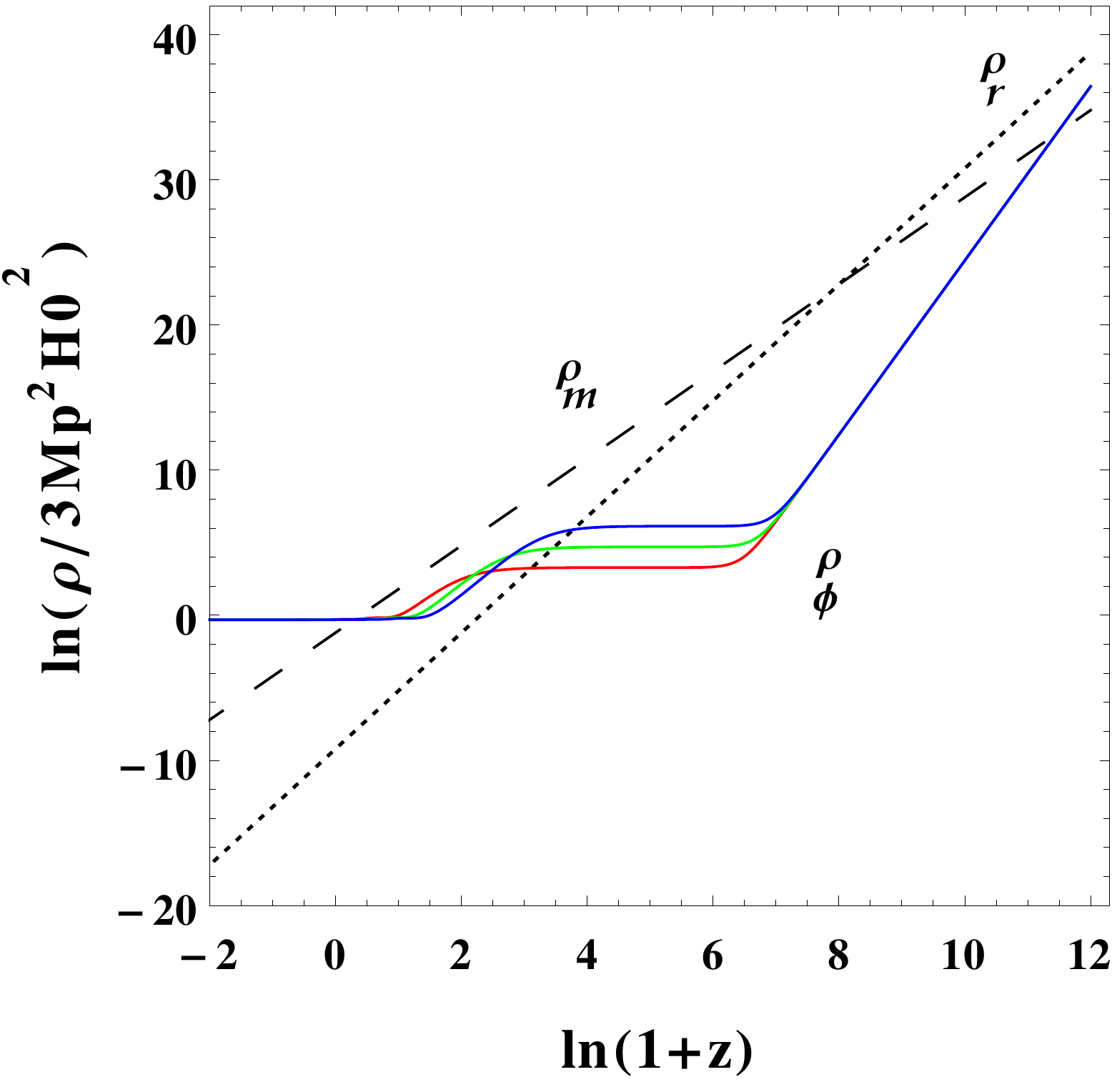}
\includegraphics[width=1.9in,height=1.9in,angle=0]{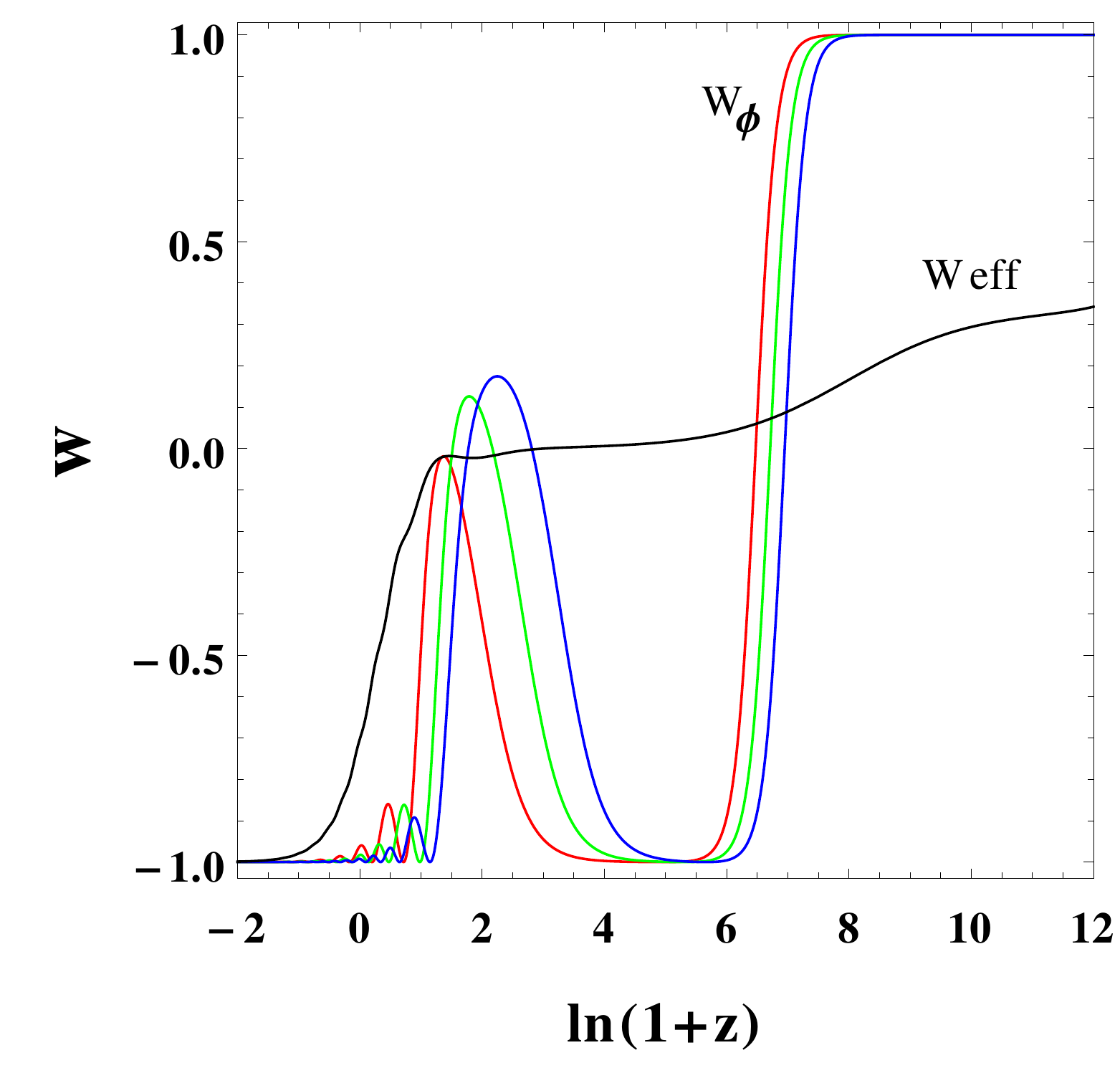}
\includegraphics[width=1.9in,height=1.9in,angle=0]{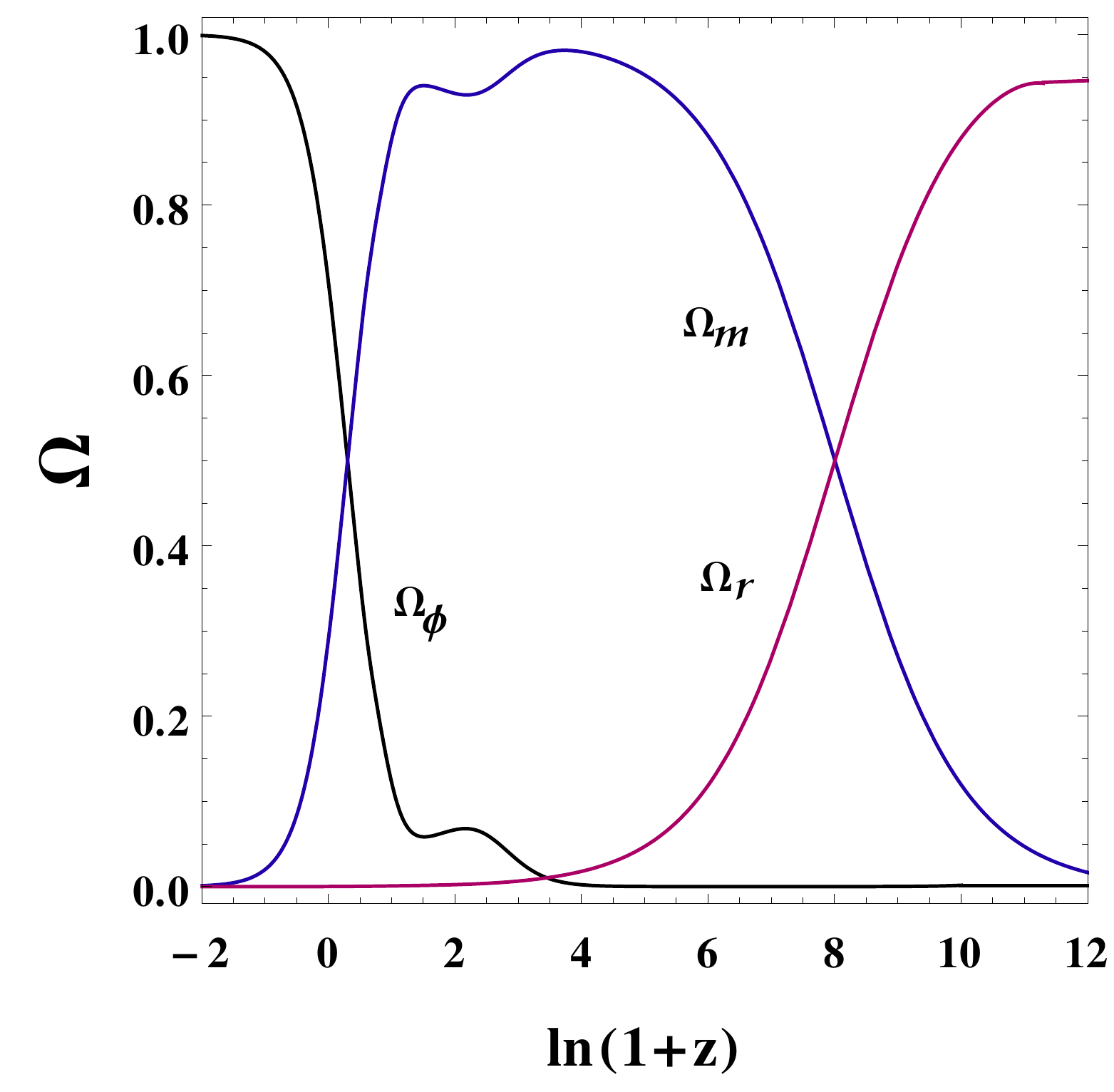}
\caption{This figure corresponds to the potential (\ref{model1}).  We show different cosmological parameters for $u=1$ taking various values of $\beta$ as $\beta=6$ (red), $8$ (green), $10$ (blue). This potential belongs to tracker class which is independent for a wide range of initial conditions. For large field values, the potential has an exponential asymptotic form where as for small field values (near minimum of potential) it is associated with quadratic term of field. Therefore, energy density of scalar field scales with background energy density around the present epoch, and exits it at late time, and correspondingly $w_\phi \simeq -1$ will be oscillatory. We note that during the analysis we have taken a specific value of $V_0$ as $V_0$=2.2 $M_p^4$. }
\label{fig:1}
\end{figure*}

Now, in order to study the qualitative behavior of this potential, we choose a specific value of $u$, the simplest one namely, $u =1$, and solve the conservation equation (\ref{K-G-eqn}) numerically. One may wonder why we fix $u =1$ since other values of $u$ are equally favorable. Thus, we 
performed similar calculations for other values of $u$ in a wide range of $u \in \mathbf{R}$, namely, $u = 2, 3$, but we did not find any significant qualitative changes in the cosmological parameters sketching the evolution of the universe. So, in this work we consider $u =1$ for this potential and proceed to its further analysis \footnote{However, concerning the observational constraints on this potential, keeping $u$ as a free parameter is certainly appealing, but this on the other hand increases the degeneracy between other parameters of the model. So, fixing $u$ to some preassigned value may reduce the degeneracy between the parameters. In this context, one can also see how the cosmological constraints change for different positive or negative values of $u$ leading to a class of quintessence potentials, such as $V = V_0 \cosh (\beta (\phi/M_p)^2), V_0 \cosh (\beta (\phi/M_p)^3),...V_0 \cosh (\beta (\phi/M_p)^{-1}), V_0 \cosh (\beta(\phi/M_p)^{-2}), ...V_0 \cosh (\beta (\phi/M_p)^{1/2}), V_0 \cosh (\beta (\phi/M_p)^{1/3}),$ etc. This might be an interesting investigation in future. }.   
So, essentially, we are interested in the original potential $V = V_0 \cosh(\beta \phi/M_p)$ of \cite{Lee:2004vm} although we keep the original $u$-dependent model (\ref{model1}) in this work for a far reaching purpose just mentioned above. The original cosh potential belongs to tracker class which is independent for a wide range of initial conditions. The qualitative evolution of this potential is presented  
in Fig. \ref{fig:1} in terms of various cosmological parameters. In the upper left plot of Fig. \ref{fig:1} we show the evolution of this potential which shows that the potential has an extremum and the parameter $\beta$ plays the role of a scaling character. The extreme right plot of the upper panel of Fig. \ref{fig:1} presents the evolution of the scalar field itself. The middle plot of the upper panel of Fig. \ref{fig:1} depicts the evolution of the deceleration parameter from which a smooth transition from the past deceleration to present acceleration is clearly displayed and the transition occurs around $z = 0$.  Moreover, from the evolution of the deceleration parameter, one can also notice that it is similar to that of the $\Lambda$CDM cosmology, however, for this case, the transition occurs after the transition for $\Lambda$CDM. The lower panel of Fig. \ref{fig:1} also displays various parameters, namely, the energy density of the fluids (lower left plot of Fig. \ref{fig:1}), equation of state of the scalar field $w_{\phi}$ and the effective equation of state of the total fluid, $w_{\rm eff}$ (lower middle plot of Fig. \ref{fig:1}); and the density parameters for the fluids (lower right plot of Fig. \ref{fig:1}). From the evolution of the energy density, one can see that  initially, the energy density of the scalar field is sub-dominant, and freezes due to large Hubble damping. Around the present epoch, it scales with the background, and at present epoch it exits from the background  and gives rise to late time acceleration. The equation of state for the scalar field as well as the effective equation of state of the total fluid, for large field values potential, behave almost in an exponential way while  for small field values since it is associated with $\phi^2$ term, hence, it suggests that at late time the scalar field behaves like a cosmological constant $w_{\phi} \simeq -1$ together with an oscillatory behavior.  

\subsection{Model 2}
\label{sec-model2}

We now consider the  model characterized by the following potential
\begin{eqnarray}\label{potential}
V (\phi) = V_0 \Bigl[ 1+ \epsilon~ {\rm sech} (\alpha \phi/M_p) \Bigr] 
\end{eqnarray}
where $\epsilon= \pm1$,
$V_0$ is a constant and $\alpha$ is the only free parameter of this model that quantifies the deviation of this potential from the constant potential $V = V_0$. The potential (\ref{potential}) is closely related to the previous potential in (\ref{model1}) since one is the inverse of the other. 
Originally the model was introduced in \cite{Pal:2009sd} (and subsequently in \cite{Pal:2012sd, Pal:2018sd}) to study cosmic inflation motivated by the models of inflation
in the framework of supergravity. Though it is a phenomenological model without having any field
theoretic origin, the model has few distinctive features in the context of inflation. Firstly, the spectral
index $r$ as predicted by this model is nearly constant with slight negative tilt. Depending on the
model parameter this model renders two types of inflationary solutions: one corresponds to small
inflaton excursion during observable inflation which predicts negligible amount of gravity waves ($r \sim \mathcal{O}(10^{-4})$) and the other describes large field inflation which is capable of large amount of gravity waves ($r \sim \mathcal{O}(10^{-1})$). This inspired us to analyze the performance of this model in the
context of late time acceleration.  However,  the authors of \cite{Pal:2009sd} discussed its cosmological features for $\epsilon = - 1$. However, concerning the cosmological importance of this model, we do not find any strong reason to exclude the model with $\epsilon= +1$. Thus, we are intended to investigate this potential for both the values of $\epsilon$ in the context of late-time acceleration of the universe. For convenience, the model with $\epsilon =1$ is denoted by Model 2a while the model with $\epsilon = -1$ is named as Model 2b. In the following we shall show that the sign of $\epsilon$ plays an interesting role in the evolution of the universe.   

\begin{figure*}[tbp]
\centering
\includegraphics[width=1.9in,height=1.9in]{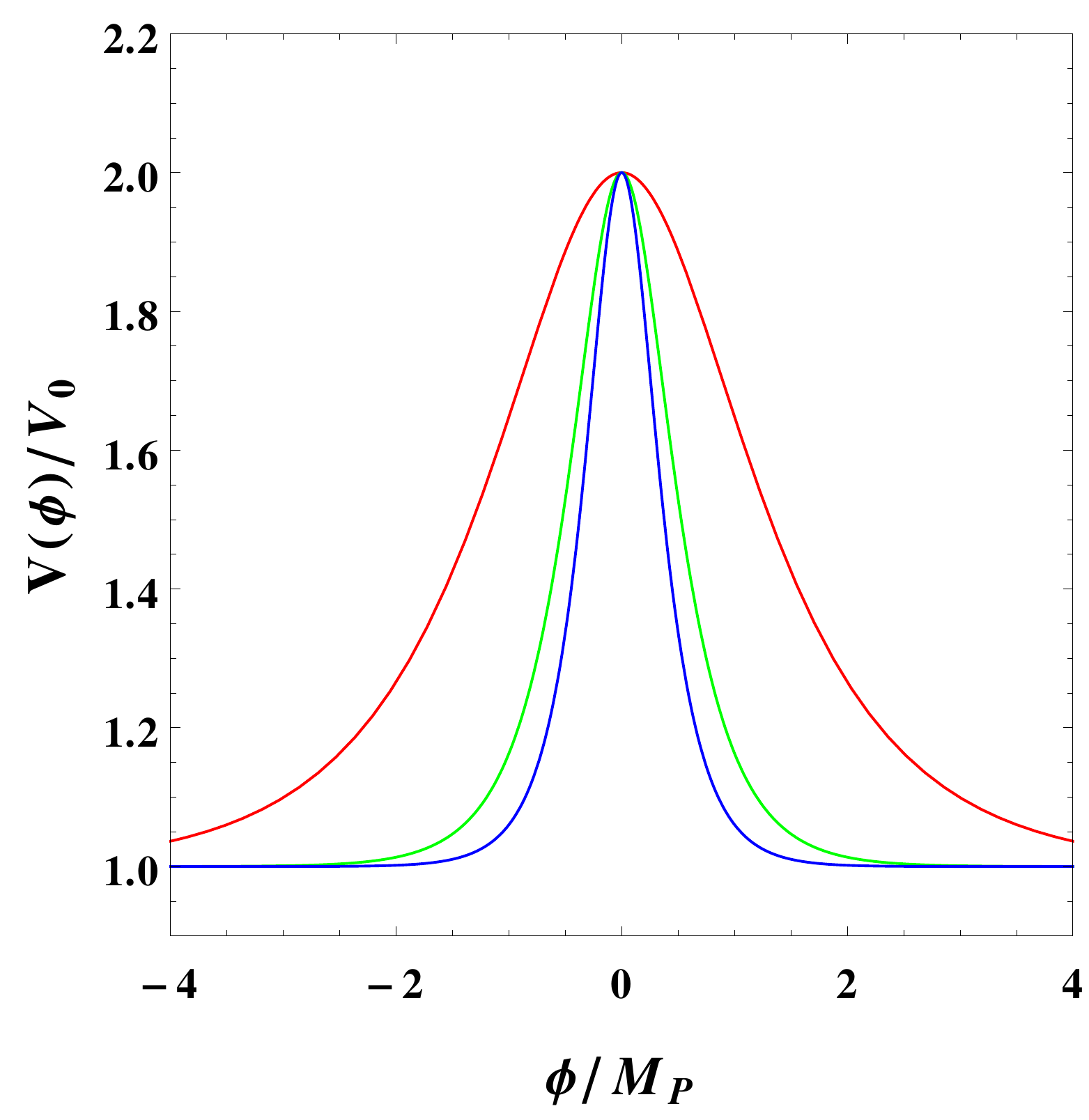} 
\includegraphics[width=1.85in,height=1.85in]{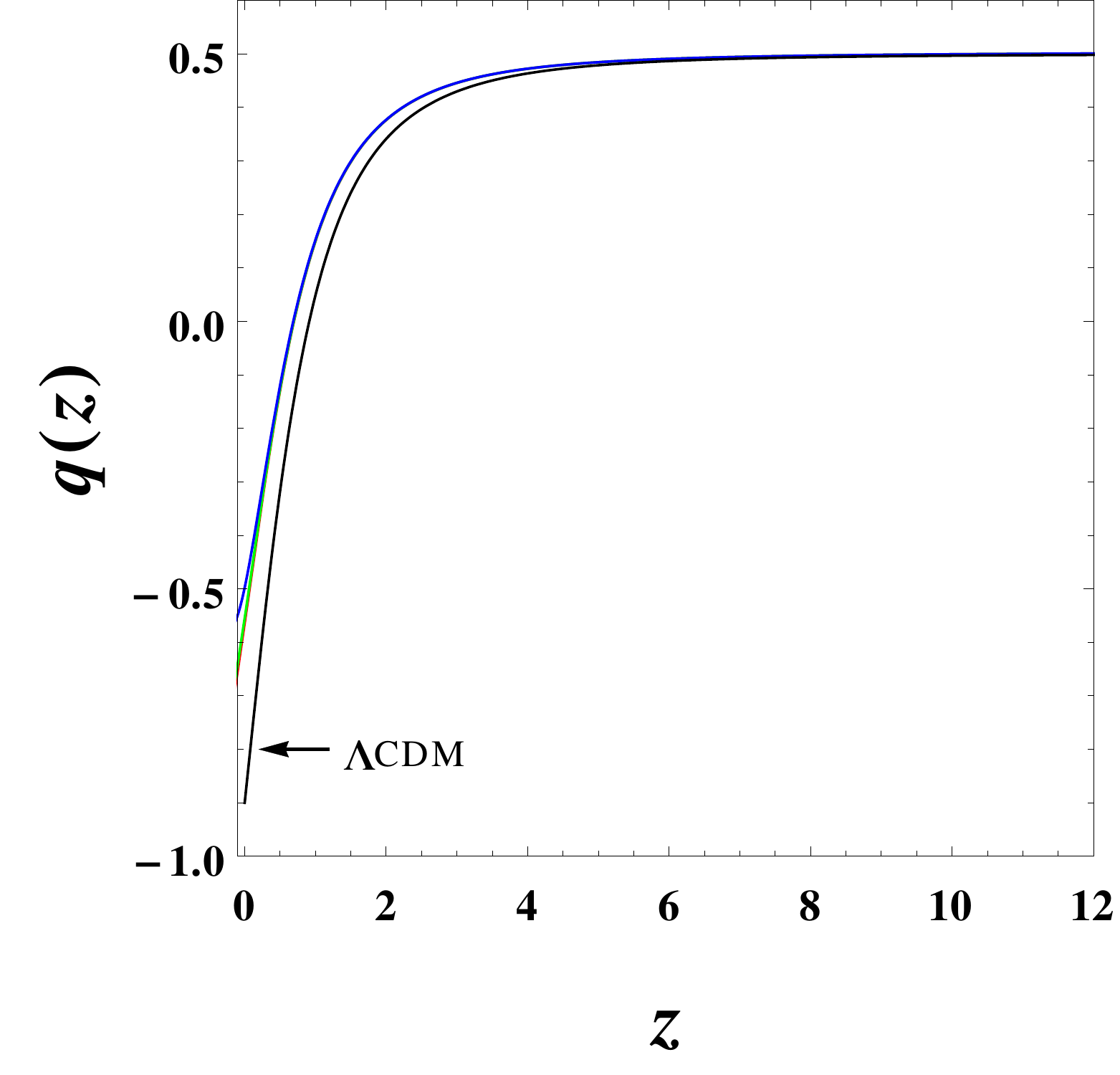} 
\includegraphics[width=1.9in,height=1.9in]{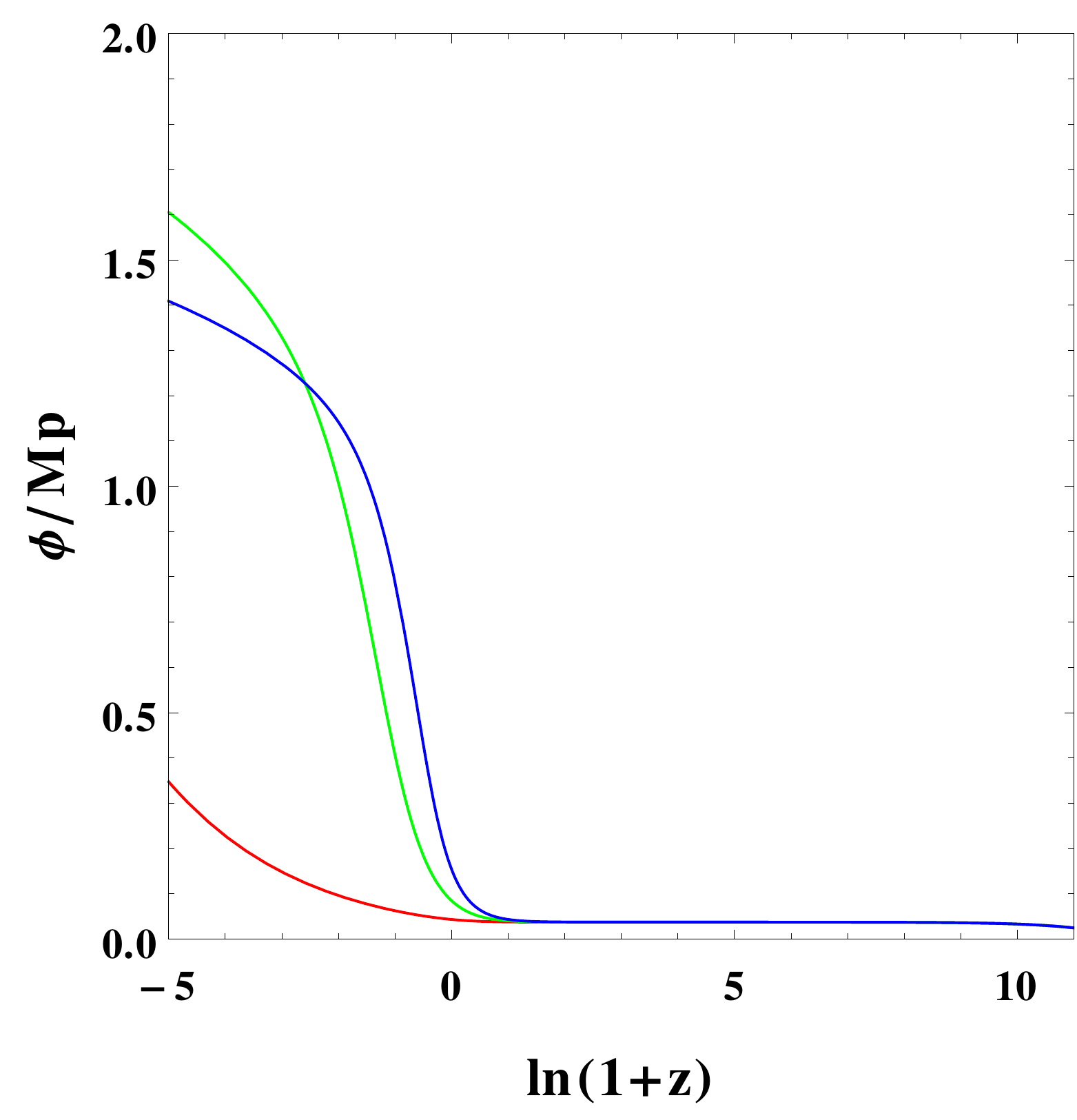} 
\includegraphics[width=1.9in,height=1.9in]{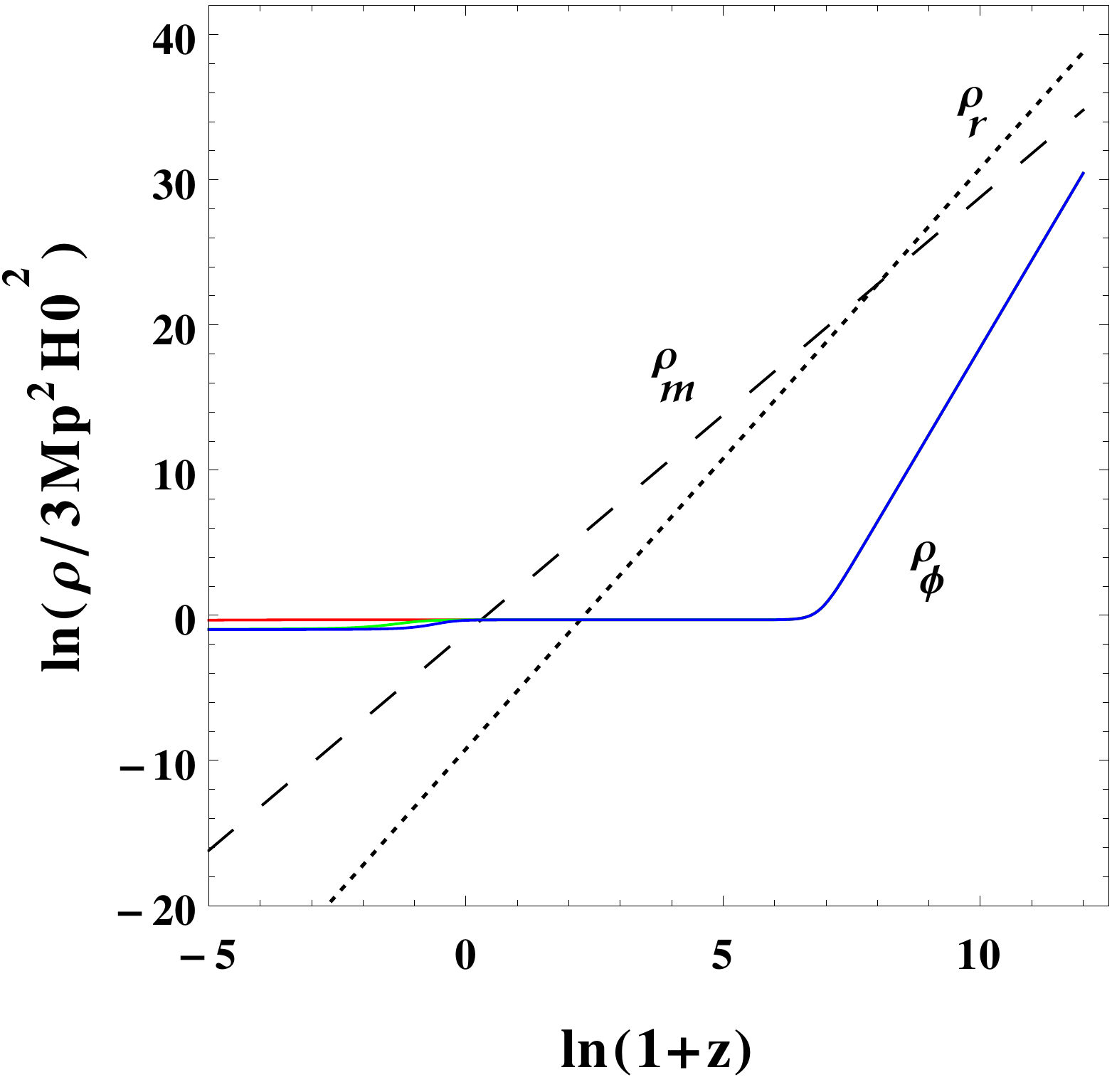}
\includegraphics[width=1.9in,height=1.9in]{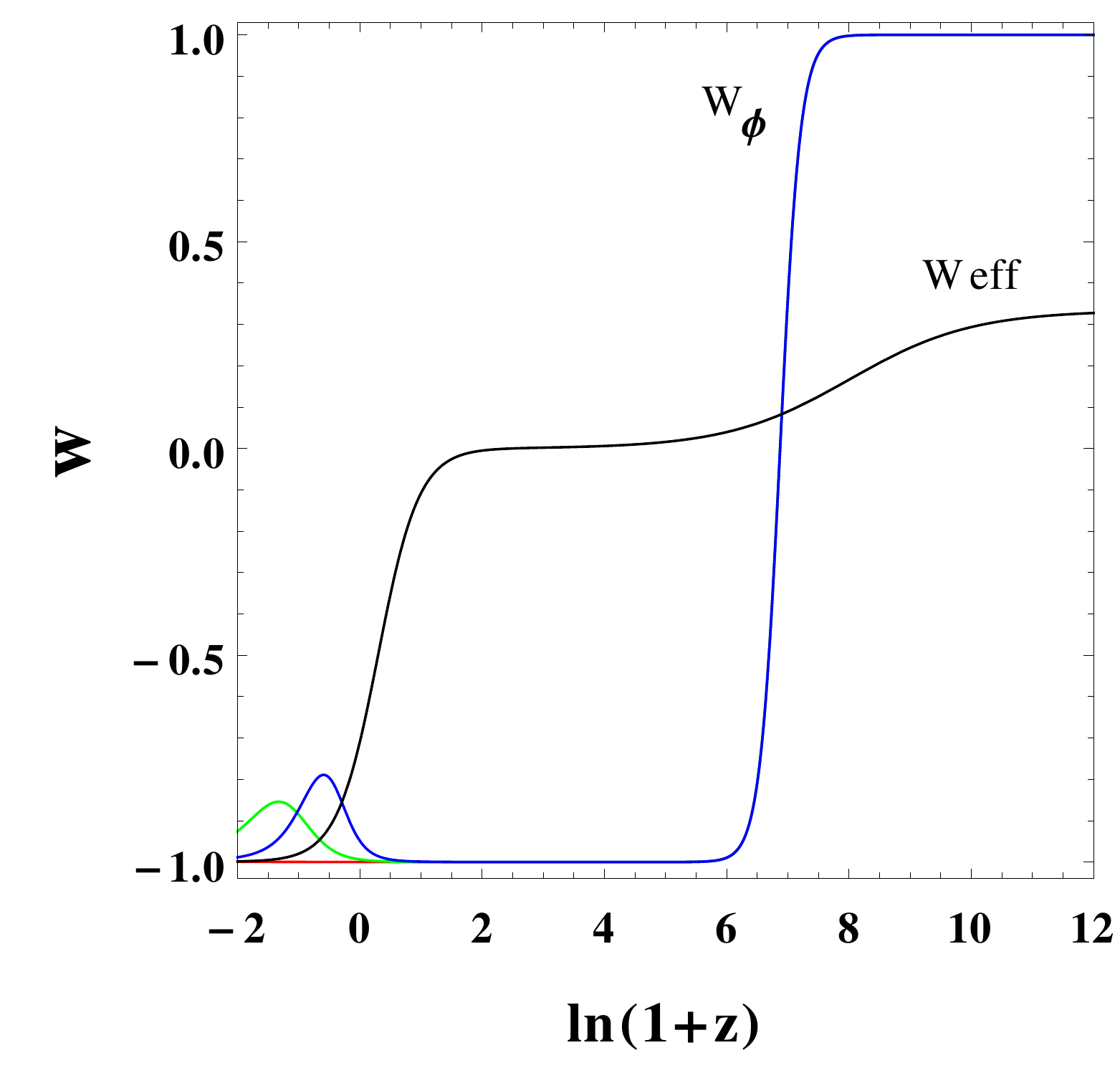} 
\includegraphics[width=1.9in,height=1.9in]{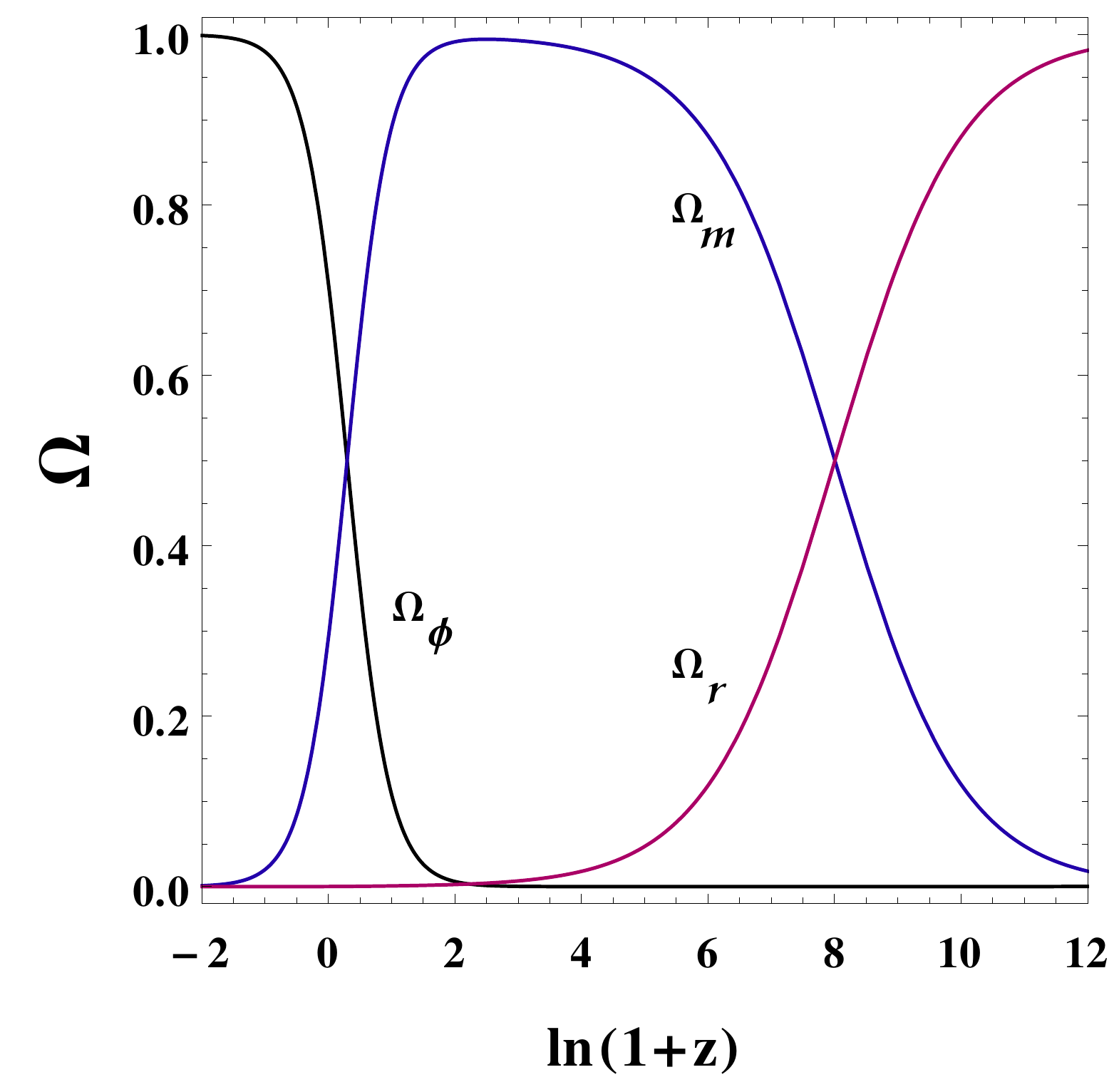} 
\caption{This figure corresponds to the potential (\ref{potential}) with $\epsilon=+1$. We choose the model parameters as $V_0=1.1 M_p^4$, $\alpha=1$ (red), $2.5$ (green) and $3.5$ (blue). From the plots one can realize that $\alpha$ works as a scaling parameter while the qualitative evolution of the model seems to be independent of $\alpha$. }
\label{fig:2a}
\end{figure*}
\begin{figure*}[tbp]
\centering
\includegraphics[width=1.9in,height=1.9in]{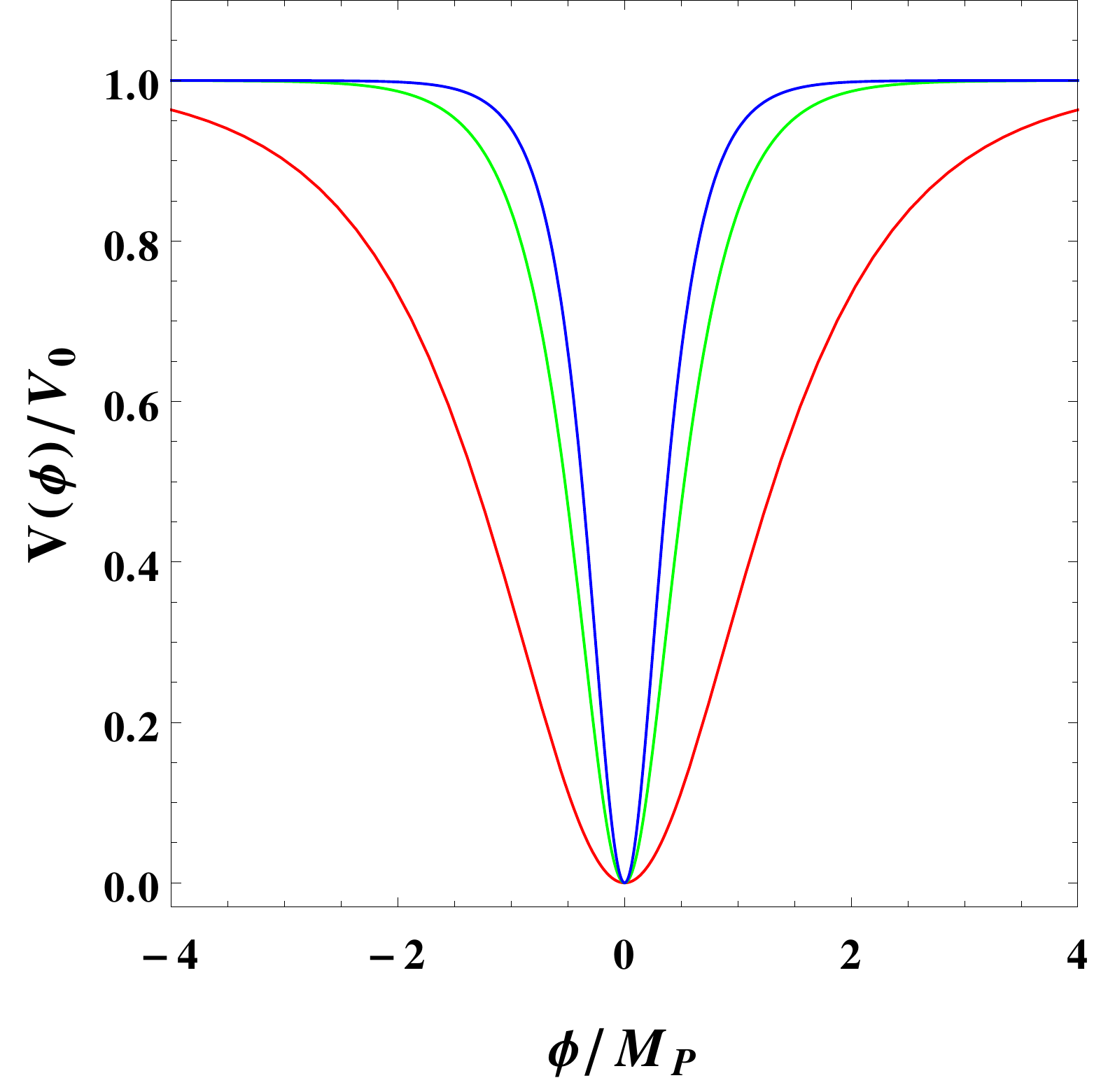} 
\includegraphics[width=1.85in,height=1.85in]{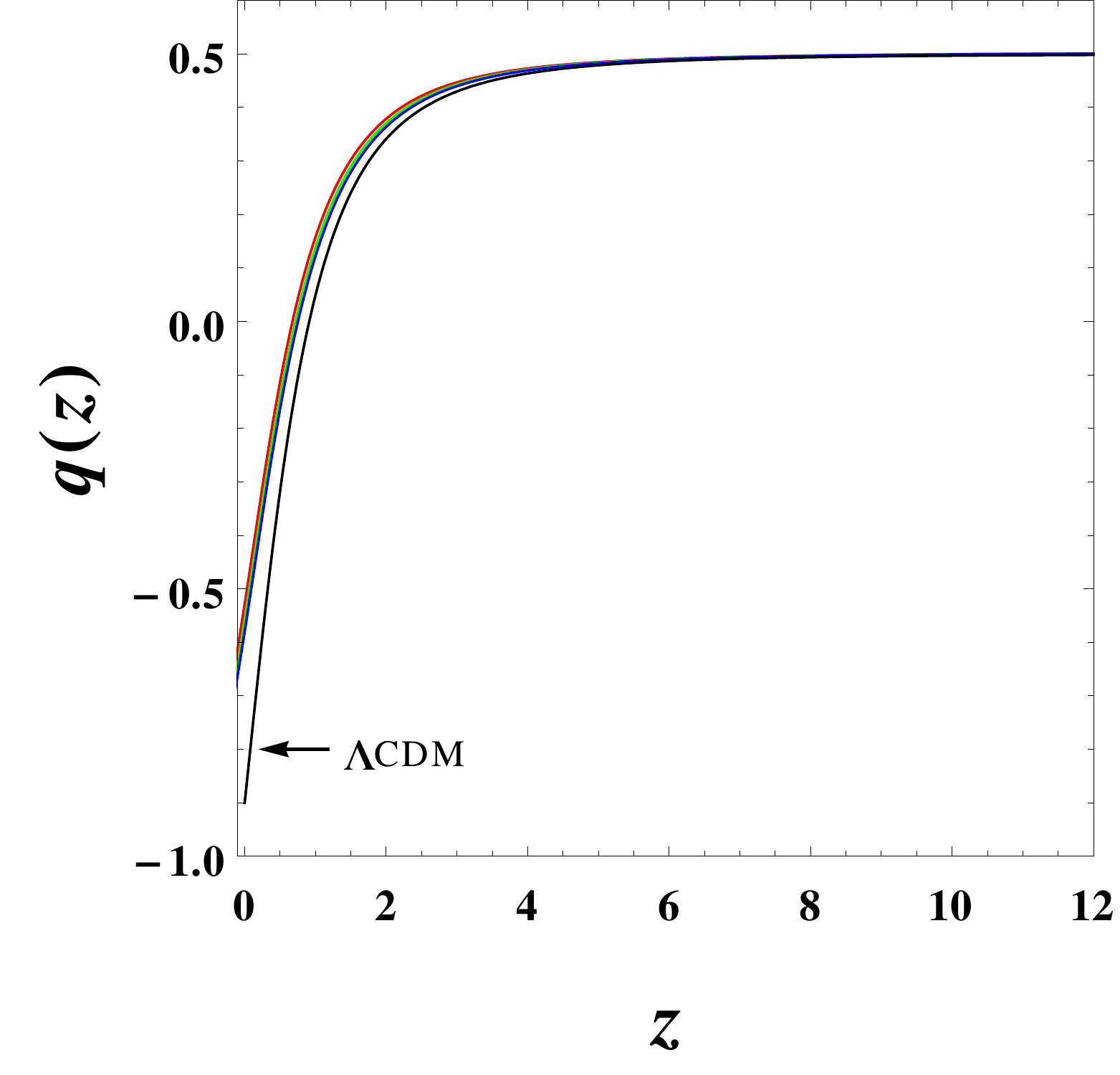} 
\includegraphics[width=1.9in,height=1.9in]{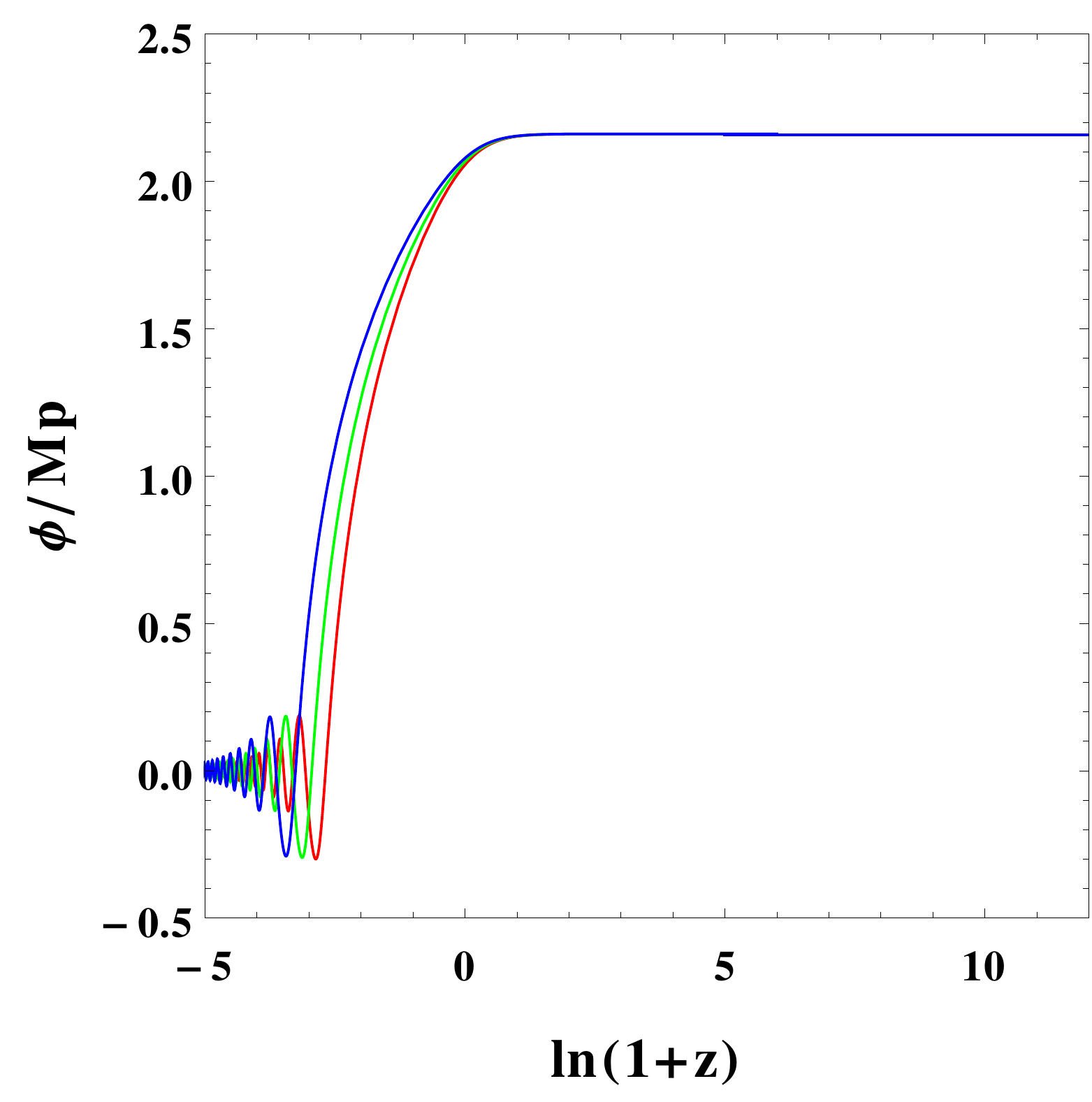} 
\includegraphics[width=1.9in,height=1.9in]{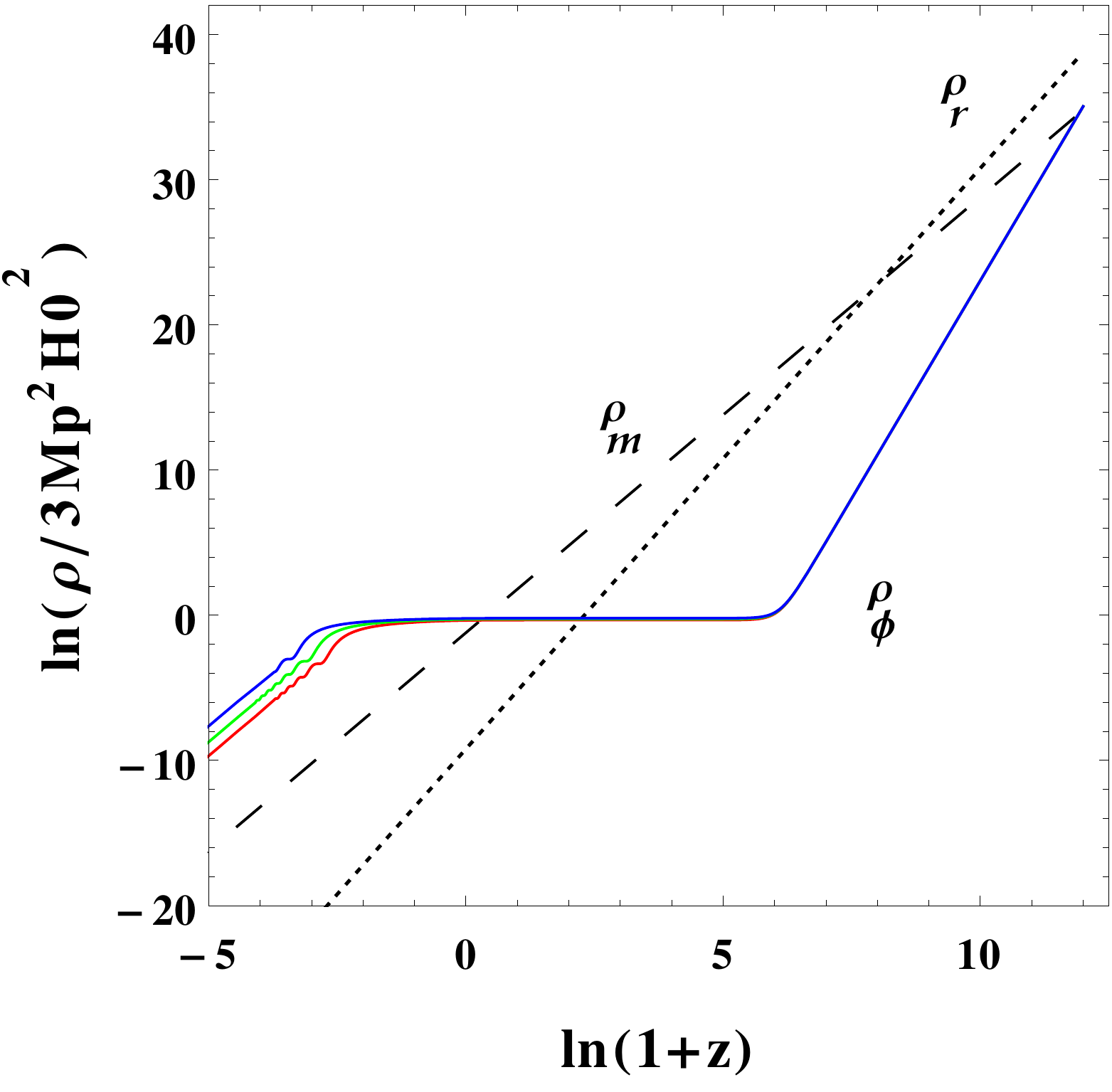}
\includegraphics[width=1.9in,height=1.9in]{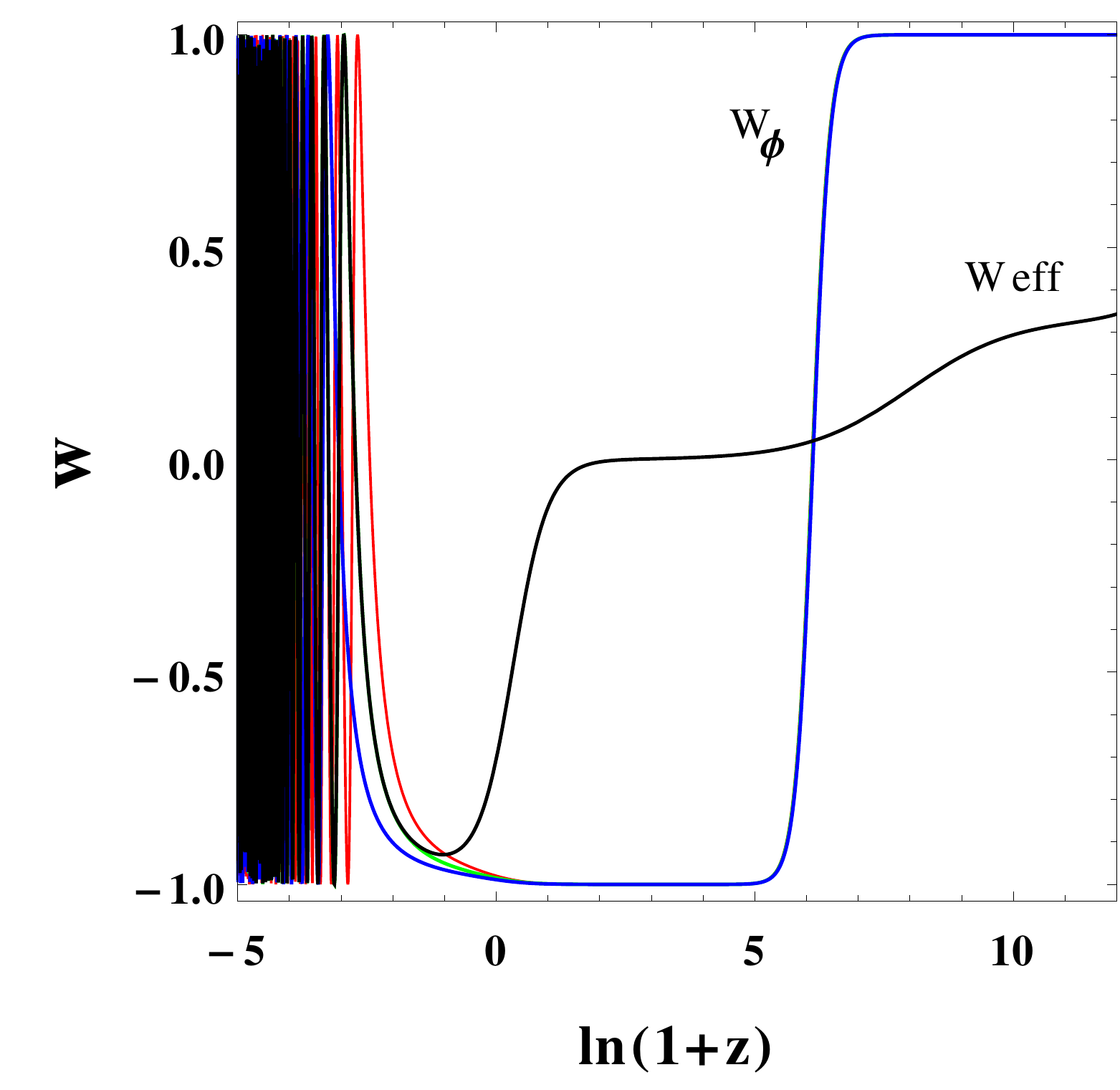} 
\includegraphics[width=1.9in,height=1.9in]{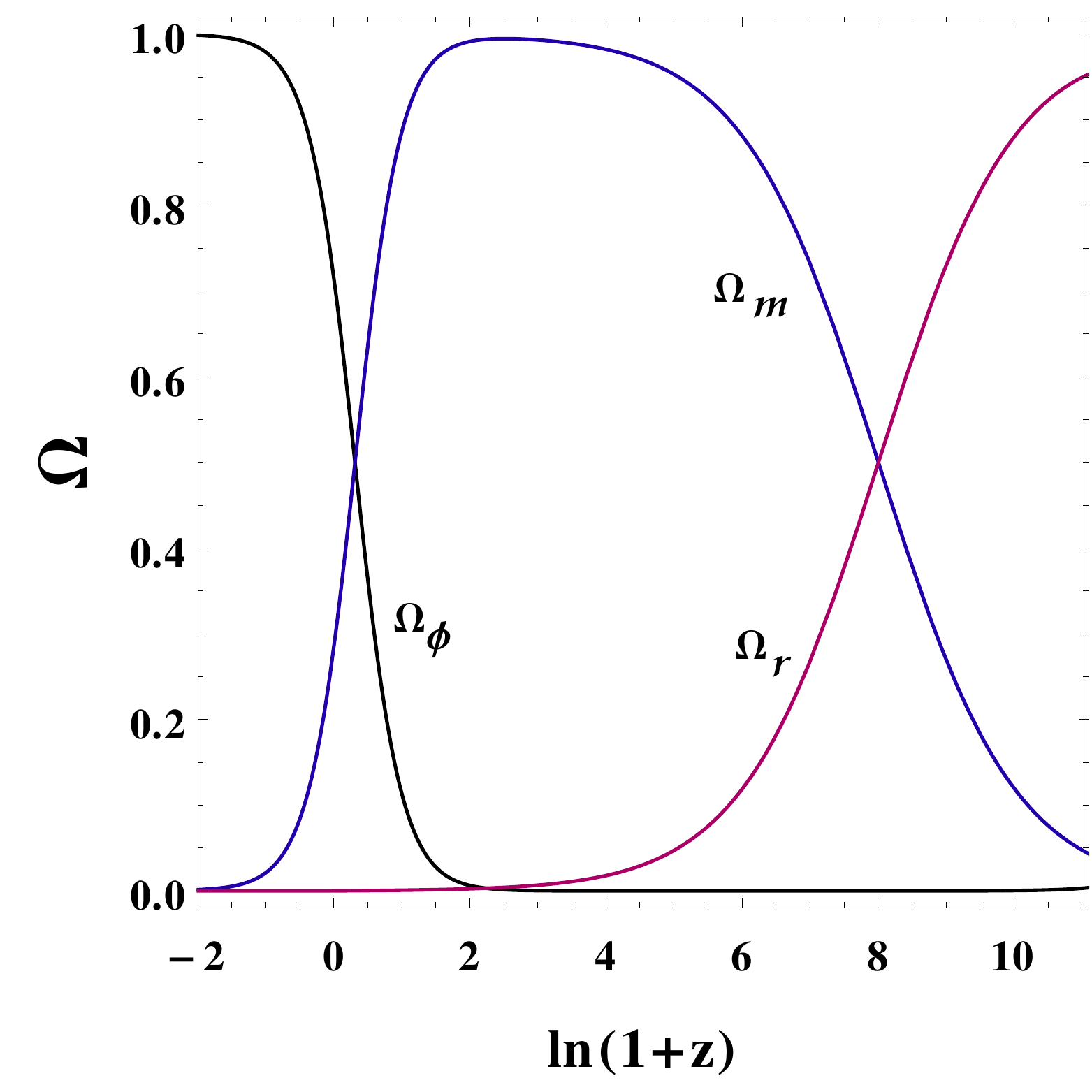} 
\caption{This figure represents the potential (\ref{potential}) with $\epsilon=-1$. We show different cosmological parameters, namely the deceleration parameter $q$, equation-of-state, $w_{\phi}$ and $w_{\rm eff}$ for different values of the characteristic parameter $\alpha$ such as  $\alpha=0.9$ (red), $1$ (green), $1.1$ (blue) and $V_0$=3 $M_p^4$ whereas the evolution of the density parameter (right) corresponds to $\alpha=1$. By looking the evolution of $w_{\phi}$, one may conclude that the $w_{\phi}$ exhibits thawing behavior at present epoch, and  oscillations in the future. }
\label{fig:2b}
\end{figure*}

In order to understand the nature of the potential and its impact on the evolution of the universe, in Fig. \ref{fig:2a} (for $\epsilon = 1$) and Fig. \ref{fig:2b} (for $\epsilon = -1$), we have displayed the  variations of $V (\phi)$ as well as the energy densities of different matter sources of the universe for different values of the free parameter $\alpha$. In the top panels of Fig. \ref{fig:2a} and Fig. \ref{fig:2b}, we have            showed the evolution of the potential (\ref{potential}) for $\epsilon = +1$ (top left of Fig. \ref{fig:2a}) and for $\epsilon = -1$ (top left of Fig. \ref{fig:2b}). One can clearly notice that here $\alpha$ does not play any significant role except that its scaling character. The evolution of the potentials are reversed due to $\epsilon = +1, -1$. But, both the  
potentials allow extremum values during its evolution. For this potential, we have numerically solved the conservation equation (\ref{K-G-eqn}) and displayed the evolution of the scalar field in the top right plots of both Fig. \ref{fig:2a} (for $\epsilon = +1$) and Fig. \ref{fig:2b} (for $\epsilon = -1$).
The energy density for this potential has been shown in the bottom left plot of Fig. \ref{fig:2a} (for $\epsilon = +1$) and Fig. \ref{fig:2b} (for $\epsilon = -1$). Here, we have some characteristic changes between these potentials as reflected from the bottom left plots of Fig. \ref{fig:2a} and Fig. \ref{fig:2b}. 
For $\epsilon = +1$, from the evolution of the energy densities, one can see that at late time, the energy density of the scalar field, is actually constant 
which corresponds to the cosmological constant, and this seems to be independent of $\alpha$. While on the other hand, a sharp fall of the scalar field energy density is observed which is also independent of $\alpha$. 

We have also presented the evolution of the deceleration parameter, $q$ (top middle plots of both Fig. \ref{fig:2a} (for $\epsilon = +1$) and Fig. \ref{fig:2b} (for $\epsilon = -1$)), equation of state for scalar field (also the effective equation of state) in the bottom middle plots of both Fig. \ref{fig:2a} (for $\epsilon = +1$) and Fig. \ref{fig:2b} (for $\epsilon = -1$)), and the evolution of the density parameters (bottom right plots of both Fig. \ref{fig:2a} (for $\epsilon = +1$) and Fig. \ref{fig:2b} (for $\epsilon = -1$)).  
We find that for both the models, a clear transition for the past decelerating phase to the current accelerating phase is suggested and $\alpha$, the quantifying parameter of the model does not play any significant role in this picture. The evolution of the deceleration parameter is found to be very close to its evolution for the $\Lambda$CDM model. Since $\alpha$ does not play any crucial role for the evolution of the scalar field model, thus, pick up $\alpha = 1$ and displays the evolution of the density parameters for the matter fluids in the top right (for $\epsilon = +1$) and bottom right (for $\epsilon = -1$) panels. However, we observe some different behavior when the equation of state of the scalar field model ($w_{\phi}$) and the effective equation of state of the models, $w_{\rm eff}$.      

For the potential with $\epsilon = +1$, we display the evolution of  $w_{\phi}$ and  $w_{\rm eff}$ in the middle plot of the top panel of Fig. \ref{fig:2a}. The evolution shows that at late time, the scalar field behaves like a cosmological constant and moreover, one can further notice that the effective equation of state also mimics a cosmological constant fluid.  For $\epsilon = -1$, similar quantities are displayed in the  middle plot of the bottom panel of Fig. \ref{fig:2b}. We find that although at present (that corresponds to $z= 0$), the equation of state for the scalar field, $w_{\phi}$ and the effective equation of state, $w_{\rm eff}$ are very close to the cosmological constant boundary, however, slight different characteristics in both $w_{\phi}$ and $w_{\rm eff}$ are observed for $z< 0$, corresponding to the evolution in the future. We find that for $z< 0$, oscillating features of both $w_{\phi}$ and $w_{\rm eff}$ are allowed.  From the analyses of the models, in particular, from the evolution of the energy density and the scalar field equation of state, $w_{\phi}$, we observe that both the models, namely,  Model 2a and Model 2b exhibit thawing behavior around the present epoch ($z \simeq 0$). However, Model 2b has an additional feature compared to Model 2a as follows: It shows oscillating nature through the equation of state, $w_{\phi}$. Thus, in summary, we see that although Model 2a exhibits only the thawing nature while Model 2b has both thawing and oscillating behavior.

\begin{figure*}[tbp]
\centering
\includegraphics[width=1.85in,height=1.85in,angle=0]{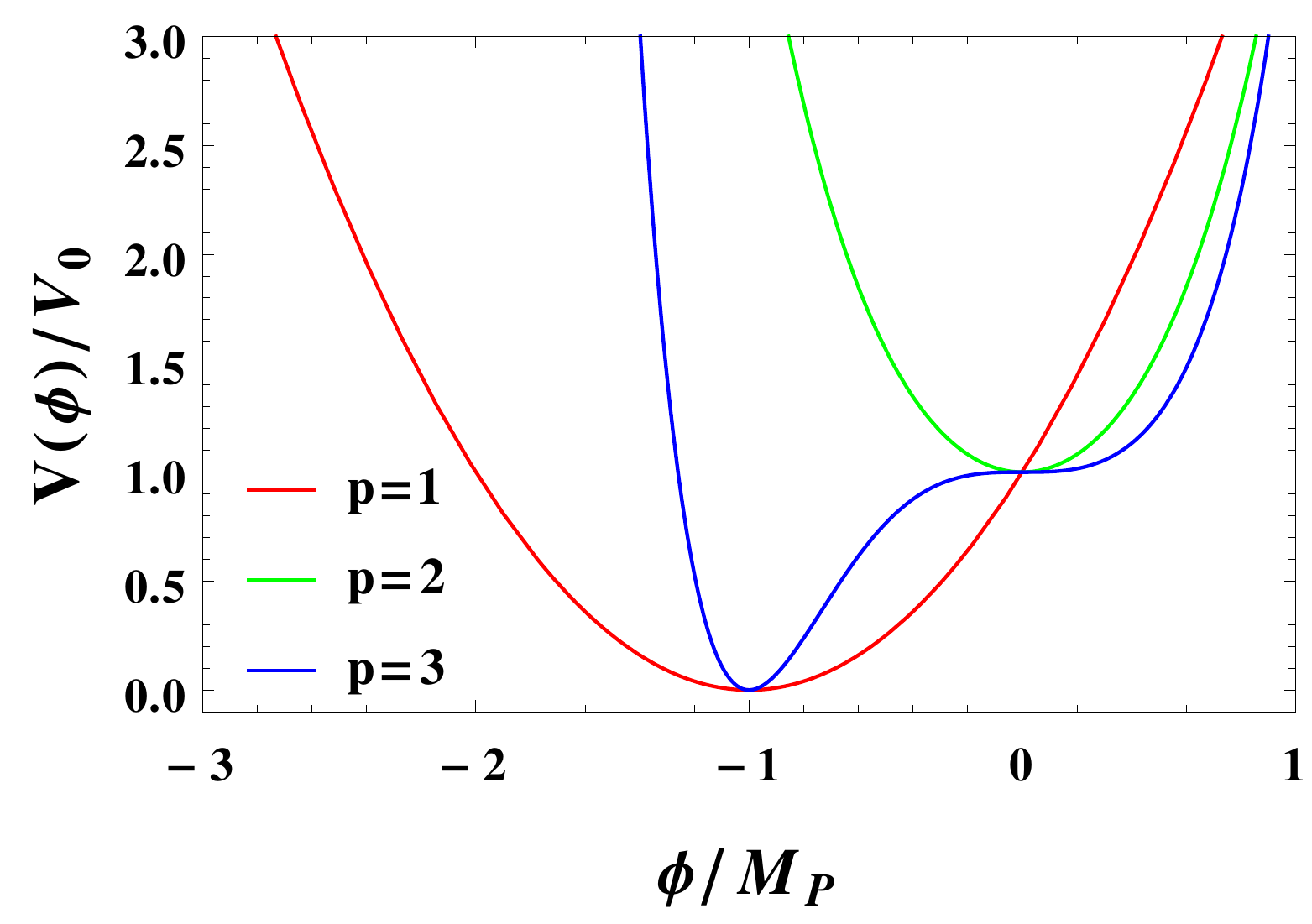}
\includegraphics[width=1.85in,height=1.85in,angle=0]{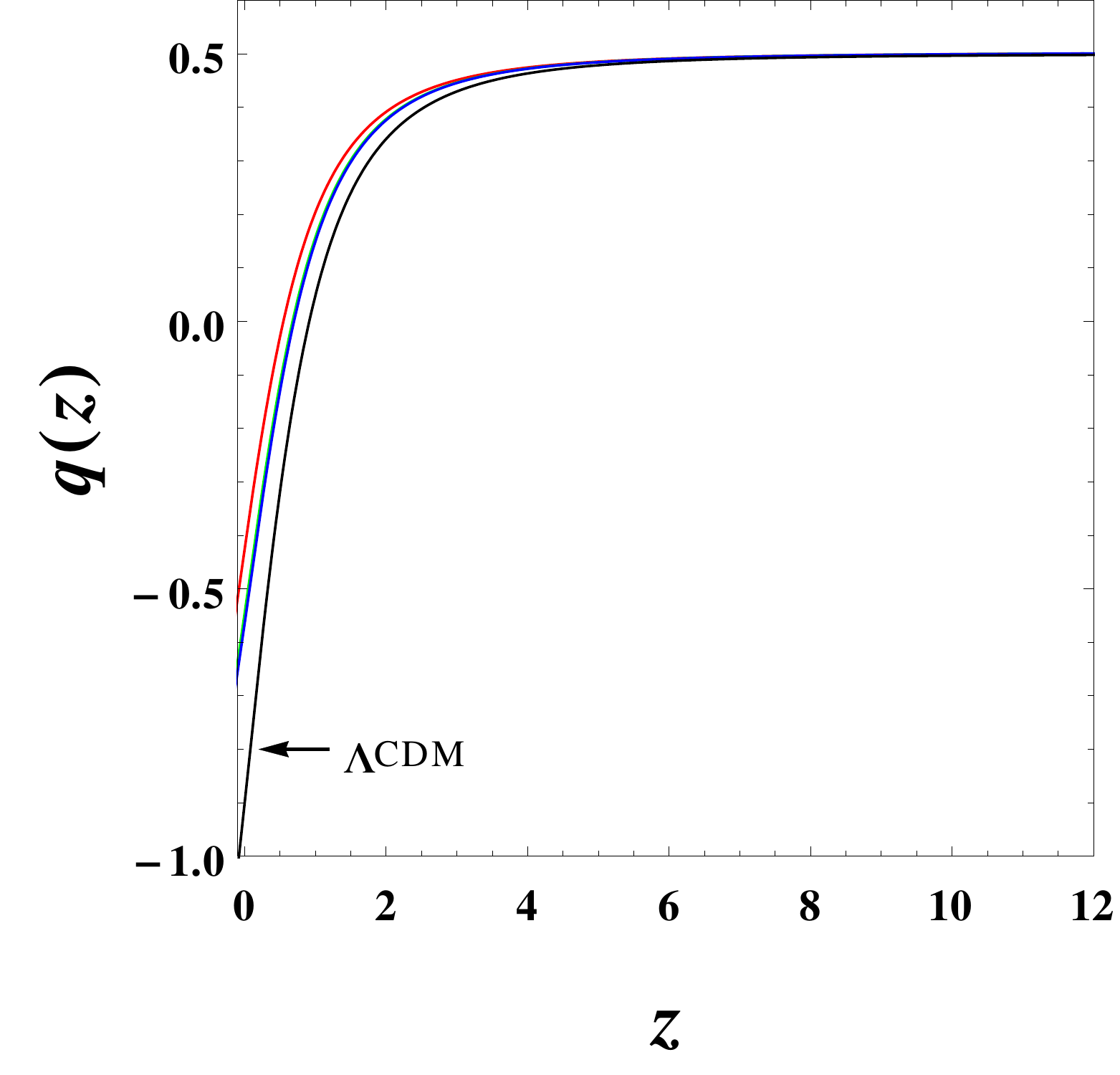}
\includegraphics[width=1.9in,height=1.9in,angle=0]{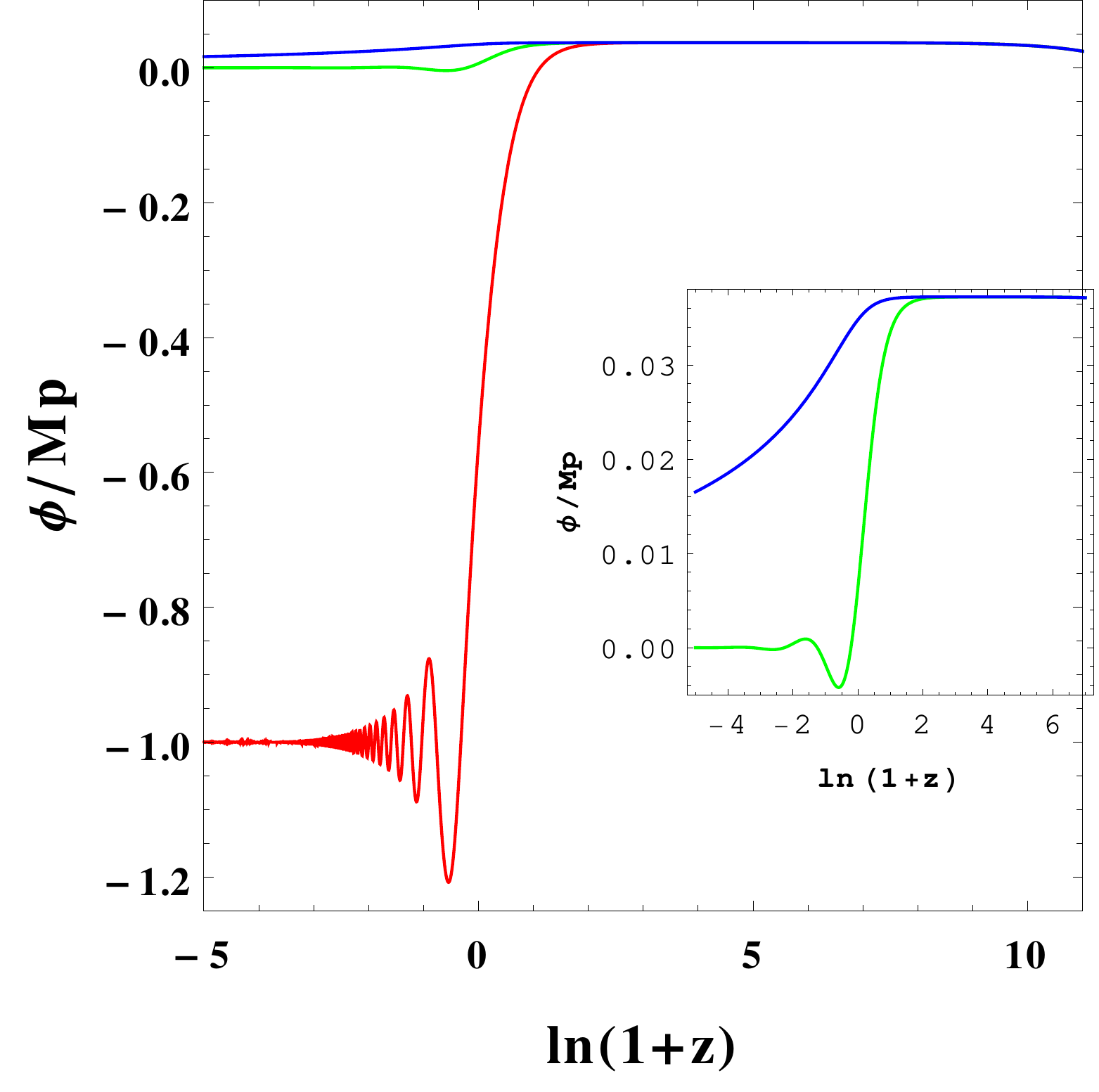}
\includegraphics[width=1.9in,height=1.9in,angle=0]{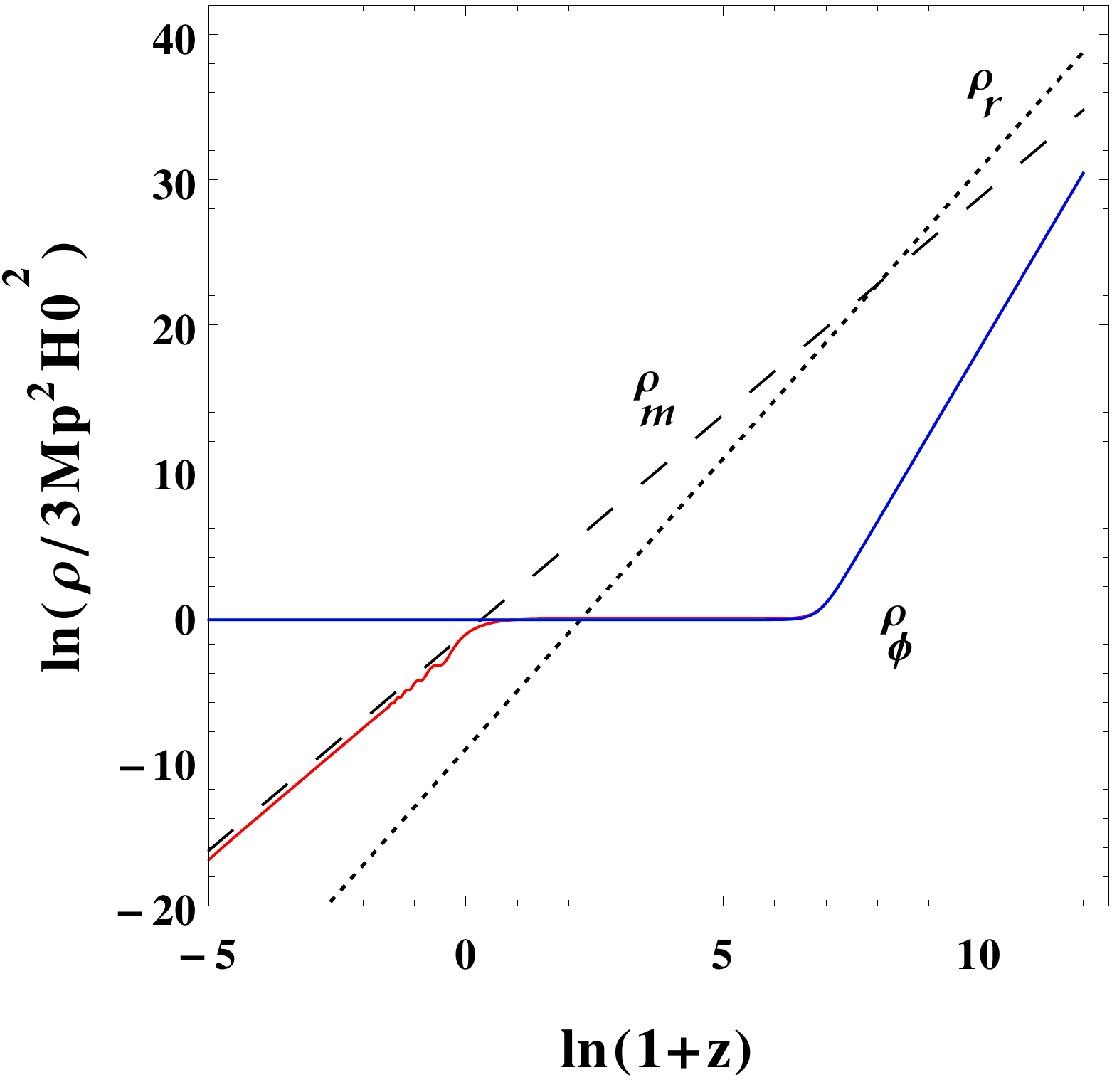}
\includegraphics[width=1.9in,height=1.9in,angle=0]{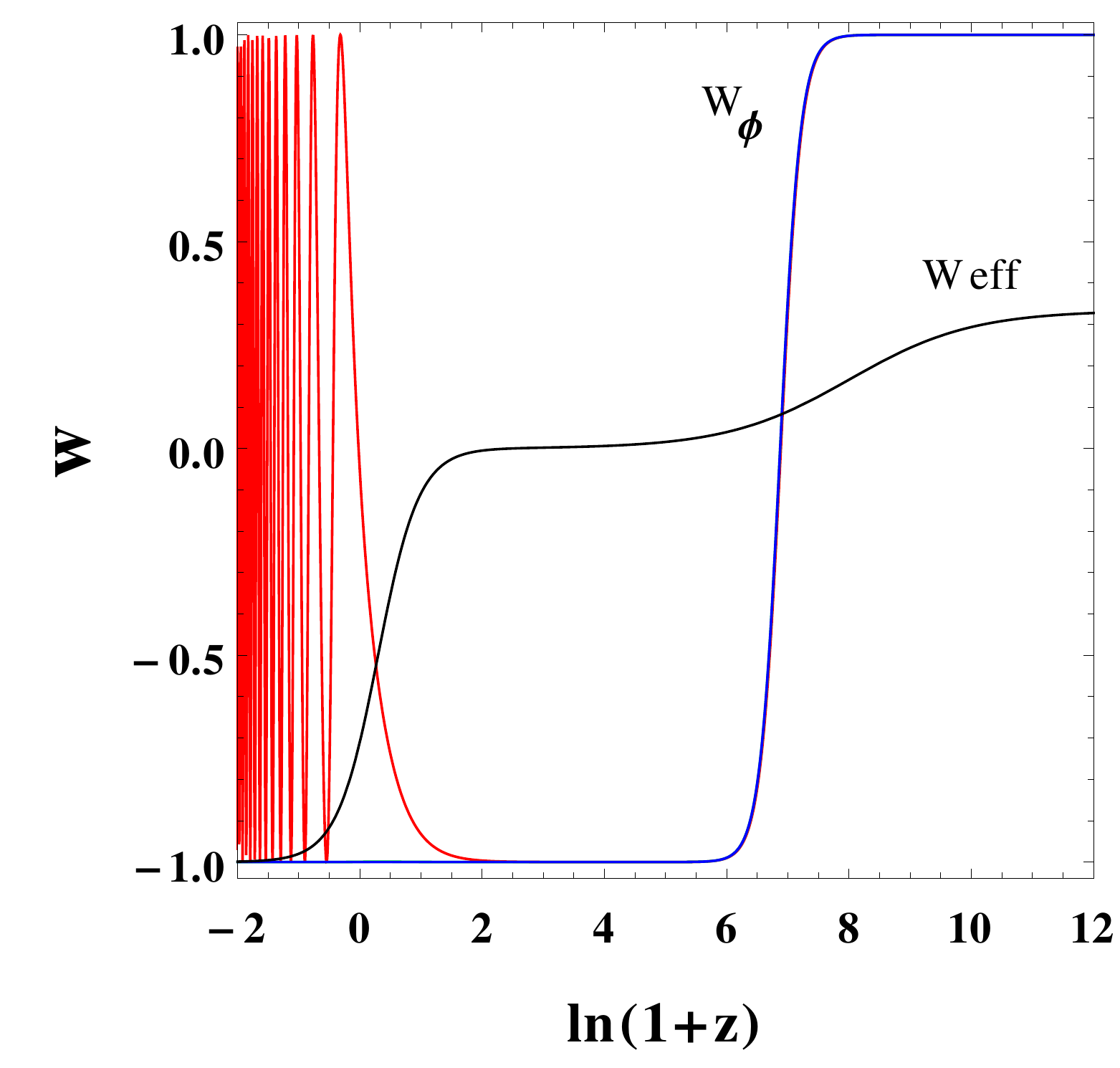}
\includegraphics[width=1.9in,height=1.9in,angle=0]{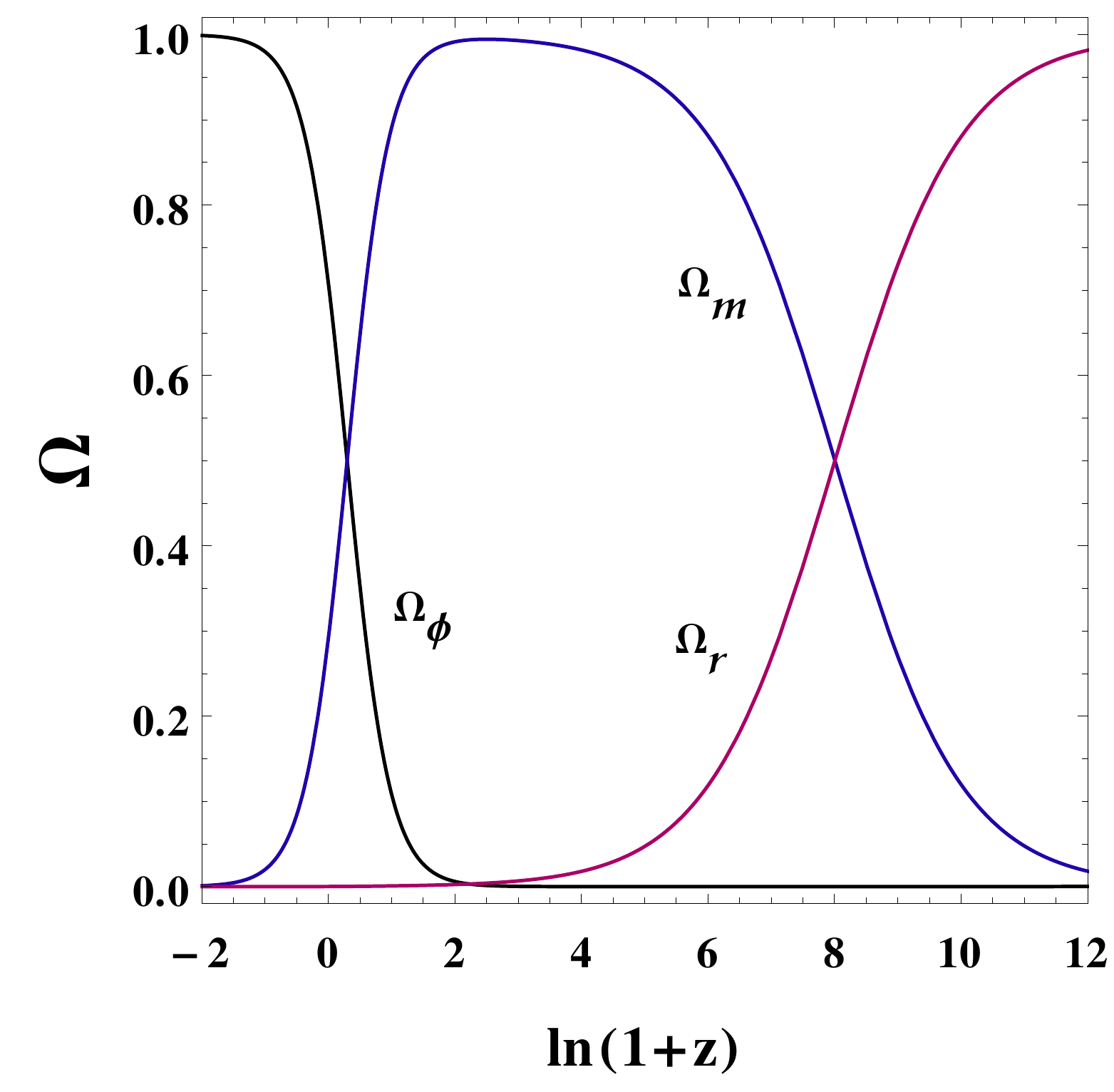}
\caption{The figure displays the evolution of various cosmological parameters associated for the potential (\ref{model3}) with $\delta=+1$. We have analyzed the model for different values of $p=1$ (red curves), $2$ (green curves), $3$ (blue curves). Let us note that for the trajectories of $w$ (middle plot of the lower panel) and the energy density for the scalar field (left plot of the lower panel), the blue curve is completely overlapped with the green curve. Let us further mention that throughout the analysis we have fixed $V_0$=2.2 $M_p^4$.}
\label{fig:3a}
\end{figure*}
\begin{figure*}[tbp]
\centering
\includegraphics[width=1.9in,height=1.9in,angle=0]{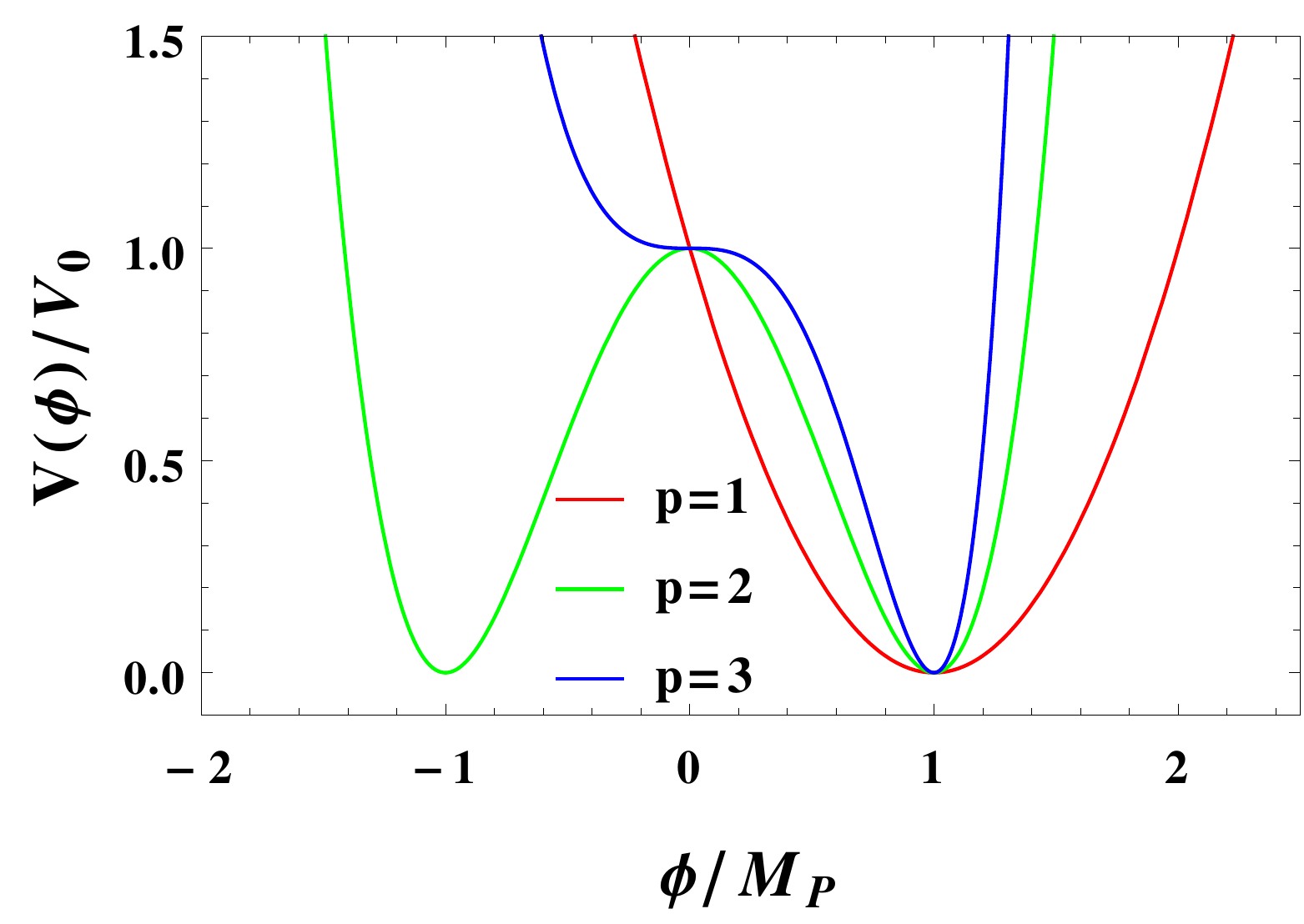}
\includegraphics[width=1.9in,height=1.9in,angle=0]{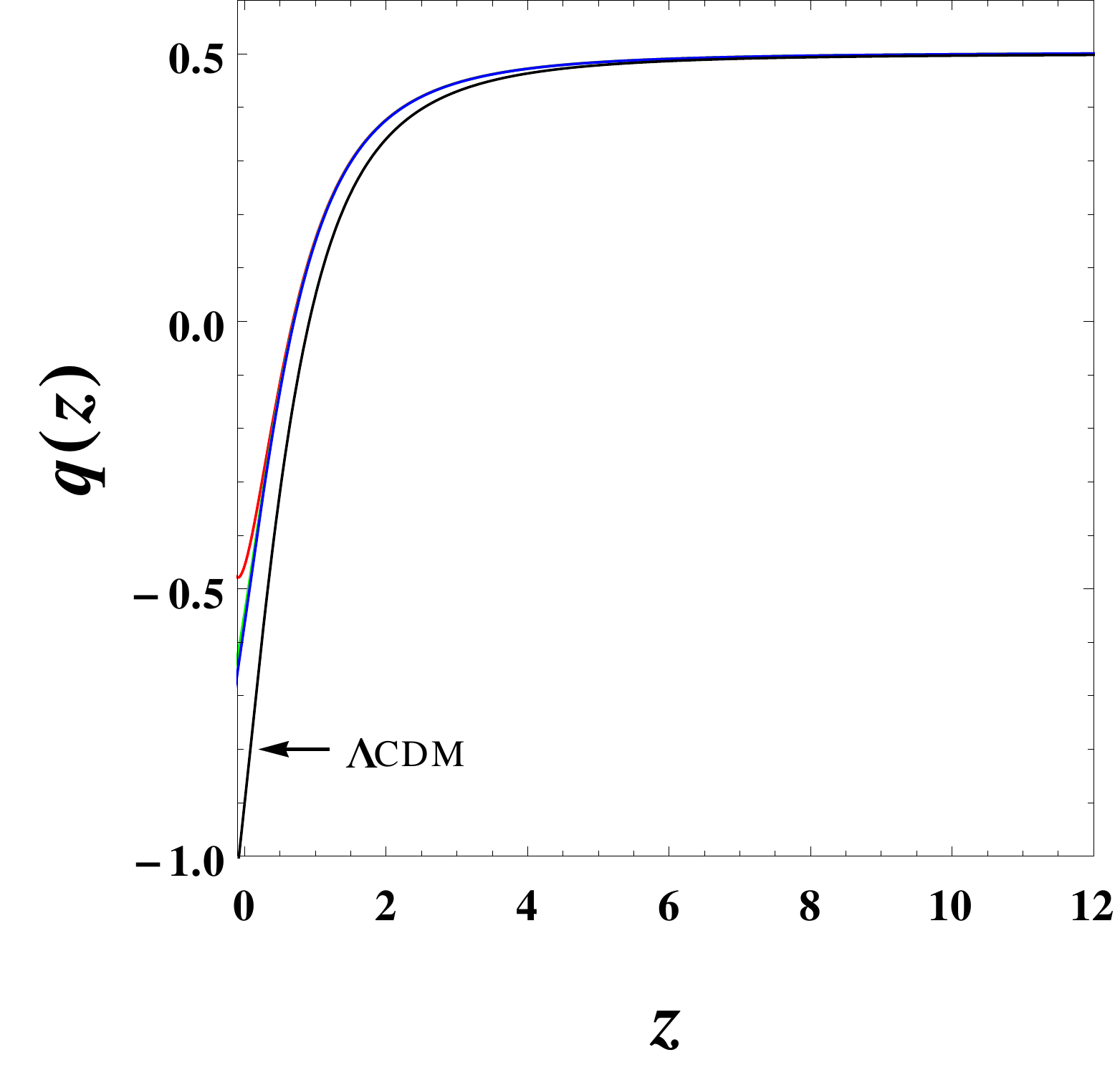}
\includegraphics[width=1.9in,height=1.9in,angle=0]{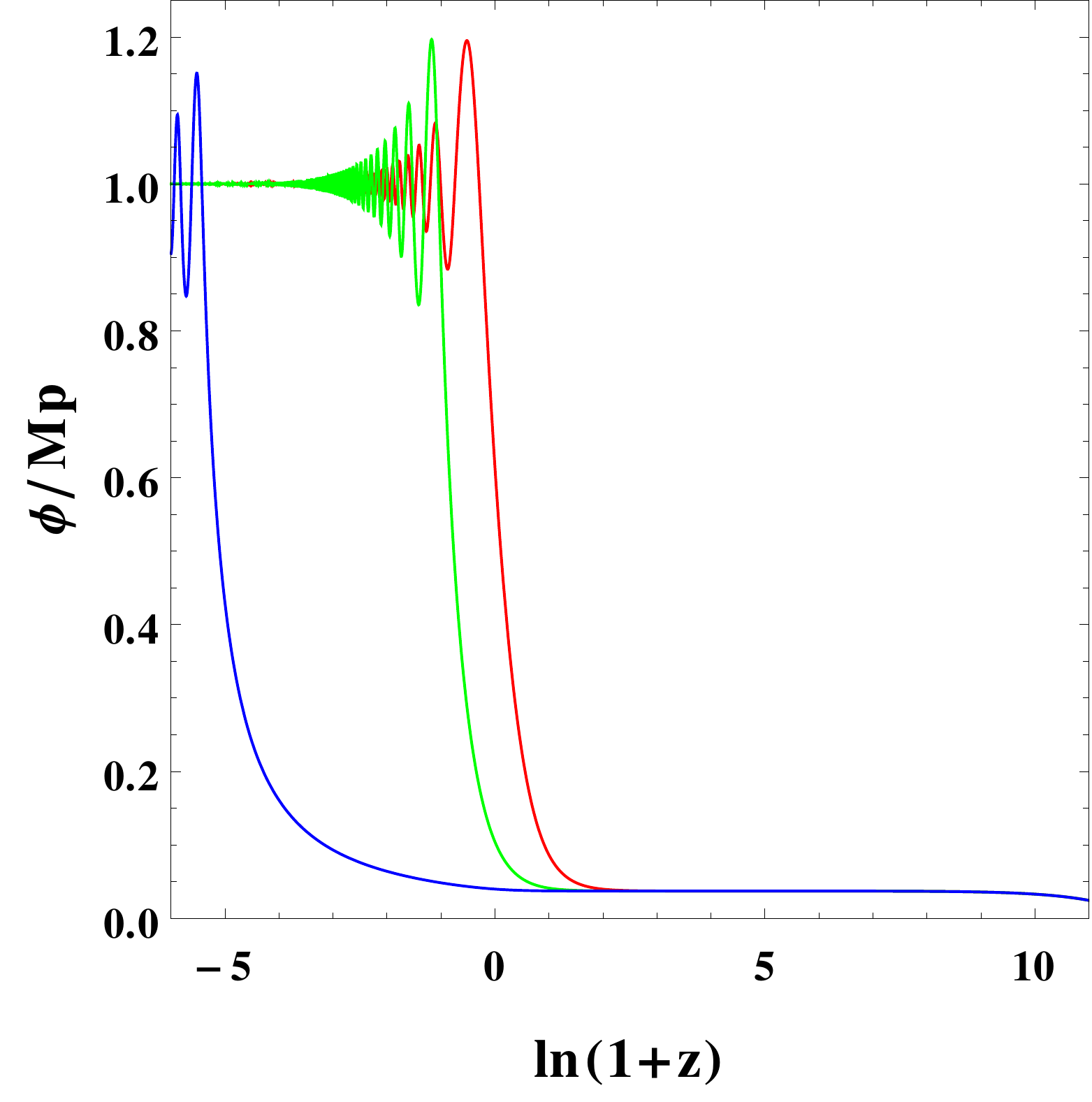}
\includegraphics[width=1.9in,height=1.9in,angle=0]{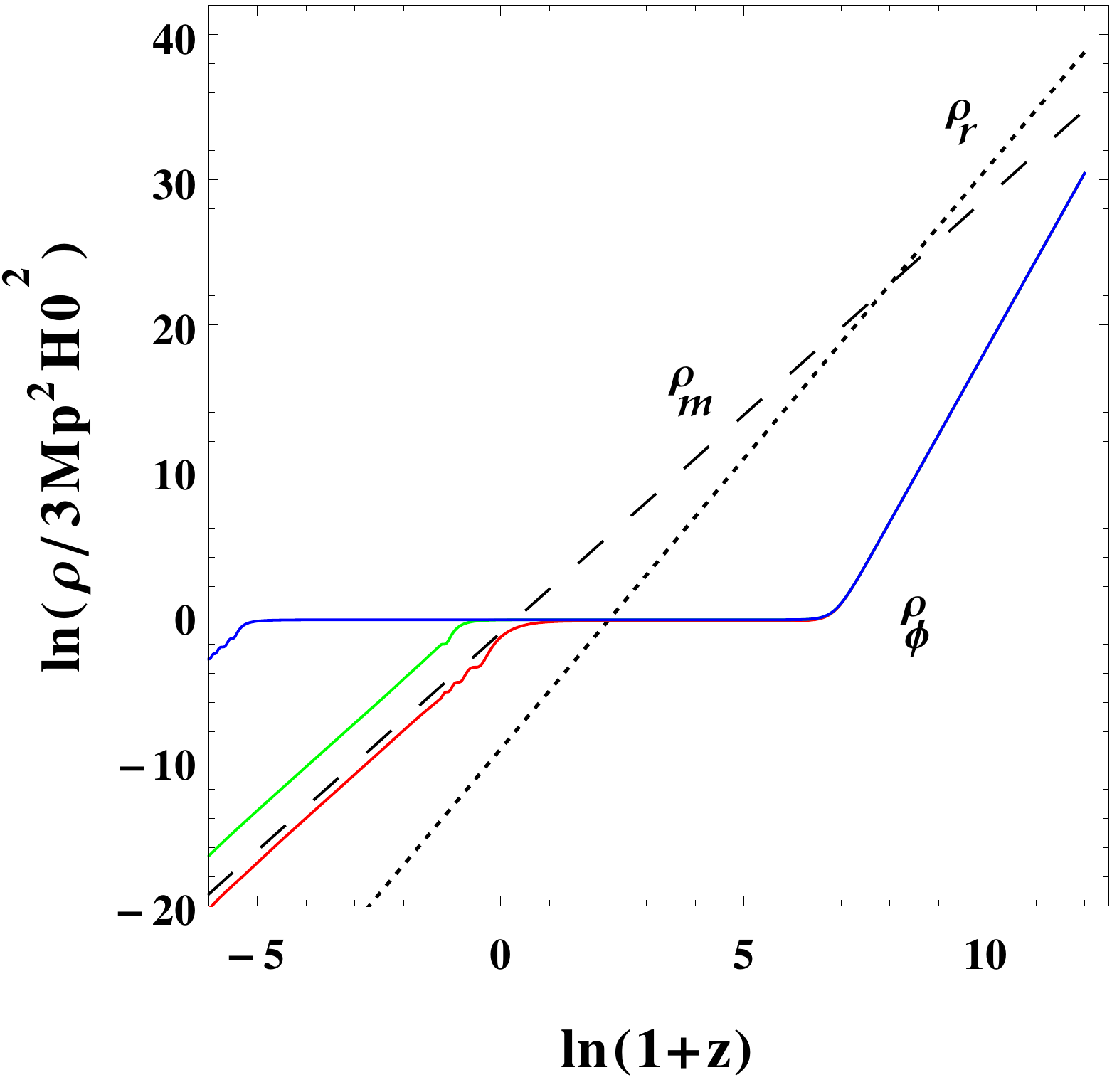}
\includegraphics[width=1.9in,height=1.9in,angle=0]{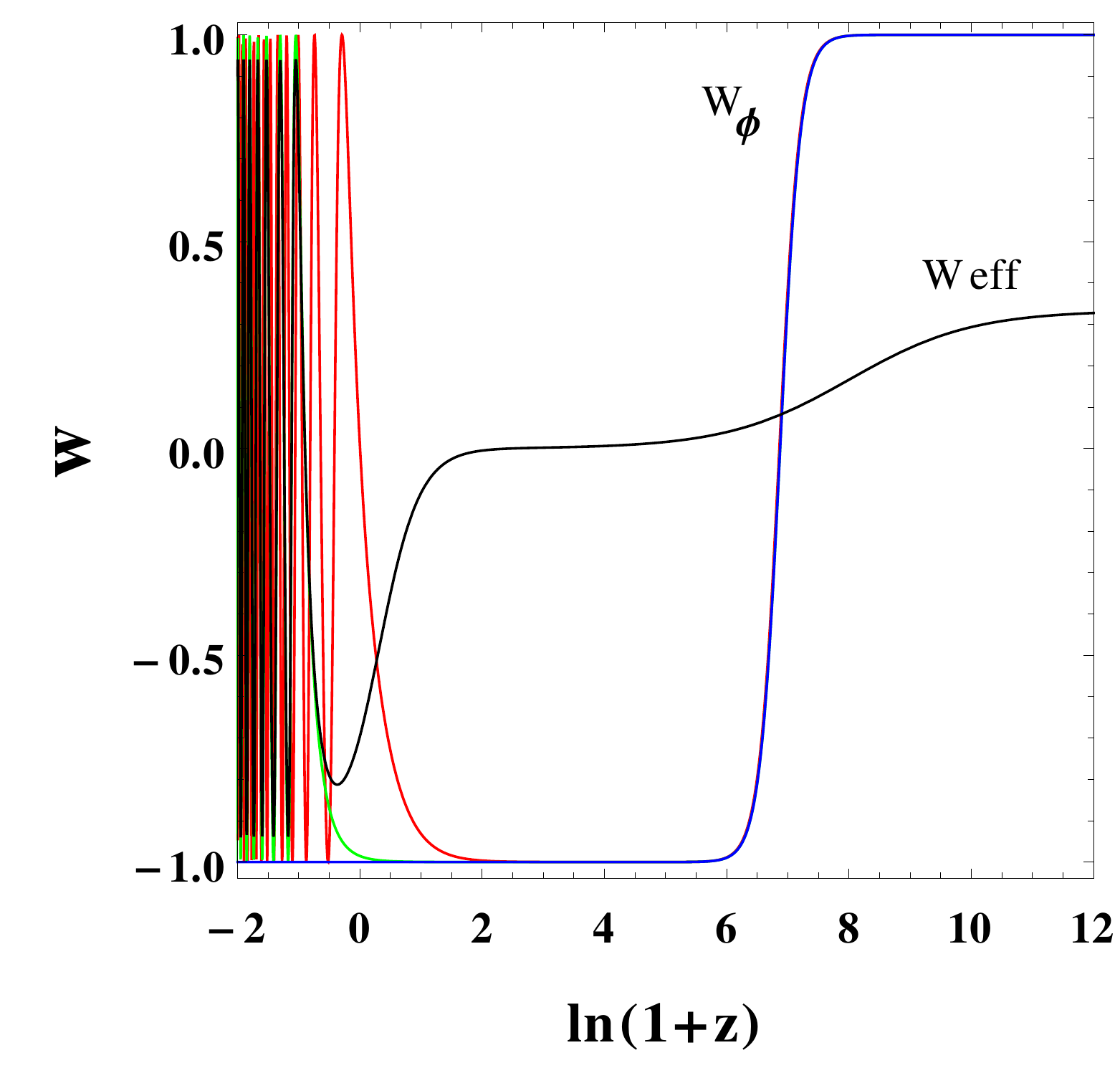}
\includegraphics[width=1.9in,height=1.9in,angle=0]{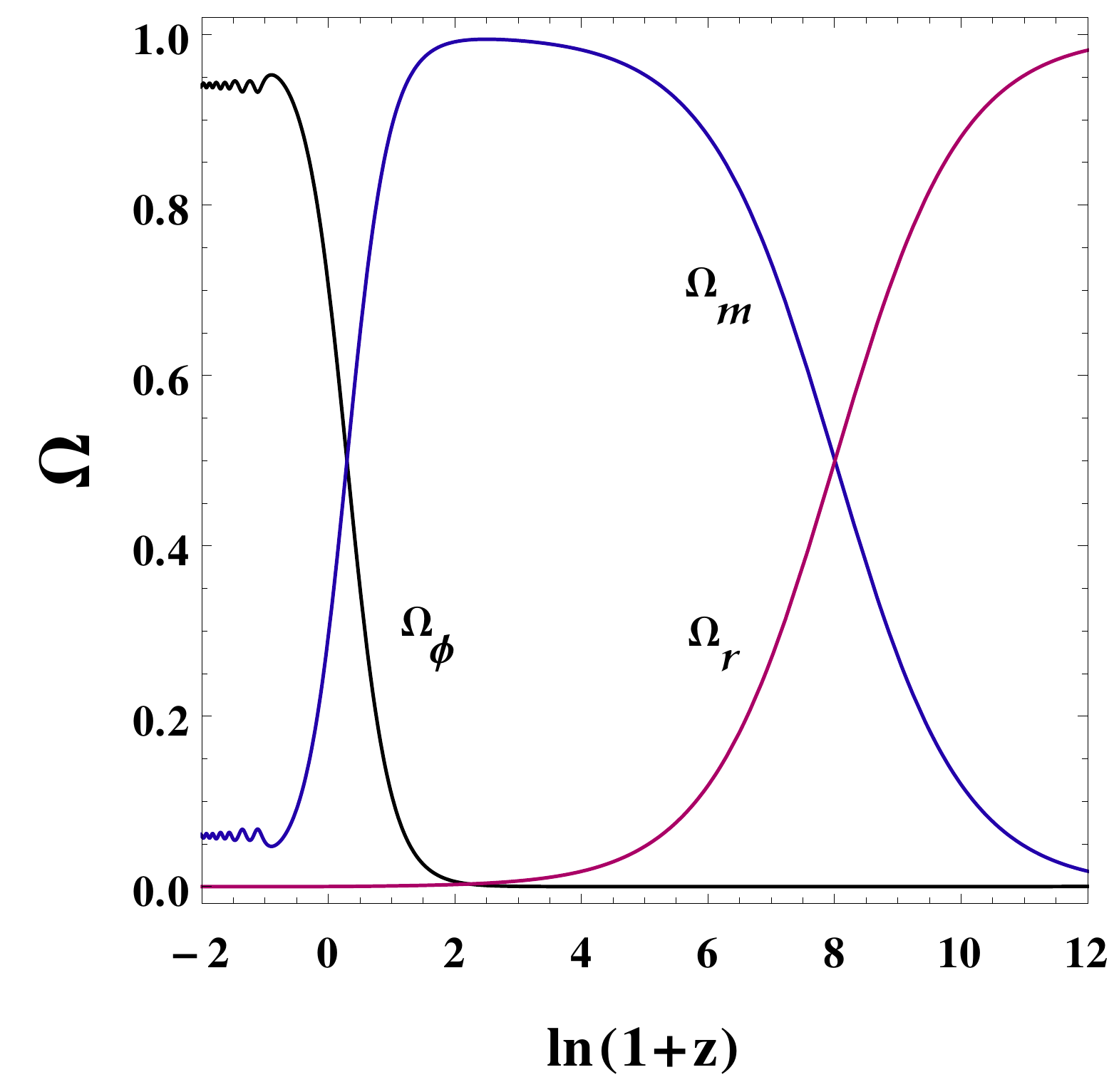}
\caption{The figure displays the evolution of various cosmological parameters associated for the potential (\ref{model3}) with $\delta= -1$. We have analyzed the model for different values of $p=1$ (red curves), $2$ (green curves), $3$ (blue curves). We note that throughout the analysis we have fixed $V_0$=2.2 $M_p^4$. Similarly, in the plot for the deceleration parameter, $q$ (middle panel of the upper plot) the blue and green curves are almost indistinguishable from one another.}
\label{fig:3b}
\end{figure*}

\subsection{Model 3} 
Finally, we consider the last model in this series: 

\begin{eqnarray}\label{model3}
V (\phi) = V_0 \left[ 1+ \delta \left(\frac{\phi}{M_P}\right)^{p}\right]^2 ,
\end{eqnarray}
that can be found in \cite{Antusch:2016con}. Here, in $\delta$, $p$, $V_0$  are the real constants and $M_P = (8 \pi G)^{-1/2}$.  Let us note that the parameter $\delta \in \mathbf{R}$ quantifies the deviation of the model from the constant potential $V = V_0$. Here, we classify Model 3 in the following manner. We denote Model 3 with $\delta \geq 0$ as Model 3a while by Model 3b we mean Model 3 with  $\delta \leq 0$. Although the potential (\ref{model3}) is one of the representatives of the collection of a large number of phenomenological potentials in the literature, however, it has a special interest in the context of cosmological observations. For $\delta  =-1$, the potential (\ref{model3}) represents hilltop inflation which is favoured by recent CMB observations. Moreover, one can further notice that for $\phi/M_p \ll 1$, the model (\ref{model3}) actually recovers the constant potential independently of the sign of $\delta$. 
Now, following  similar trend, we have numerically solved the conservation equation (\ref{K-G-eqn}) for this potential and discuss various quantities graphically in Fig. \ref{fig:3a} and Fig. \ref{fig:3b}. Let us describe the qualitative behaviour of Model 3a and Model 3b. 

The Fig. \ref{fig:3a} completely describes the qualitative evolution of Model 3a.  The Fig. \ref{fig:3a} has two panels, namely, the upper and lower panels where each panel contains three different plots.  In the upper panel of Fig. \ref{fig:3a}, we show the evolution of the potential with respect to the scalar field $\phi$ (upper left panel of Fig. \ref{fig:3a}) , the deceleration parameter as a function of the redshift (upper middle panel of Fig. \ref{fig:3a}) and the scalar field itself with respect to $\ln (1+z)$ (upper right panel of Fig. \ref{fig:3a}).  In all three plots, we have used three different values of $p$, namely, $p =1$ (red curves), $p =2$ (green curves) and $p= 3$ (blue curves). From the evolution of the potential we find that that model allows an extremum during the evolution of the universe. From the upper middle panel (Fig. \ref{fig:3a}), a clear transition of the deceleration parameter from its past decelerating phase to the present accelerating phase of the universe is clearly seen and this evolution is quite similar to that of the $\Lambda$CDM cosmology. The scalar field appears to be sharply decrease (upper right panel of Fig. \ref{fig:3a}) with the evolution of the universe. The lower panel of Fig. \ref{fig:3a} has three different plots, namely, the evolution of the energy densities of the component fluids (lower left panel of Fig. \ref{fig:3a}), the equation of state of the scalar field model and the effective equation of state (lower middle panel of Fig. \ref{fig:3a}), the evolution of the density parameters of the component fluids (lower left panel of Fig. \ref{fig:3a}). From the evolution of different energy densities, we see that at late time, $\rho_{\phi}$ does not seem to evolve with the cosmic time, so effectively, $\rho_{\phi} $ behaves like a cosmological constant fluid. From the equation of state of both scalar field as well as the effective fluid, we see that at late time, both of them has a converging nature to `$-1$'.

For Model 3b, we have described various cosmological parameters in Fig. \ref{fig:3b}. Similarly, we have made two panels $-$ the upper panel containing three distinct plots, namely, the evolution of the potential (upper left panel of Fig. \ref{fig:3b}), deceleration parameter (upper middle panel of Fig. \ref{fig:3b}), evolution of the scalar field (upper right panel of Fig. \ref{fig:3b}) $-$ 
and one the lower panel containing another three distinct plots, namely, the evolution of the energy densities of different fluids (lower left panel of Fig. \ref{fig:3b}), equation of state for the scalar field as well as for the effective fluid (lower middle panel of Fig. \ref{fig:3b}) and the density parameters of the fluids (lower right panel of Fig. \ref{fig:3b}). The evolution of this model (i.e., Model 3b) has similarities to Model 3a.

\section{Observational data, fitting methodology and the results}
\label{sec-data}

In order to constrain the model, we use several cosmological data with latest origin. In the following we describe the observational data and then we describe the analysis of the model. 
\begin{itemize}

\item {\bf Cosmic Microwave Background}: The data from cosmic microwave background (CMB) radiation are
very effective for the analyses with dark energy models. In this work we make use of the CMB temperature and polarization anisotropies, as well as their cross-correlations, from the Planck \cite{Adam:2015rua}. To be precise, our measurements include the combination of high- and low-$\ell$ TT likelihoods in the overall multipole range $2\leq \ell \leq 2508$, and the combination of high- and low-$\ell$ polarization likelihoods~\cite{Aghanim:2015xee}. To constrain the present quintessence models, we use the freely available Planck likelihood~\cite{Aghanim:2015xee}, which marginalizes over several nuisance parameters describing the uncertainties related to the calibration, residual foreground contamination, residual beam-leakage, and several more.

\item {\bf Baryon Acoustic Oscillations}:  
We have used four particular measurements of the Baryon Acoustic Oscillations (BAO) distance measurements, namely, the data from the 6dF Galaxy Survey (6dFGS) ($z_{\emph{\emph{eff}}}=0.106$)~\cite{Beutler:2011hx}, the Main Galaxy Sample of Data Release 7 of the Sloan Digital Sky Survey (SDSS-MGS) ($z_{\emph{\emph{eff}}}=0.15$)~\cite{Ross:2014qpa}, the CMASS sample of Data Release 12 (DR12) of the Baryon Oscillation Spectroscopic Survey (BOSS) ($z_{\mathrm{eff}}=0.57$)~\cite{Gil-Marin:2015nqa} and finally the LOWZ sample ($z_{\mathrm{eff}}=0.32$) from the same BOSS data release (DR12)~\cite{Gil-Marin:2015nqa}.

\item {\bf Redshift Space Distortion}: 
We also consider the Redshift Space Distortion (RSD) data in our analysis. In particular, we consider RSD data from the CMASS sample ($z_{\rm eff} = 0.57$) \cite{Gil-Marin:2016wya}  and the LOWZ sample ($z_{\rm eff} = 0.32$) \cite{Gil-Marin:2016wya}. Let us note that when we use these two RSD data in the analysis, the BOSS DR12 results are not used in order to avoid
double-counting of data.

\item {\bf Supernovae Type Ia}: The Supernovae Type Ia (SNIa) data are the first observational indicators for an accelerating universe, that means, for the existence of dark energy in the universe sector. In this work we have used the  joint light curve analysis (JLA) sample \cite{Betoule:2014frx}, an updated list of SNIa  data, distributed over the redshift range $z \in [0.01, 1.30]$ with $740$ number of data points. The $\chi^2$ function for this data is given by

\item {\bf Cosmic chronometers}:
Finally, our analyses also include the Hubble 
parameter measurements from the most old and passively evolving galaxies, known as cosmic chronometers (CC). The measurements from the cosmic chronometers are believed to be the most potential and model independent measurements and hence, they could able to provide the most important information about the expansion history of the universe.  The total number of data points spanned in $0< z < 2$ are $30$ (see Table 4 of Ref. \cite{Moresco:2016mzx}), see the details of the data and more discussions 
in \cite{Moresco:2016mzx}.

\end{itemize}

In order to constrain the cosmological parameters, we use the markov chain monte carlo package {\it cosmomc} \cite{Lewis:2002ah}, an efficient 
sampling method. The convergence of the MCMC chains are assessed through the Gelman-Rubin statistics $R-1$~\cite{Gelman-Rubin}. 
In addition to that we also include the Bayesian model comparison which provides a goodness of fit of the model under consideration compared to some reference model, usually taken to be the $\Lambda$CDM cosmological model. Now, for the statistical analysis, we consider the following parameter space 

\begin{equation*}
P\equiv \{\Omega _{b}h^{2},\Omega _{c}h^{2},\Theta _{S}, \tau, 
,n_{s},\log [10^{10}A_{S}], \mu \},
\end{equation*}
where the above symbols have the following meanings: $\Omega _{b}h^{2}$, $\Omega _{c}h^{2}$ are respectively the density of
baryons and the dark matter; $\Theta _{S}=100\theta _{MC}$ is
the ratio of sound horizon to the angular diameter distance; $\tau $ is the optical depth; $n_{s}$ is the scalar spectral index; $%
A_{s} $ is the amplitude of the initial power spectrum. Here, $\mu$ is the free parameter coming from the scalar field models under consideration. That means when we consider Model 1, $\mu = \beta$, when Model 2 is considered, $\mu = \alpha$ and for Model 3, $\mu  = \delta$. So, overall, during the statistical analyses, we consider a seven dimensional parameter space. Finally, in Table \ref{tab:priors} we enlist the priors on the model parameters for our analysis.

\begin{table}
\begin{center}
\begin{tabular}{ccccccccccccccc}
\hline
Parameter  &~ Prior (Model 1) &~ Prior (Model 2a) &~ Prior (Model 2b) &~ Prior (Model 3a) &~ Prior (Model 3b)\\
\hline 
$\Omega_{b} h^2$    & $[0.005,0.1]$  & $[0.005,0.1]$ & $[0.005,0.1]$ & $[0.005,0.1]$ & $[0.005,0.1]$ & \\
$\Omega_{c} h^2$    & $[0.01,0.99]$ & $[0.01,0.99]$ & $[0.01,0.99]$ & $[0.01,0.99]$ & $[0.01,0.99]$ \\
$\tau$  & $[0.01,0.8]$ &  $[0.01,0.8]$ &  $[0.01,0.8]$ &  $[0.01,0.8]$ &  $[0.01,0.8]$ \\
$n_s$    & $[0.5, 1.5]$ & $[0.5, 1.5]$ & $[0.5, 1.5]$ & $[0.5, 1.5]$ & $[0.5, 1.5]$ \\
$\log[10^{10}A_{s}]$  & $[2.4,4]$ & $[2.4,4]$ & $[2.4,4]$ & $[2.4,4]$ & $[2.4,4]$ \\
$100\theta_{MC}$  & $[0.5,10]$ & $[0.5,10]$ & $[0.5,10]$ & $[0.5,10]$ & $[0.5,10]$ & \\
$\beta$     & $[0, 10]$ & $-$ & $-$ & $-$ & $-$ \\
$\alpha$    & $-$ & $[0, 10]$ & $[0, 10]$ & $-$ & $-$ \\
$\delta$   & $-$ & $-$ & $-$ & $[0, 10]$ & $[-10, 0]$ \\ 
\hline 
\hline
\end{tabular}
\end{center}
\caption{The table summarizes the flat priors on various free parameters for all the quintessence scalar field models.  }
\label{tab:priors}
\end{table}

\subsection{Observational Constraints}
\label{sec-observational}

Here we present the  observational constraints of the scalar field 
models using the following cosmological data. 

\begin{itemize}

\item CMB (Planck TTTEEE+lowTEB)

\item CMB+BAO 

\item CMB+BAO+JLA  

\item CMB+RSD+JLA

\item CMB+BAO+JLA+RSD+CC

\end{itemize}
For each of the above scalar field models, we sample the posterior distribution of the parameters by running MCMC chains with \texttt{cosmomc}, using various observational datasets. We determine the convergence of the chains by computing the Gelman-Rubin statistic $R -1$, and requiring $R-1 < 0.1$. In particular, all continued the running of all the chains until $R-1 \lesssim 0.03$ was achieved.  In what follows, we separately describe the observational constraints imposed on the scalar field models.  

\subsubsection{Model 1}
\label{results-model1}

\begin{figure*}[h]
\includegraphics[width=0.6\textwidth]{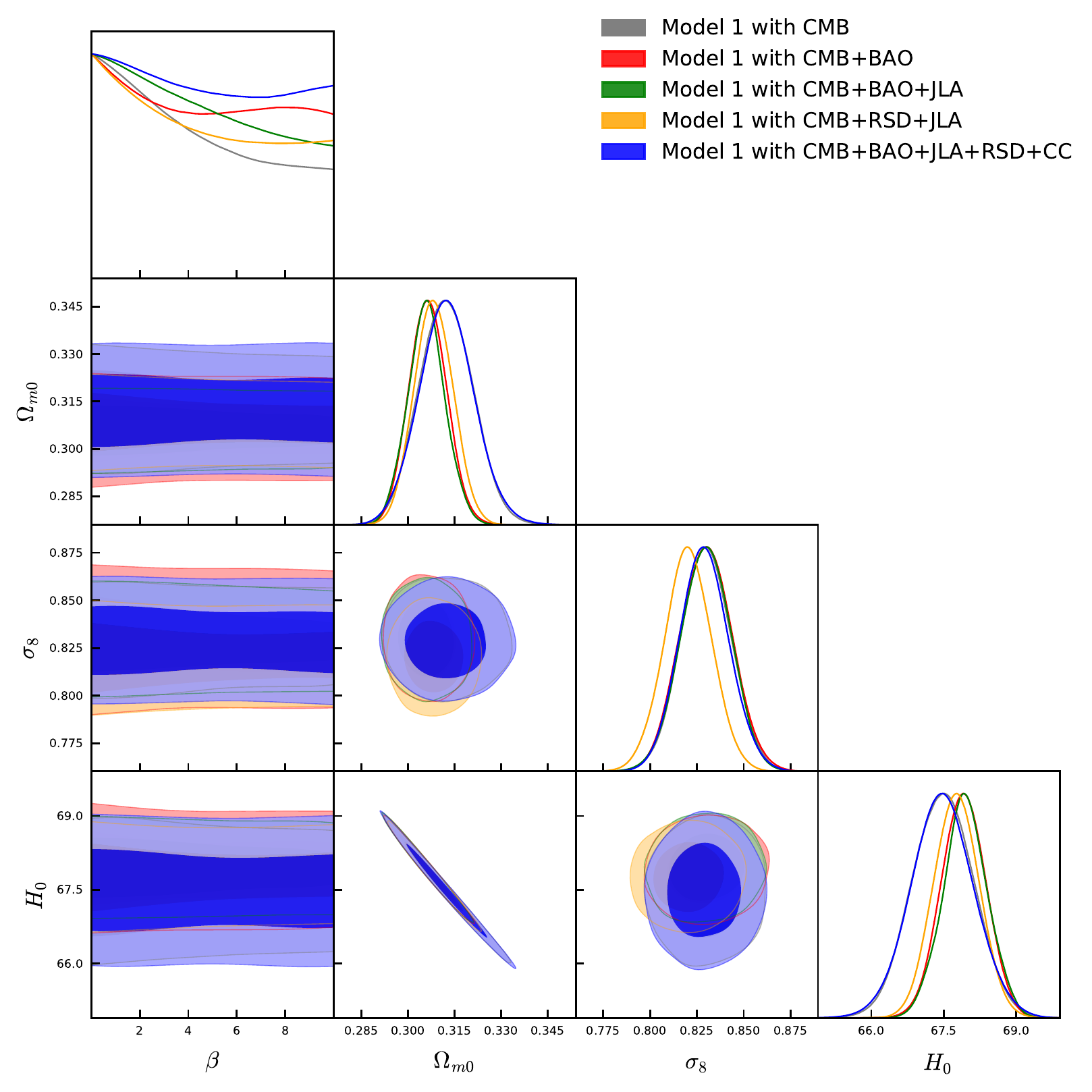}
\caption{1D marginalized posterior distributions for ($\beta$, $\Omega_{m0}$, $\sigma_8$, $H_0$) and 2D contour plots of  several combinations of the above model parameters have been presented for Model 1: $V (\phi) = V_0 \cosh(\beta \phi^{u})$ with $u=1$, using various astronomical datasets CMB, CMB+BAO, CMB+BAO+JLA, CMB+RSD+JLA, CMB+BAO+JLA+RSD+CC. }
\label{fig-model1}
\end{figure*}

As mentioned earlier, for this model we focus on a specific value of $u$, namely, $u =1$ and proceed towards its observational constraints using a number of cosmological datasets  mentioned in section \ref{sec-data}.  
The observational summary for this model has been shown in Table \ref{tab:results-model1} where we present the constraints on the model parameters at 68\% and 95\% CL. In Fig. \ref{fig-model1}, we display the 1D marginalized posterior distributions for some selected parameters of the model as well as 2D joint contours considering several combinations of the model parameters.

From the constraints on $H_0$ by all the observational datasets, we see that it takes almost similar mean values  to the $\Lambda$CDM based Planck's estimation \cite{Ade:2015xua}, but the $68\%$ error bars on $H_0$ for this scalar field model (considering all the datasets) are slightly lower compared to its error bars from Planck  \cite{Ade:2015xua}. Precisely, the estimations of $H_0$ for this scalar field model considering the employed observational datasets are many standard deviations apart from the local estimation of $H_0$ 
by Riess et al. \cite{Riess:2016jrr}, thus, the tension on $H_0$ is not released for this quintessence scalar field model. However, this is not a new result and it  is quite expected in this context because the alleviation of $H_0$ needs a phantom dark energy equation of state in the universe sector as already discussed in  \cite{DiValentino:2016hlg,Vagnozzi:2018jhn}. Actually the presence of phantom dark energy makes the expansion of the universe faster at late times, and hence decreases the distance to
last scattering.  In fact, in \cite{Vagnozzi:2018jhn}, the authors have argued that the dark energy models with equation of state $w \geq -1$ cannot  alleviate the $H_0$ tension. Thus, naturally quintessence scalar field models are included in this class and hence they are unable to release the $H_0$ tension.  
Furthermore, the constraints on $\Omega_{m0}$, are similar to the Planck's $\Lambda$CDM based estimation \cite{Ade:2015xua} while the estimated values of $\sigma_8$ are slightly higher (although mildly) compared to the Planck's $\Lambda$CDM based estimation \cite{Ade:2015xua}.

Concerning the free key parameter $\beta$ of this scalar field model,  our analyses actually show that it is unconstrained for CMB data alone and also by other observational datasets, such as CMB+BAO, CMB+BAO+JLA, CMB+RSD+JLA and CMB+BAO+JLA+RSD+CC. This is clear if one looks at the 1D marginalized distributions of $\beta$ as well as the joint contours (see Fig. \ref{fig-model1}). 
Thus, one thing is clear from this context that the sensitivity of CMB is not strong enough to test the parameter, $\beta$, and hence this parameter remains unconstrained.
When the external datasets, such as BAO, JLA, RSD, CC, are added to CMB, the nature of the parameter, $\beta$, does not alter. That means, none of the above external datasets (BAO, JLA, RSD, CC), add much to the CMB data in order to constrain $\beta$. In fact, this could be a general statement for this work if we closely look at the constraints and their error bars. From Table \ref{tab:results-model1}, one could clearly see the error bars on almost all the parameters (except $\beta$ since this is unconstrained) remain same for all the observational datasets. That means the addition of any external dataset to CMB alone does not actually improve the error bars on the parameters. For instance one could study the error bars on $H_0$, $\Omega_{m0}$ and even $\sigma_8$ for all the datasets. This again  means that these external datasets (BAO, JLA, RSD, CC) do not add much to CMB alone, and it likely means that in quintessence models $H_0$ as well as other parameters are pretty much determined by the CMB data alone. 
Perhaps it is worth noting that the unconstrained nature of $\beta$ is independent on its prior considered during the statistical analysis. Even if we increase the upper limit of the prior on $\beta$ which for this case we set to be $[0, 10]$ (see Table \ref{tab:priors}), the result on $\beta$ does not alter. That means $\beta $ remains unconstrained independent of the observational datasets.

Finally, we conclude this section with the presence and absence of correlations between the model parameters. Looking at Fig. \ref{fig-model1}, one can see that a strong negative correlation between $\Omega_{m0}$ and $H_0$ exists which is a consequence of the geometric degeneracy, while the remaining parameters shown in Fig. \ref{fig-model1} do not have any correlation amongst them. In particular, the parameters $H_0$ and $\sigma_8$ are not correlated. This is one of the interesting observations in this work. A possible explanation towards such uncorrelated feature is as follows. The important thing to notice is that $H_0$ and $\sigma_8$ are respectively more background and perturbation quantities. Even though,  $\sigma_8$ is a derived parameter and is related to $A_s$ (namely $\sigma_8 \propto A_s^2$). So, they govern different physics, even in the simplest $\Lambda$CDM model. The Hubble constant at present, $H_0$, governs the distance to last-scattering and thus $\Theta_s$ ($= 100 \theta_{MC}$). Changing $H_0$ changes the position of all the peaks in the CMB spectrum, especially the first one. Whereas changing $\sigma_8$ alters the amplitude of the CMB peaks. The effects are distinct and thus can be well distinguished between each other. So, in principle we don't expect a strong correlation between $H_0$ and $\sigma_8$ to begin with. The only effect is an `indirect' correlation between $H_0$ and $\Omega_{m0}$.

\begingroup                                                                                                                     
\squeezetable                                                                                                                   
\begin{center}                                                                                                                  
\begin{table}[h]                                                                                                                   
\begin{tabular}{ccccccccccccccccccc}                                                                                                            
\hline\hline                                                                                                                    
Parameters & CMB & CB & CBJ & CRJ & CBJRC\\ \hline
$\Omega_c h^2$ & $    0.12_{-    0.0014-    0.0027}^{+    0.0014+    0.0027}$ & $    0.12_{-    0.0010-    0.0020}^{+    0.0010+    0.0020}$ & $    0.12_{-    0.0010-    0.0020}^{+    0.0010+    0.0019}$ &  $    0.12_{-    0.0010-    0.0020}^{+    0.0010+    0.0020}$ & $    0.12_{-    0.0014-    0.0028}^{+    0.0014+    0.0028}$  \\

$\Omega_b h^2$ & $    0.022_{-    0.00015-    0.00031}^{+    0.00015+    0.00031}$ & $    0.022_{-    0.00014-    0.00027}^{+    0.00014+    0.00027}$ & $    0.022_{-    0.00014-    0.00028}^{+    0.00014+    0.00027}$ & $    0.022_{-    0.00014-    0.00027}^{+    0.00014+    0.00028}$ & $    0.022_{-    0.00016-    0.00030}^{+    0.00015+    0.00030}$ \\

$100\theta_{MC}$ & $    1.041_{-    0.00032-    0.00065}^{+    0.00033+    0.00063}$ & $    1.041_{-    0.00030-    0.00057}^{+    0.00030+    0.00061}$ & $    1.041_{-    0.00030-    0.00059}^{+    0.00030+    0.00060}$ & $    1.041_{-    0.00030-    0.00057}^{+    0.00030+    0.00058}$ & $    1.041_{-    0.00033-    0.00064}^{+    0.00032+    0.00063}$ \\

$\tau$ & $    0.081_{-    0.017-    0.033}^{+    0.017+    0.034}$ & $    0.085_{-    0.017-    0.032}^{+    0.017+    0.033}$ & $    0.085_{-    0.016-    0.034}^{+    0.016+    0.031}$ & $    0.072_{-    0.016-    0.031}^{+    0.016+    0.031}$ & $    0.079_{-    0.017-    0.032}^{+    0.017+    0.033}$ \\

$n_s$ & $    0.97_{-    0.0045-    0.0088}^{+    0.0045+    0.0092}$ & $    0.97_{-    0.0038-    0.0076}^{+    0.0038+    0.0076}$ & $    0.97_{-    0.0038-    0.0076}^{+    0.0038+    0.0076}$ & $    0.97_{-    0.0037-    0.0072}^{+    0.0037+    0.0076}$ & $    0.97_{-    0.0047-    0.0092}^{+    0.0045+    0.0092}$ \\

${\rm{ln}}(10^{10} A_s)$ & $    3.095_{-    0.034-    0.065}^{+    0.034+    0.065}$ & $    3.101_{-    0.034-    0.065}^{+    0.033+    0.066}$ & $    3.101_{-    0.032-    0.066}^{+    0.032+    0.061}$ & $    3.076_{-    0.031-    0.063}^{+    0.031+    0.062}$ &  $    3.092_{-    0.033-    0.064}^{+    0.032+    0.065}$ \\

$\beta$ & unconstrained & unconstrained & unconstrained & unconstrained & unconstrained \\

$\Omega_{m0}$ & $    0.31_{-    0.009-    0.016}^{+    0.009+    0.017}$ & $    0.31_{-    0.006-    0.012}^{+    0.006+    0.012}$ & $    0.31_{-    0.006-    0.012}^{+    0.006+    0.012}$ & $    0.31_{-    0.006-    0.012}^{+    0.006+    0.012}$ & $    0.31_{-    0.009-    0.017}^{+    0.009+    0.018}$ \\

$\sigma_8$ & $   0.83_{-    0.013-    0.026}^{+    0.014+    0.026}$ & $    0.83_{-    0.014-    0.026}^{+    0.014+    0.027}$ & $    0.83_{-    0.013-    0.026}^{+    0.013+    0.025}$ & $    0.82_{-    0.012-    0.025}^{+    0.013+    0.025}$ & $    0.83_{-    0.013-    0.025}^{+    0.013+    0.026}$ \\

$H_0$ & $   67.48_{-    0.63-    1.21}^{+    0.63+    1.22}$ & $   67.91_{-    0.46-    0.89}^{+    0.46+    0.89}$ & $   67.94_{-    0.44-    0.90}^{+    0.46+    0.90}$ & $   67.76_{-    0.47-    0.90}^{+    0.47+    0.93}$ & $   67.47_{-    0.63-    1.26}^{+    0.64+    1.28}$ \\
\hline\hline                                                                                                                    
\end{tabular}                                                                                                                   
\caption{Summary of 68\% and 95\% CL constraints on various model parameters of the scalar field model for Model 1: $V (\phi) = V_0 \cosh(\beta \phi^{u})$ with $u =1$ using different observational datasets. We note that $\Omega_{m0}$ is the present value of $\Omega_m = \Omega_c + \Omega_b$ and $H_0$ is in the units of km/s/Mpc.  
Here, we have shortened the notations as follows: CB = CMB+BAO, CBJ = CMB+BAO+JLA, CRJ = CMB+RSD+JLA, and CBJRC = CMB+BAO+JLA+RSD+CC.}
\label{tab:results-model1}                                                                                                   
\end{table}                                                                                                                     
\end{center}                                                                                                                    
\endgroup

\subsubsection{Model 2}
\label{results-model2}

In order to extract the cosmological constraints of the scalar field model (\ref{potential}), we have separately analyzed two cases, one with $\epsilon =1$ (Model 2a) and the other with $\epsilon = -1$ (Model 2b).

\begin{figure*}[h]
\includegraphics[width=0.6\textwidth]{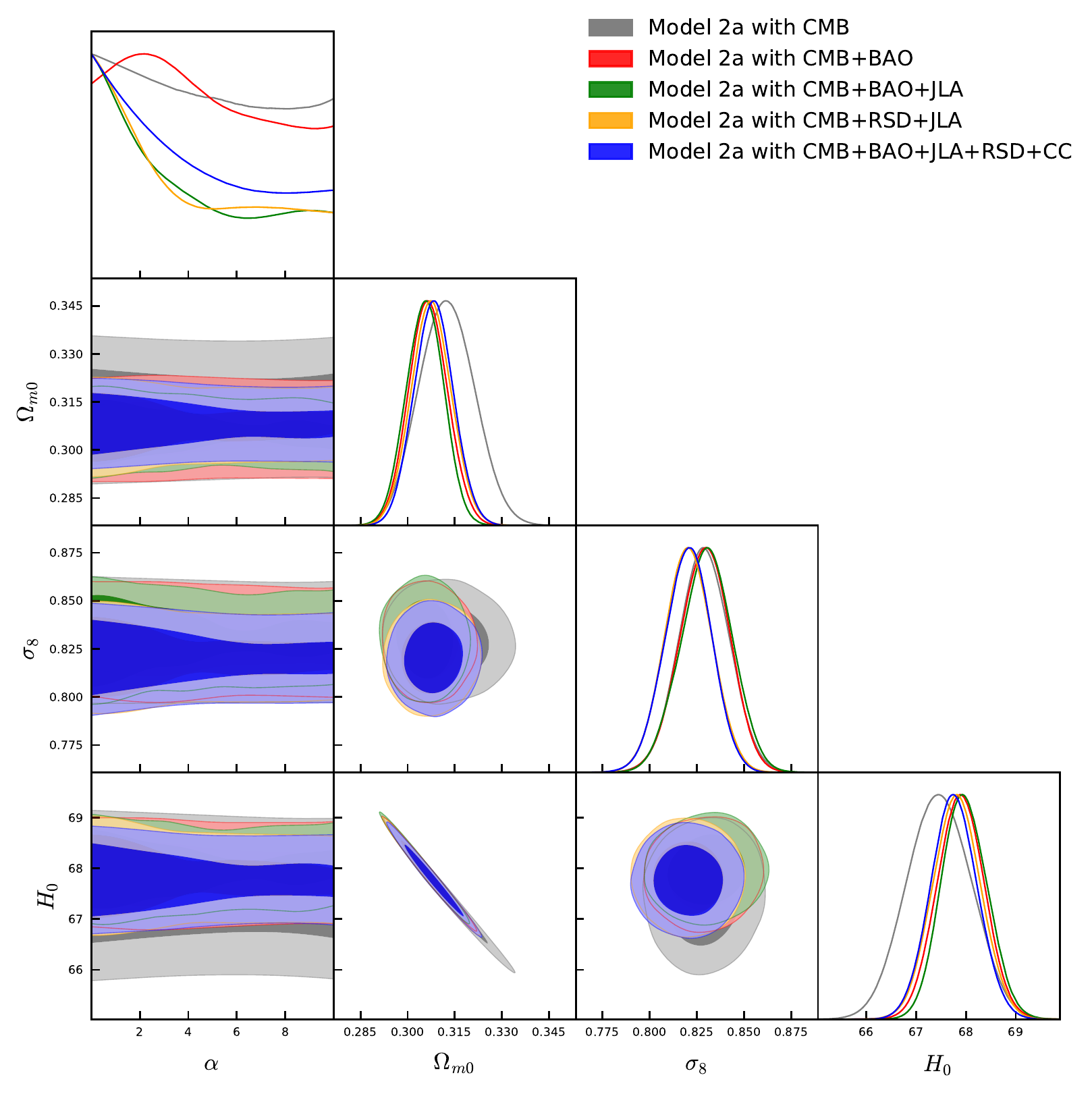}
\caption{1D marginalized posterior distributions for ($\alpha$, $\Omega_{m0}$, $\sigma_8$, $H_0$) and 2D contour plots of  several combinations of the above model parameters have been presented for Model 2a: $V (\phi) = V_0 \left[ 1+ {\rm s}ech (\alpha \phi) \right] $ ($\alpha \geq 0$), using various {\bf cosmological} datasets CMB, CMB+BAO, CMB+BAO+JLA, CMB+RSD+JLA, CMB+BAO+JLA+RSD+CC. }
\label{fig-model2a}
\end{figure*}

Let us now summarize the observational constraints for Model 2a, that means for the scalar field model: $V (\phi) = V_0 [1+ {\rm s}ech (\alpha \phi)]$. In Table \ref{tab:constraints} we have summarized the observational constraints on free and derived parameters of this model for different cosmological datasets at 68\% CL and 95\% CL. Additionally, in Fig. \ref{fig-model2a}, we present the 1D marginalized posterior distributions for some selected model parameters and 2D joint contours at 68\% and 95\% CL considering various combinations of the model parameters.
We first concentrate on the constraints on some derived parameters, such as, $H_0$, $\Omega_{m0}$ and $\sigma_8$. From the analyses,  we find that the estimated values of the Hubble parameter for different observational datasets are almost similar to the $\Lambda$CDM based Planck's estimation, see \cite{Ade:2015xua} with mildly lower error bars. So, it means the tension on $H_0$ is not released by this scalar field model which actually supports the previous results \cite{DiValentino:2016hlg,Vagnozzi:2018jhn}. On the other hand, the estimated values of $\Omega_{m0}$ are mildly larger with slightly larger error bars compared to Planck \cite{Ade:2015xua}. This is expected because as one can see from Fig. \ref{fig-model2a} that there is a negative correlation between $H_0$ and $\Omega_{m0}$. About the $\sigma_8$ parameter we have similar conclusion as we already have found with Model 1, that means the values of $\sigma_8$ for different combinations are slightly larger compared to Planck \cite{Ade:2015xua}. The important point that we like to note here that, when we add any external dataset to CMB, we see that the error bars on the parameters, do not actually improve. So, the addition of any other external dataset does not actually add anything to the constraints obtained from CMB alone dataset. Thus, similar to what we have observed in Model 1, here too, one can safely conclude that the constraints on this model are pretty much determined by the CMB data alone. This observation is also supported if we now  focus on the observational constraints on the free parameter $\alpha$. Our analyses show that the parameter $\alpha$ is unconstrained. It is much clear if one looks at the 1D posterior distributions for $\alpha$ (see Fig. \ref{fig-model2a}).  Perhaps it should mentioned that the observational nature of this free parameter does not change even if we increase the upper limit of the prior imposed on it (see Table  \ref{tab:priors}). So, similar to the previous model (i.e. Model 1), here too, we can comment that the CMB sensitivity is not enough to constrain this parameter. And when any one of the external datasets, such as BAO, JLA, RSD and CC, are added to CMB, the results do not change at all. That means the unconstrained nature of this parameter does not actually change. Finally, we again see that similar to Model 1, the parameter $\sigma_8$ is not correlated to $H_0$. Now, this is becoming an interesting issue since such a relation remains same irrespective of the change of the potential. The absence of such correlation between these two parameters follows similar explanation as already given in section \ref{results-model1}.

\begin{figure*}[h]
\includegraphics[width=0.6\textwidth]{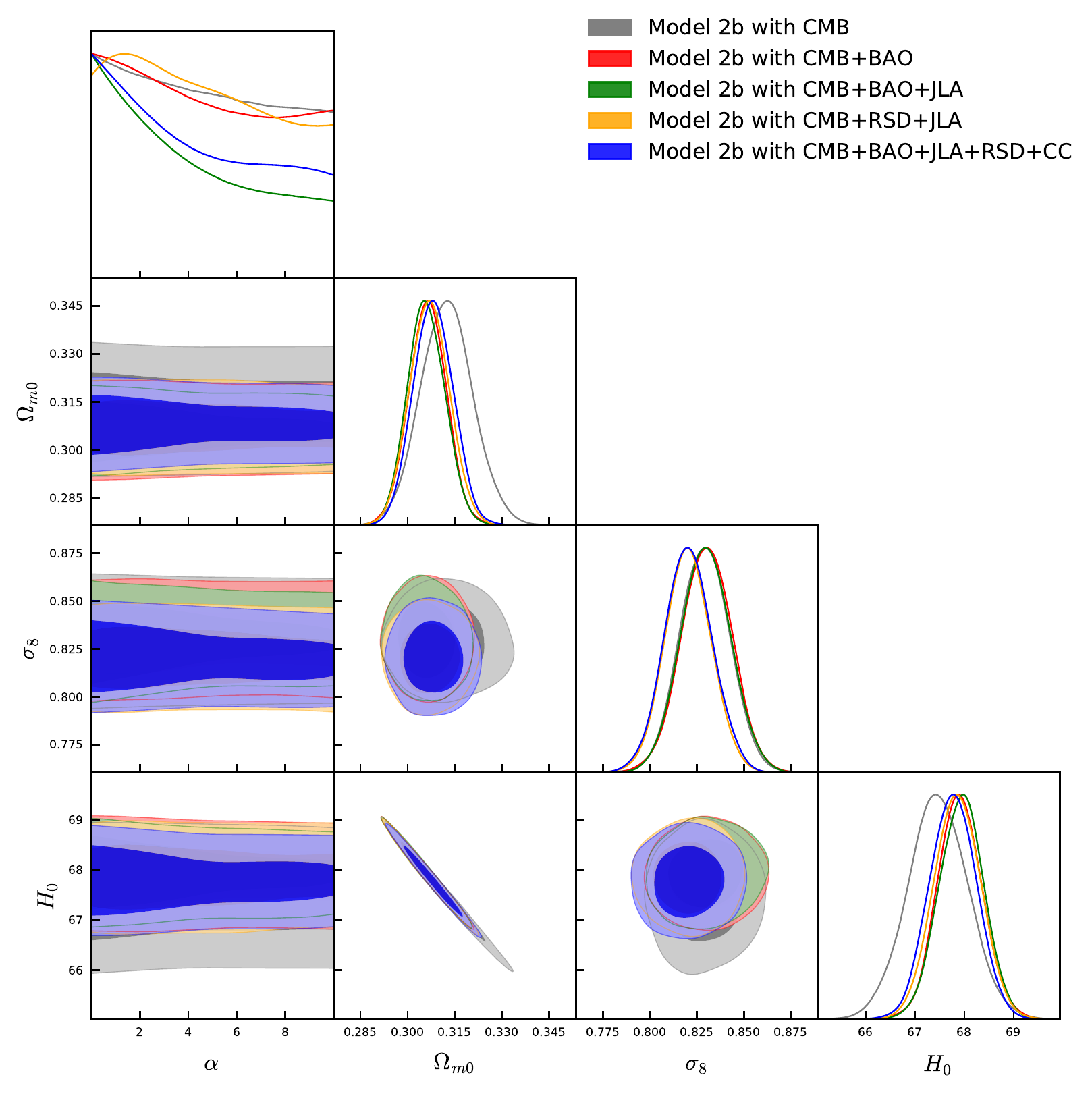}
\caption{1D marginalized posterior distributions for ($\alpha$, $\Omega_{m0}$, $\sigma_8$, $H_0$) and 2D contour plots of  several combinations of the above model parameters have been presented for Model 2b using various cosmological datasets CMB, CMB+BAO, CMB+BAO+JLA, CMB+RSD+JLA, CMB+BAO+JLA+RSD+CC. }
\label{fig-model2b}
\end{figure*}

We now discuss the observational constraints of Model 2b: $V (\phi) = V_0 [1- {\rm s}ech (\alpha \phi)]$. The observational summary for this model has been displayed in Table \ref{tab:constraints2} where we have shown 68\% and 95\% CL constraints on the model parameters. And in Fig. \ref{fig-model2b}, we show the 1D marginalized posterior distributions for some selected model parameters and 2D joint contours  at 68\% and 95\% CL considering various combinations of the model parameters.
From the analysis summarized in Table \ref{tab:constraints2}, we notice that this model returns almost similar fit to Model 2a, and in a similar fashion, the parameter $\alpha$ is again found to be unconstrained. Similar to the previous case (i.e., the potential (\ref{potential}) with $\epsilon = +1 $), here too, we observe that the estimated values of the Hubble constant for all the combinations are very close to the $\Lambda$CDM based Planck's estimation \cite{Ade:2015xua}. So, both the models, as one can see, cannot solve the $H_0$-tension.  This agrees with two earlier conclusions, namely for Model 1 and Model 2a and supports the arguments of  \cite{DiValentino:2016hlg,Vagnozzi:2018jhn}. And moreover we also comment that only CMB alone data are enough to test this model since the addition of any other external datasets do not add anything to the constraints obtained from CMB alone. However, the analyses with other external datasets to CMB are important in order to verify whether this model returns similar conclusion or not. Concerning the absence of the correlation between $H_0$ and $\sigma_8$, we follow similar arguments as already given in section \ref{results-model1} and in the upper half of this section.

\begingroup                                                                                                                     
\squeezetable
\tiny
\begin{center}                                                                                                                  
\begin{table*}[h]
\begin{tabular}{ccccccccccccccc}                                                                                                            
\hline\hline  

Parameters & CMB & CB & CBJ & CRJ & CBJRC\\ \hline
$\Omega_c h^2$ & $    0.12_{-    0.0014-    0.0027}^{+    0.0014+    0.0028}$ & $    0.12_{-    0.0010-    0.0020}^{+    0.0010+    0.0020}$ & $    0.12_{-    0.0010-    0.0019}^{+    0.0010+    0.0019}$ &  $    0.12_{-    0.0010-    0.0021}^{+    0.0011+    0.0021}$ & $    0.12_{-    0.0010-    0.0020}^{+    0.0010+    0.0020}$ \\

$\Omega_b h^2$ & $    0.022_{-    0.00016-    0.00030}^{+    0.00016+    0.00031}$ & $    0.022_{-    0.00015-    0.00026}^{+    0.00014+    0.00027}$ & $    0.022_{-    0.00014-    0.00027}^{+    0.00014+    0.00028}$ & $    0.022_{-    0.00014-    0.00027}^{+    0.00014+    0.00027}$ & $    0.022_{-    0.00014-    0.00027}^{+    0.00014+    0.00027}$ \\

$100\theta_{MC}$ &  $    1.041_{-    0.00033-    0.00065}^{+    0.00032+    0.00065}$ & $    1.041_{-    0.00030-    0.00060}^{+    0.00031+    0.00058}$ & $    1.041_{-    0.00030-    0.00059}^{+    0.00030+    0.00059}$ & $  1.041_{-    0.00030-    0.00060}^{+    0.00030+    0.00058}$ & $    1.041_{-    0.00031-    0.00061}^{+    0.00031+    0.00059}$ \\

$n_s$ & $    0.97_{-    0.0046-    0.0089}^{+    0.0046+    0.0089}$ &  $    0.97_{-    0.0040-    0.0077}^{+    0.0039+    0.0080}$ & $    0.97_{-    0.0041-    0.0074}^{+    0.0038+    0.0076}$ & $    0.97_{-    0.0039-    0.0077}^{+    0.0038+    0.0074}$ & $    0.97_{-    0.0037-    0.0073}^{+    0.0037+    0.0074}$ \\

$\tau$ & $    0.079_{-    0.017-    0.033}^{+    0.017+    0.034}$ & $    0.084_{-    0.016-    0.032}^{+    0.016+    0.031}$ & $    0.086_{-    0.016-    0.033}^{+    0.018+    0.032}$  & $    0.073_{-    0.015-    0.030}^{+    0.015+    0.030}$  & $    0.072_{-    0.015-    0.030}^{+    0.015+    0.029}$ \\

${\rm{ln}}(10^{10} A_s)$ &   $    3.092_{-    0.033-    0.064}^{+    0.034+    0.065}$ &   $    3.099_{-    0.032-    0.064}^{+    0.032+    0.061}$ & $    3.10_{-    0.032-    0.065}^{+    0.036+    0.063}$  & $    3.077_{-    0.030-    0.060}^{+    0.031+    0.059}$ & $    3.076_{-    0.030-    0.060}^{+    0.030+    0.058}$ \\

$\alpha$ & unconstrained & unconstrained & unconstrained & unconstrained & unconstrained \\

$\Omega_{m0}$ &  $    0.31_{-    0.0088-    0.016}^{+    0.0087+    0.017}$ & $    0.31_{-    0.0061-    0.012}^{+    0.0062+    0.012}$ & $    0.31_{-    0.0060-    0.012}^{+    0.0059+    0.011}$ & $    0.31_{-    0.0063-    0.013}^{+    0.0065+    0.013}$ & $    0.31_{-    0.0062-    0.012}^{+    0.0061+    0.012}$ \\

$\sigma_8$ & $    0.83_{-    0.013-    0.026}^{+    0.013+    0.026}$ & $    0.83_{-    0.013-    0.026}^{+    0.013+    0.025}$  & $    0.83_{-    0.014-    0.026}^{+    0.014+    0.026}$ & $    0.82_{-    0.012-    0.024}^{+    0.012+    0.025}$ & $    0.82_{-    0.012-    0.024}^{+    0.012+    0.024}$ \\

$H_0$ & $   67.46_{-    0.64-    1.22}^{+    0.65+    1.23}$  &  $   67.89_{-    0.47-    0.91}^{+    0.46+    0.92}$ & $   67.97_{-    0.49-    0.86}^{+    0.44+    0.90}$ & $   67.81_{-    0.48-    0.93}^{+    0.48+    0.94}$ & $   67.76_{-    0.46-    0.90}^{+    0.46+    0.92}$ \\

\hline\hline                                                                                                                    
\end{tabular}                                                                                                                   
\caption{Summary of 68\% and 95\% CL constraints on various model parameters of Model 2a: $V (\phi) = V_0 \left[ 1+ {\rm s}ech (\alpha \phi) \right] $ ($\alpha \geq 0$) using different observational datasets.  We note that $\Omega_{m0}$ is the present value of $\Omega_m = \Omega_c + \Omega_b$ and $H_0$ is in the units of km/s/Mpc. Here, we have shortened the notations as follows: CB = CMB+BAO, CBJ = CMB+BAO+JLA, CRJ = CMB+RSD+JLA, and CBJRC = CMB+BAO+JLA+RSD+CC.}
\label{tab:constraints}                                                                                                   
\end{table*}                                                                                                                     
\end{center}                                                                                                                    
\endgroup  
\begingroup                                                                                                                     
\squeezetable                                                                                                                   
\begin{center}                                                                                                                  
\begin{table}[h]
\begin{tabular}{cccccccccccccccccccc} 
\hline\hline                                                                                                                    
Parameters & CMB &  CB & CBJ & CRJ & CBJRC\\ \hline
$\Omega_c h^2$ & $    0.12_{-    0.0014-    0.0027}^{+    0.0013+    0.0026}$ & $    0.12_{-    0.0010-    0.0020}^{+    0.0010+    0.0019}$ & $    0.12_{-    0.0010-    0.0019}^{+    0.0010+    0.0020}$ & $    0.12_{-    0.0010-    0.0020}^{+    0.0011+    0.0021}$ & $    0.12_{-    0.0011-    0.0020}^{+    0.0010+    0.0020}$\\

$\Omega_b h^2$ & $    0.022_{-    0.00015-    0.00030}^{+    0.00015+    0.00031}$ & $    0.022_{-    0.00014-    0.00028}^{+    0.00014+    0.00027}$ & $    0.022_{-    0.00014-    0.00027}^{+    0.00014+    0.00028}$ & $    0.022_{-    0.00015-    0.00028}^{+    0.00014+    0.00028}$ & $    0.022_{-    0.00014-    0.00028}^{+    0.00014+    0.00027}$ \\

$100\theta_{MC}$ & $    1.041_{-    0.00032-    0.00063}^{+    0.00031+    0.00063}$ & $    1.041_{-    0.00030-    0.00059}^{+    0.00030+    0.00059}$ & $    1.041_{-    0.00030-    0.00060}^{+    0.00030+    0.00057}$ & $    1.041_{-    0.00029-    0.00060}^{+    0.00029+    0.00058}$ & $    1.041_{-    0.00030-    0.00059}^{+    0.00030+    0.00060}$\\

$n_s$ & $    0.97_{-    0.0044-    0.0088}^{+    0.0044+    0.0087}$ & $    0.97_{-    0.0037-    0.0075}^{+    0.0037+    0.0074}$ & $    0.97_{-    0.0038-    0.0076}^{+    0.0040+    0.0075}$ & $    0.97_{-    0.0039-    0.0074}^{+    0.0038+    0.0076}$ & $    0.97_{-    0.0037-    0.0074}^{+    0.0038+    0.0074}$ \\

$\tau$ & $    0.080_{-    0.017-    0.032}^{+    0.017+    0.032}$ & $    0.085_{-    0.016-    0.032}^{+    0.016+    0.031}$ & $    0.085_{-    0.017-    0.032}^{+    0.017+    0.032}$ & $    0.074_{-    0.016-    0.030}^{+    0.015+    0.031}$ & $    0.073_{-    0.016-    0.031}^{+    0.016+    0.031}$\\

${\rm{ln}}(10^{10} A_s)$ & $    3.092_{-    0.033-    0.063}^{+    0.032+    0.063}$ & $    3.10_{-    0.032-    0.065}^{+    0.032+    0.062}$ & $    3.10_{-    0.032-    0.065}^{+    0.032+    0.064}$ & $    3.077_{-    0.030-    0.059}^{+    0.030+    0.060}$ & $    3.076_{-    0.031-    0.062}^{+    0.031+    0.062}$\\

$\alpha$ & unconstrained & unconstrained & unconstrained & unconstrained & unconstrained \\

$\Omega_{m0}$ & $    0.31_{-    0.0083-    0.016}^{+    0.0081+    0.017}$ & $    0.31_{-    0.0060-    0.012}^{+    0.0060+    0.012}$ & $    0.31_{-    0.0060-    0.011}^{+    0.0061+    0.012}$ & $    0.31_{-    0.0063-    0.012}^{+    0.0063+    0.013}$ & $    0.31_{-    0.0063-    0.012}^{+    0.0063+    0.012}$\\

$\sigma_8$ & $    0.83_{-    0.013-    0.026}^{+    0.013+    0.026}$ & $    0.83_{-    0.013-    0.026}^{+    0.013+    0.026}$ & $    0.83_{-    0.013-    0.026}^{+    0.013+    0.026}$ & $    0.82_{-    0.012-    0.024}^{+    0.012+    0.024}$ & $    0.82_{-    0.013-    0.025}^{+    0.012+    0.025}$\\

$H_0$ & $   67.46_{-    0.59-    1.21}^{+    0.60+    1.21}$ & $   67.91_{-    0.45-    0.86}^{+    0.45+    0.91}$ & $   67.93_{-    0.45-    0.91}^{+    0.45+    0.89}$ & $   67.86_{-    0.48-    0.94}^{+    0.48+    0.94}$ & $   67.77_{-    0.48-    0.89}^{+    0.47+    0.94}$\\

\hline\hline                                                                                                                    
\end{tabular}                                                                                                                   
\caption{Summary of 68\% and 95\% CL constraints on various model parameters of the scalar field model with potential  $V (\phi) = V_0 \left[ 1- {\rm s}ech (\alpha \phi) \right] $ ($\alpha \geq 0$) using different observational datasets. We note that $\Omega_{m0}$ is the present value of $\Omega_m = \Omega_c + \Omega_b$ and $H_0$ is in the units of km/s/Mpc.   Here, we have shortened the notations as follows: CB = CMB+BAO, CBJ = CMB+BAO+JLA, CRJ = CMB+RSD+JLA, and CBJRC = CMB+BAO+JLA+RSD+CC. }\label{tab:constraints2}                                                                                                   
\end{table}                                                                                                                     
\end{center}                                                                                                                    
\endgroup 
\begin{figure*}[h]
\includegraphics[width=0.6\textwidth]{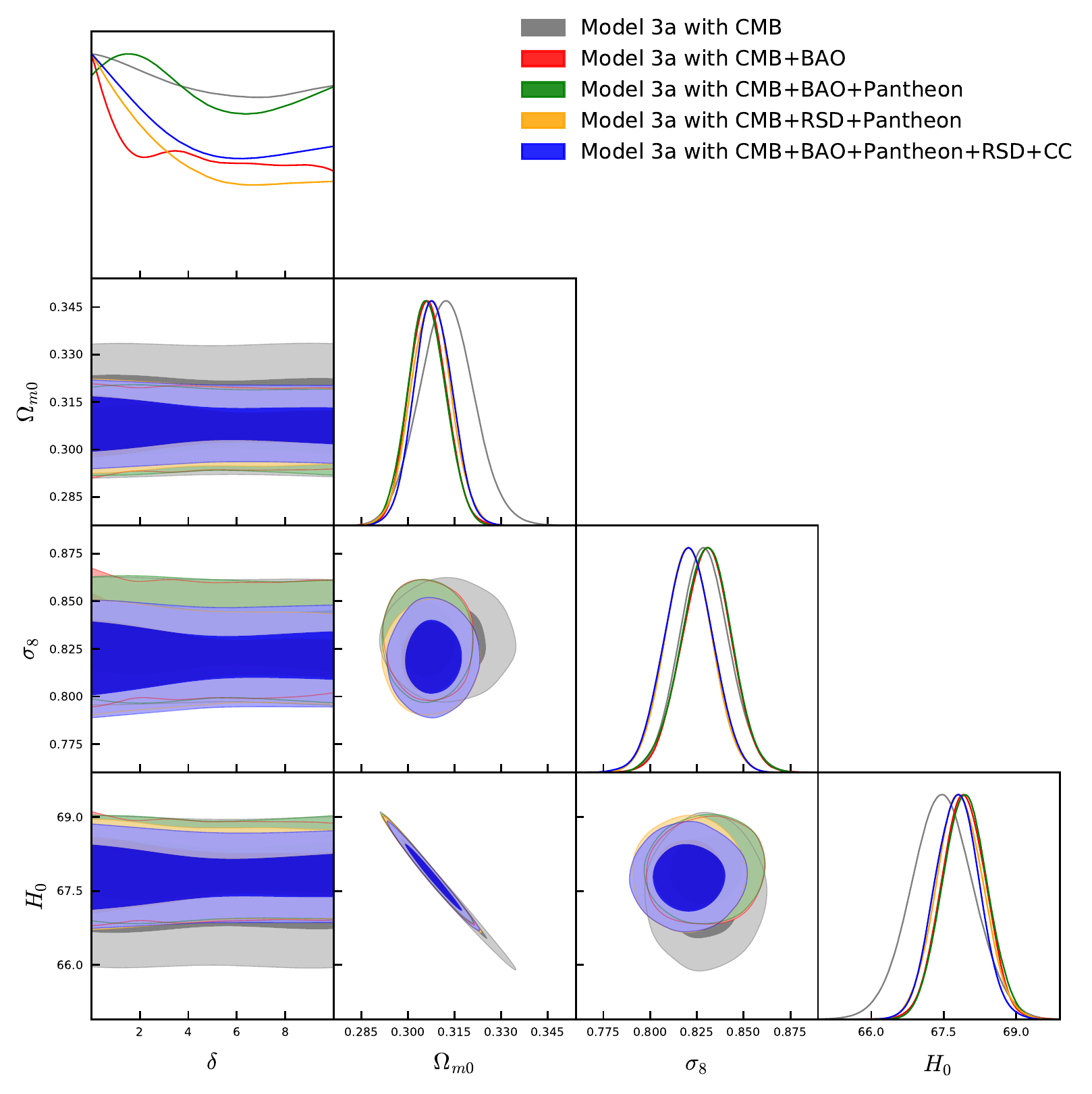}
\caption{1D marginalized posterior distributions for ($\delta$, $\Omega_{m0}$, $\sigma_8$, $H_0$) and 2D contour plots of  several combinations of the above model parameters have been presented for Model 3a using various cosmological  datasets CMB, CMB+BAO, CMB+BAO+JLA, CMB+RSD+JLA, CMB+BAO+JLA+RSD+CC. }
\label{fig-model3a}
\end{figure*}
\begin{figure*}[h]
\includegraphics[width=0.6\textwidth]{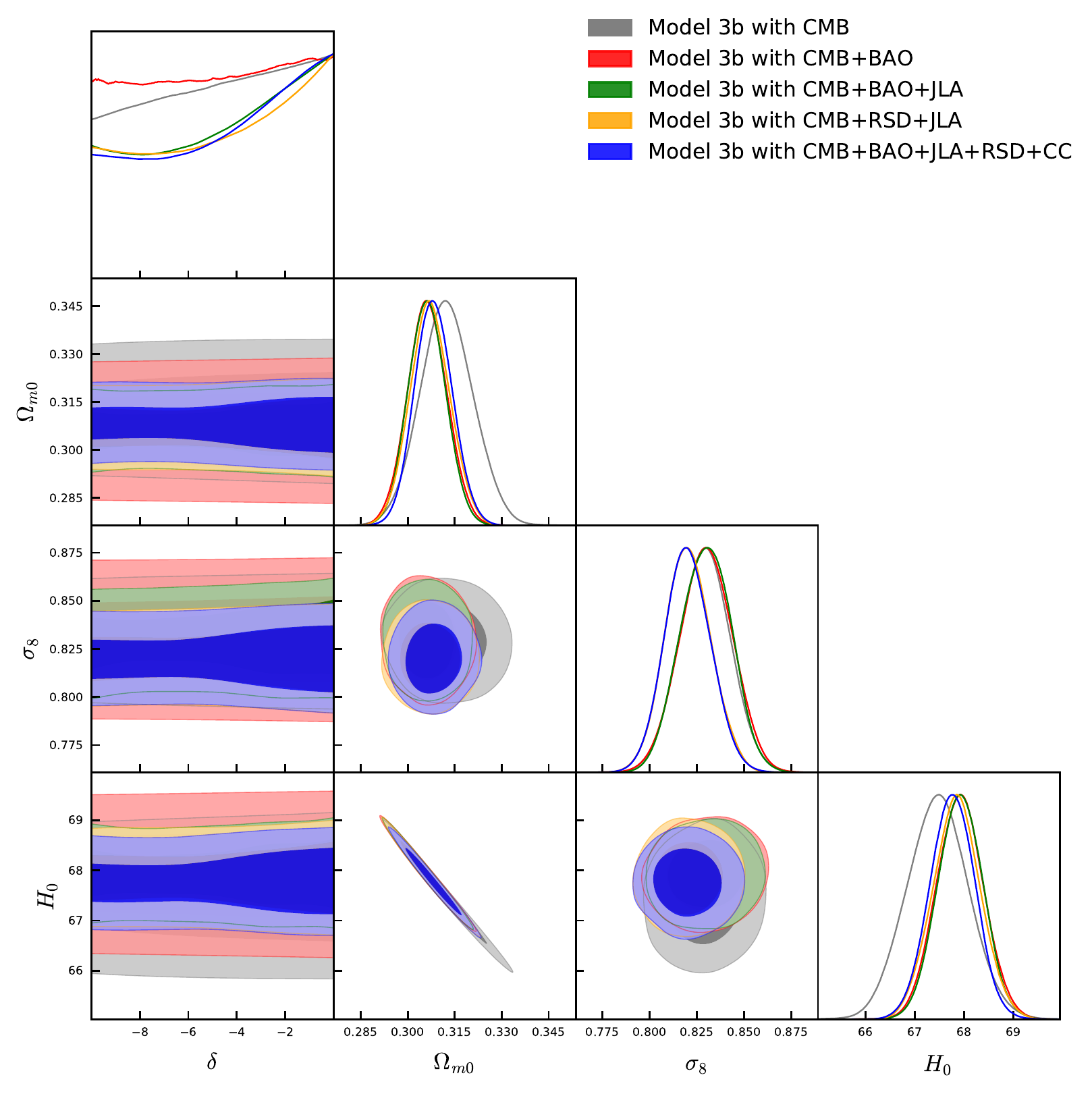}
\caption{1D marginalized posterior distributions for ($\delta$, $\Omega_{m0}$, $\sigma_8$, $H_0$) and 2D contour plots of  several combinations of the above model parameters have been presented for Model 3b using various cosmological  datasets CMB, CMB+BAO, CMB+BAO+JLA, CMB+RSD+JLA, CMB+BAO+JLA+RSD+CC. }
\label{fig-model3b}
\end{figure*}

\subsubsection{Model 3}
\label{results-model3}

We now present the observational constraints for the last scalar field model in this series, namely, the model of eqn. (\ref{model3}). We have analyzed the potential (\ref{model3}) for two different regions of the parameter $\delta$, namely for $\delta \geq 0$ (identified as Model 3a) and for $\delta \leq 0$ (identified as Model 3b) using exactly similar observational datasets that we have already used to study the previous models. 
In both the cases, we have analyzed the model for a fixed value of $p$, namely $p =1$. 
In what follows we describe the observational constraints for each model separately.

Table \ref{tab:results-model3A} summarizes the observational constraints for Model 3a (i.e. when $\delta \geq 0$ in eqn. (\ref{model3})) at 68\% and 95\% CL using different observational datasets. Similarly, in Fig. \ref{fig-model3a} we display the 1D marginalized posterior distributions for some selected parameters of this model plus the 2D joint contour plots at 68\% and 95\% CL considering various combinations of the model parameters.
We first focus on some specific derived parameters and their constraints. Concerning the constraints on the  Hubble constant $H_0$ from different data, we again come up with the same conclusion as found in previous scalar field models. That means, for this model too, the estimated values of $H_0$ are almost similar to what the Planck ($\Lambda$CDM based experiments) \cite{Ade:2015xua}
reports but the error bars on $H_0$ derived for this scalar field  model are mildly higher than Planck \cite{Ade:2015xua}. So, naturally, the tension in $H_0$ still exists in this model and strongly agrees the previous works \cite{DiValentino:2016hlg,Vagnozzi:2018jhn}. This now seems to be a generic feature where $H_0$ tension still alives in quintessence scalar field models. Focusing on the next two important parameters, $\Omega_{m0}$ and $\sigma_8$, we have exactly similar observation to the previous cases. That means the error bars on $\Omega_{m0}$ are slightly higher to what we see in Planck \cite{Ade:2015xua} and the estimated values of $\sigma_8$ are also slightly higher (mildly although) compared to Planck \cite{Ade:2015xua}. But, the interesting observation that we have already seen in other two quintessence scalar field models, is that, the addition of other external dataset such as BAO, JLA, RSD or CC, does not actually play any crucial role in constraining the parameters. Precisely, as we have already noticed from the error bars on different parameters, that, these external datasets do not add any constraining power to CMB dataset because after the inclusion of these external datasets to CMB dataset, there is basically no such improvements in their error bars. So, for this model we again found that CMB data alone pretty much determine the constraints.  This statement is further strengthened when one considers the free parameter $\delta$ and its constraining nature. From the analyses (precisely we refer to the 1D plots of $\delta$ shown in Fig. \ref{fig-model3a}), it is clear that the free parameter, $\delta$, of this model remains unconstrained for CMB alone and maintains the same feature after the addition of the external datasets, such as BAO, JLA, RSD, CC. To be more conclusive we tested this parameter taking its prior in the closed interval $[0, 10]$, but we have seen that its constraining nature does not depend on the choice of the prior. That means if we set the prior in the closed interval $[0, 30]$ or even increase the upper limit of this interval this does not work out.  One final remark is as follows. The absence of correlation between $H_0$ and $\sigma_8$, similar to other quintessence models, is again confirmed for this model. 
Therefore, in summary we see that $H_0$ tension is not reconciled for this model too, the parameter $\delta$ remains unconstrained, no correlation is present between $H_0$ and $\sigma_8$ similar to what we have observed with other quintessence scalar field models, and more importantly, CMB data alone can determine the constraints for this model.

We  perform similar analyses for Model 3b (i.e., when $\delta \leq 0$ in eqn. (\ref{model3})) and summarize the results in Table \ref{tab:results-model3B}. Also, in Fig. \ref{fig-model3b} we show the 1D marginalized posterior distributions for some selected parameters as well as the 2D contour plots for some combinations of the model parameters. From the analyses, we again find that, $\delta $ is again unconstrained for all the observational datasets, irrespective of its prior (see the 1D posterior distribution of $\delta$ shown in Fig. \ref{fig-model3b}). Additionally, other parameters, namely, $H_0$, $\Omega_{m0}$ and $\sigma_8$ follow exactly similar pattern (concerning their constraining behaviour) as observed in case of Model 3a and other models in this work. So, $H_0$ tension remains alive for this model; $\sigma_8$ and $H_0$ also remains uncorrelated; and CMB data alone are sufficient to study this model since the external datasets do not add any constraining power to CMB.

\begingroup                                                                                                                     
\squeezetable                                                                                                                   
\begin{center}                                                                                                                  
\begin{table}[h]                                                                                                                   
\begin{tabular}{cccccccccccccccc}                                                                                                            
\hline\hline                                                                                                                    
Parameters & CMB & CB & CBJ & CRJ & CBJRC \\ \hline

$\Omega_c h^2$ & $    0.12_{-    0.0014-    0.0028}^{+    0.0014+    0.0028}$ & $    0.12_{-    0.0010-    0.0019}^{+    0.0010+    0.0020}$ & $    0.12_{-    0.0010-    0.0020}^{+    0.0010+    0.0020}$ & $    0.12_{-    0.0011-    0.0021}^{+    0.0010+    0.0021}$ & $    0.12_{-    0.0010-    0.0020}^{+    0.0010+    0.0020}$ \\

$\Omega_b h^2$ & $    0.022_{-    0.00016-    0.00030}^{+    0.00015+    0.00030}$ & $    0.022_{-    0.00014-    0.00027}^{+    0.00014+    0.00027}$ & $    0.022_{-    0.00015-    0.00026}^{+    0.00014+    0.00027}$ & $    0.022_{-    0.00014-    0.00027}^{+    0.00013+    0.00027}$ & $    0.022_{-    0.00013-    0.00027}^{+    0.00013+    0.00027}$ \\

$100\theta_{MC}$ & $    1.041_{-    0.00033-    0.00064}^{+    0.00032+    0.00063}$ & $    1.041_{-    0.00030-    0.00059}^{+    0.00031+    0.00058}$ & $    1.041_{-    0.00030-    0.00061}^{+    0.00030+    0.00058}$  & $    1.041_{-    0.00031-    0.00059}^{+    0.00031+    0.00060}$ & $    1.041_{-    0.00030-    0.00059}^{+    0.00031+    0.00058}$ \\

$\tau$ & $    0.079_{-    0.017-    0.032}^{+    0.017+    0.033}$ & $    0.085_{-    0.016-    0.033}^{+    0.017+    0.031}$ & $    0.085_{-    0.016-    0.032}^{+    0.017+    0.032}$ & $    0.073_{-    0.015-    0.031}^{+    0.015+    0.030}$ & $    0.073_{-    0.016-    0.031}^{+    0.016+    0.030}$ \\

$n_s$ & $    0.97_{-    0.0047-    0.0092}^{+    0.0045+    0.0092}$ & $    0.97_{-    0.0038-    0.0075}^{+    0.0038+    0.0073}$ & $    0.97_{-    0.0039-    0.0079}^{+    0.0039+    0.0076}$ & $    0.97_{-    0.0039-    0.0075}^{+    0.0040+    0.0075}$ & $    0.97_{-    0.0038-    0.0073}^{+    0.0037+    0.0074}$ \\

${\rm{ln}}(10^{10} A_s)$ & $    3.092_{-    0.033-    0.064}^{+    0.032+    0.065}$ & $    3.10_{-    0.031-    0.064}^{+    0.034+    0.061}$ & $    3.10_{-    0.032-    0.065}^{+    0.032+    0.063}$ & $    3.077_{-    0.030-    0.061}^{+    0.030}$ & $    3.077_{-    0.031-    0.062}^{+    0.032+    0.061}$ \\

$\delta$ & unconstrained & unconstrained & unconstrained & unconstrained & unconstrained \\

$\Omega_{m0}$ & $    0.31_{-    0.0087-    0.017}^{+    0.0086+    0.018}$ & $    0.31_{-    0.0064-    0.011}^{+    0.0058+    0.012}$ & $    0.31_{- 0.0060-  0.012}^{+    0.0061+    0.012}$ & $    0.31_{-0.0064-    0.012}^{+    0.0062+    0.013}$ & $    0.31_{-    0.0060-    0.012}^{+    0.0061+    0.012}$ \\

$\sigma_8$ & $    0.83_{-    0.013-    0.025}^{+    0.013+    0.026}$ & $    0.83_{-    0.013-    0.025}^{+    0.013+    0.025}$ & $    0.83_{-    0.013-    0.027}^{+    0.013+    0.025}$ & $    0.82_{-    0.012-    0.024}^{+    0.012+    0.024}$ & $    0.82_{-    0.013-    0.025}^{+    0.013+    0.025}$ \\

$H_0$ & $   67.47_{-    0.63-    1.26}^{+    0.64+    1.28}$ & $   67.91_{-    0.46-    0.87}^{+    0.45+    0.87}$ & $   67.94_{-    0.46-    0.88}^{+    0.46+    0.91}$ & $   67.83_{-    0.47-    0.94}^{+    0.48+    0.96}$ & $   67.78_{-    0.46-    0.87}^{+    0.45+    0.88}$ \\

\hline\hline                                                                                                                    
\end{tabular}                                                                                                                   
\caption{Summary of 68\% and 95\% CL constraints on various model parameters of Model 3a: $V (\phi) = V_0 \left[ 1+ \delta \left(\frac{\phi}{M_P}\right)^{p}\right]^2$ (where $\delta \geq 0$) with $p=1$,  using different observational datasets. We note that $\Omega_{m0}$ is the present value of $\Omega_m = \Omega_r + \Omega_b$ and $H_0$ is in the units of km/s/Mpc. Here, we have shortened the notations as follows: CB = CMB+BAO, CBJ = CMB+BAO+JLA, CRJ = CMB+RSD+JLA, and CBJRC = CMB+BAO+JLA+RSD+CC. }\label{tab:results-model3A}                                                                                                   
\end{table}                                                                                                                     
\end{center}                                                                                                                    
\endgroup  
\begingroup                                                                                                                     
\squeezetable                                                                                                                   
\begin{center}                                                                                                                  
\begin{table}[h]                                                                                                                   
\begin{tabular}{cccccccccccccccc}                                                                                                            
\hline\hline                                                                                                                    
Parameters & CMB & CB & CBJ & CRJ & CBJRC\\ \hline

$\Omega_c h^2$ & $    0.12_{-    0.0013-    0.0027}^{+    0.0014+    0.0026}$ & $    0.12_{-    0.0010-    0.0021}^{+    0.0010+    0.0021}$ & $    0.12_{-    0.0010-    0.0019}^{+    0.0010+    0.0019}$ & $    0.12_{-    0.0011-    0.0021}^{+    0.0011+    0.0021}$ & $    0.12_{-    0.0010-    0.0019}^{+    0.0010+    0.0020}$ \\

$\Omega_b h^2$ & $    0.022_{-    0.00016-    0.00030}^{+    0.00015+    0.00030}$ & $    0.022_{-    0.00014-    0.00027}^{+    0.00014+    0.00028}$ & $    0.022_{-    0.00014-    0.00028}^{+    0.00014+    0.00027}$ &  $    0.022_{-    0.00014-    0.00028}^{+    0.00014+    0.00028}$ &  $    0.022_{-    0.00013-    0.00026}^{+    0.00013+    0.00026}$ \\

$100\theta_{MC}$ & $    1.041_{-    0.00033-    0.00066}^{+    0.00033+    0.00066}$ & $    1.041_{-    0.00030-    0.00061}^{+    0.00031+    0.00060}$  & $    1.041_{-    0.00030-    0.00060}^{+    0.00030+    0.00060}$ & $    1.041_{-    0.00030-    0.00060}^{+    0.00031+    0.00060}$ & $    1.041_{-    0.00030-    0.00059}^{+    0.00030+    0.00058}$ \\

$\tau$ & $    0.080_{-    0.017-    0.033}^{+    0.017+    0.033}$ & $    0.085_{-    0.017-    0.034}^{+    0.017+    0.033}$ & $    0.085_{-    0.016-    0.032}^{+    0.017+    0.031}$ & $    0.073_{-    0.015-    0.030}^{+    0.015+    0.031}$ & $    0.072_{-    0.016-    0.029}^{+    0.015+    0.030}$ \\

$n_s$ & $    0.97_{-    0.0045-    0.0087}^{+    0.0045+    0.0088}$ & $    0.97_{-    0.0039-    0.0074}^{+    0.0039+    0.0077}$ & $ 0.97_{-    0.0039-    0.0074}^{+    0.0039+    0.0076}$ & $    0.97_{-    0.0039-    0.0074}^{+    0.0039+    0.0076}$ & $    0.97_{-    0.0038-    0.0074}^{+    0.0038+    0.0073}$ \\

${\rm{ln}}(10^{10} A_s)$ & $    3.092_{-    0.034-    0.066}^{+    0.034+    0.065}$ & $    3.101_{-    0.034-    0.067}^{+    0.034+    0.065}$ & $    3.101_{-    0.032-    0.063}^{+    0.035+    0.060}$ & $    3.077_{-    0.030-    0.058}^{+    0.030+    0.060}$ & $    3.076_{-    0.029-    0.057}^{+    0.030+    0.059}$ \\

$\delta$ & unconstrained & unconstrained & unconstrained & unconstrained & unconstrained \\

$\Omega_{m0}$ & $    0.31_{-    0.0084-    0.016}^{+    0.0084+    0.017}$ & $    0.31_{-    0.0062-    0.012}^{+    0.0063+    0.013}$ & $    0.31_{-    0.0060-    0.011}^{+    0.0059+    0.012}$ & $    0.31_{-    0.0063-    0.012}^{+    0.0064+    0.013}$ & $    0.31_{-    0.0060-    0.012}^{+    0.0060+    0.012} $\\

$\sigma_8$ & $    0.83_{-    0.013-    0.026}^{+    0.013+    0.026}$ & $    0.83_{-    0.014-    0.027}^{+    0.014+    0.027}$ & $    0.83_{-    0.013-    0.026}^{+    0.014+    0.025}$ & $    0.82_{-    0.013-    0.023}^{+    0.012+    0.024}$ & $    0.82_{-    0.013-    0.023}^{+    0.012+    0.024}$ \\

$H_0$ & $   67.47_{-    0.62-    1.20}^{+    0.61+    1.23}$ & $   67.92_{-    0.47-    0.92}^{+    0.47+    0.94}$ & $   67.92_{-    0.46-    0.89}^{+    0.45+    0.89}$ & $   67.85_{-    0.48-    0.95}^{+    0.48+    0.96}$ & $   67.76_{-    0.45-    0.88}^{+    0.45+    0.89}$  \\

\hline\hline                                                                                                                    
\end{tabular}                                                                                                                   
\caption{Summary of 68\% and 95\% CL constraints on various model parameters of Model 3b: $V (\phi) = V_0 \left[ 1+ \delta \left(\frac{\phi}{M_P}\right)^{p}\right]^2 $ (where $\delta \leq 0$) with $p=1 $,  using different observational datasets. We note that $\Omega_{m0}$ is the present value of $\Omega_m = \Omega_r + \Omega_b$ and $H_0$ is in the units of km/s/Mpc. Here, we have shortened the notations as follows: CB = CMB+BAO, CBJ = CMB+BAO+JLA, CRJ = CMB+RSD+JLA, and CBJRC = CMB+BAO+JLA+RSD+CC. }\label{tab:results-model3B}                                                                                                 
\end{table}                                                                                                                     
\end{center}                                                                                                                    
\endgroup

\subsection{Geometrical test: The $Om$ diagnostic}
\label{subsec-om}

\begin{figure*}[tbp]
\includegraphics[width=0.52\textwidth]{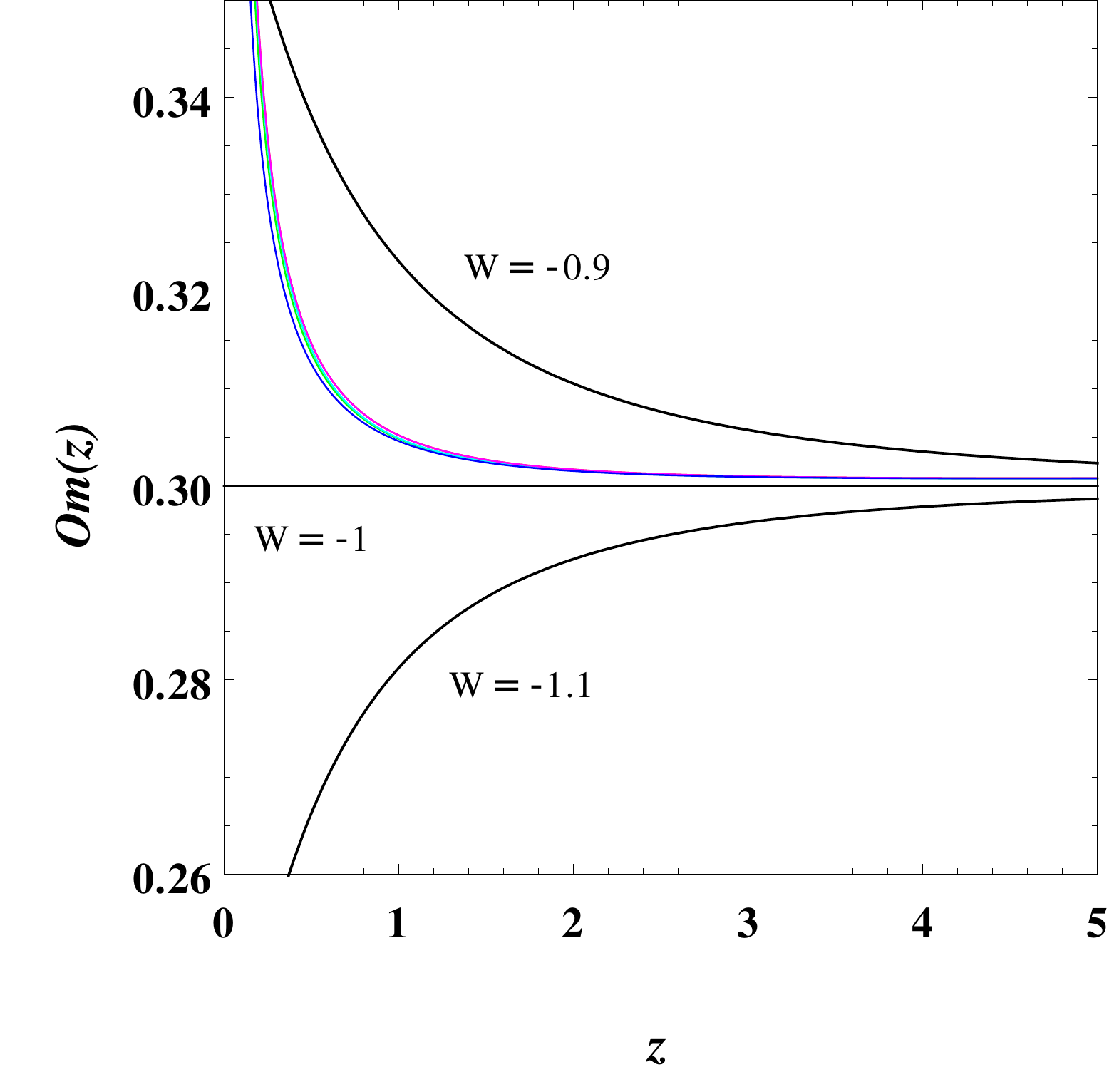}
\caption{Qualitative evolution of the $Om (z)$ function for the scalar field models (\ref{model1}), (\ref{potential}) and (\ref{model3}). The different trajectories of $Om(z)$ correspond to Model 1  with $\beta=2$ and $u=1$ (red curve), Model 2  with $\alpha=2$ (green curve for $\epsilon=+1$ and cyan curve for $\epsilon=-1$), and Model 3  with $p=2$ (magenta curve for $\delta=+1$ and blue curve for $\delta=-1$). We note that the curves (red, green, cyan, magenta, and blue) describing the $Om (z)$ function for different scalar field models are very close to each other and hence it is very hard to distinguish from one another.   
The horizontal solid black line depicts the $Om(z)$ function (which is constant as shown in the text) for the spatially flat $\Lambda$CDM model. The remaining black lines exhibit the evolution of $Om(z)$ for the dark energy models with constant equation of state  $w = -0.9, -1$ and $-1.1$, respectively from the top to bottom.  The dark energy models having $w > -1$ (quintessence) show negative curvature whereas the model with $w < -1$ (phantom) designates positive curvature which are generic features of quintessence and phantom models, respectively.}
\label{fig:om}
\end{figure*}

According to the current literature, a large number of cosmological models are introduced to understand the late-time accelerated phase of the universe. Amongst them, sometimes from the statistical ground, finding the   
differences between the cosmological models becomes very difficult and not sound. Thus, a geometrical way that enables us to distinguish the cosmological models is welcome and very appealing. One of such geometrical tests that we are interested in this work is the $Om$ diagnostic \cite{Sahni:2008xx,Zunckel2008}. This diagnostic is also considered to be the simplest one in compared to the statefinder and cosmographic parameters since the only geometrical parameter involved in this method is the Hubble parameter, in other words, only the first order derivative of the cosmic time appears, while the cosmographic constructions  
involve the higher order derivatives of the cosmic time. We refer to some works on cosmographic parameters used to distinguish the dark energy models in the literature  \cite{Sahni:2002fz,Visser:2004bf, Pan:2014afa,alam:2012PRD,alam:2013JCAP,alam:2015JCAP}.
We begin our analysis with the definition of $Om$ as follows \cite{Sahni:2008xx,Zunckel2008}

\begin{eqnarray}\label{defn-om}
Om (z) = \frac{E(z)^2-1 }{(1+z)^3 -1}
\end{eqnarray}
where $E (z) = H(z)/H_0$. Looking at eqn. (\ref{defn-om}), One can easily see that in a spatially flat  $\Lambda$CDM driven universe, where $E(z)^2 = \Omega_{m0}(1+z)^3 + (1-\Omega_{m0})$, the definition for $Om (z)$ returns, $Om (z) = \Omega_{m0}$, that means, for the $\Lambda$CDM universe, $Om (z)$ is time independent. Conversely, if for any cosmological model, we are given that, $Om (z) = \Omega_{m0}$, during the evolution of the universe, then using equation (\ref{defn-om}), one may conclude that, the model is basically $\Lambda$CDM where the background is described by the usual flat FLRW universe, so mathematically, one can say that, $Om (z) = \Omega_{m0}$, \textit{iff} the model is the $\Lambda$CDM. 
This statement actually works to distinguish the cosmological models from the $\Lambda$CDM and it has been used extensively in the literature. 
Thus, for any cosmological model, any deviation of $Om (z)$ from $\Omega_{m0}$ actually signals the difference of the toy model from the base $\Lambda$CDM cosmological model. In addition to that, the $Om$ diagnostic also offers a characterization of the dark energy models whether they belong to the quintessence class or the phantom class \cite{alam:2015MNRAS,alam:2017EPJC,shibesh:2017,alam:2015GRG}. Now, in order to understand the evolutions of $Om (z)$ for the scalar field models, we have numerically solved the Hubble function for each scalar field model and present the variations of $Om (z)$ in Fig. \ref{fig:om}. From Fig. \ref{fig:om}, one can easily find that the models actually are very close to each other and at large redshifts,  
the evolution of $Om (z)$ for the scalar field models is same to that of the $\Lambda$CDM model (represented by the solid horizontal line), while for $z \lesssim 2$, the deviation of $Om (z)$ for each scalar field model compared to the $\Lambda$CDM are clearly pronounced.

\subsection{Bayesian Evidence: A statistical tool for model comparison}
\label{sec-bayesian}

In this section we present a statistical comparison of the scalar field models using the Bayesian analysis, an analysis that quantifies the support of the cosmological model with the observational data. The analysis needs a reference model with respect to which the comparison should be made, and without any doubt,    
$\Lambda$CDM model will be the best choice based on its performance with the observational evidences. Let us first describe how the Bayesian evidence is calculated. 
In the Bayesian analysis we need the posterior probability of the model parameters (denoted by the symbol $\theta$) of any preassigned model $M$, given a particular data set $x$ which is employed to analyze the model, and any prior information. Now, recalling the Bayes theorem, one may easily write: 
\begin{eqnarray}\label{BE}
p(\theta|x, M) = \frac{p(x|\theta, M)\,\pi(\theta|M)}{p(x|M)}
\end{eqnarray}
where $p(x|\theta, M)$ is the likelihood function  that entirely depends on the model parameters ($\theta$) with the fixed data set ($x$); and $\pi(\theta|M)$ is the prior information that is supplied during the analysis. The quantity $p(x|M)$ placed in the denominator in the right hand side of eqn. (\ref{BE}) 
is the Bayesian evidence for the model comparison and this is nothing but the integral 
over the unnormalised posterior $\tilde{p} (\theta|x, M) \equiv p(x|\theta,M)\,\pi(\theta|M)$ taking the following expression 

\begin{eqnarray}\label{sp-be01}
E \equiv p(x|M) = \int d\theta\, p(x|\theta,M)\,\pi(\theta|M),
\end{eqnarray}
which is also referred to as the marginal likelihood. Now, for any particular model $M_i$ and the reference model $M_j$ (this is the $\Lambda$CDM model under consideration),  the posterior probability is given by

\begin{eqnarray}
\frac{p(M_i|x)}{p(M_j|x)} = \frac{\pi(M_i)}{\pi(M_j)}\,\frac{p(x| M_i)}{p(x|M_j)} = \frac{\pi(M_i)}{\pi(M_j)}\, B_{ij}.
\end{eqnarray}
where the quantity $B_{ij} = \frac{p(x| M_i)}{p(x|M_j)}$, is the Bayes factor of the model $M_i$ relative to the reference model $M_j$ (here it is $\Lambda$CDM). This factor essentially tells us how the observational data support the cosmological model under consideration. For $B_{ij} > 1 $, we say that the observational data support the model $M_i$ more strongly than the model $M_j$. For different values of $B_{ij}$ (or alternatively, $\ln B_{ij}$) we quantify the models. The quantification is generally adopts the widely accepted revised Jeffreys scales \cite{Kass:1995loi} (see Table \ref{tab:jeffreys} for the details). If $B_{ij}$ (or $\ln B_{ij}$) assumes negative values, the result is reversed, that means, the negative values of $\ln B_{ij}$ indicate that the 
model $M_j$ is preferred over the model $M_i$.

\begingroup                                                                                                                     
\begin{center}                                                                                                                  
\begin{table}[!h]                                                                                                                
\begin{tabular}{cc}                                                                                                            
\hline\hline                                                                                                                    
$\ln B_{ij}$ & Strength of evidence for model ${M}_i$ \\ \hline
$0 \leq \ln B_{ij} < 1$ & Weak \\
$1 \leq \ln B_{ij} < 3$ & Definite/Positive \\
$3 \leq \ln B_{ij} < 5$ & Strong \\
$\ln B_{ij} \geq 5$ & Very strong \\
\hline\hline                                                                                                                    
\end{tabular}                                                                                                                   
\caption{Revised Jeffreys scale used to quantify the observational support of any  model $M_i$ with respect to one another model (reference model) $M_j$. } \label{tab:jeffreys}                                                                                                   
\end{table}                                                                                                                     
\end{center}                                                                                                                    
\endgroup 
\begingroup                                                                                                                     
\begin{center}                                                                                                                  
\begin{table}[!h]                                                                                                                
\begin{tabular}{cccccc}                                                                                                            
\hline\hline                                                                                                                    
Dataset & Model & $\ln B_{ij}$ & Strength of evidence for model $\Lambda$CDM \\ \hline
CMB  & Model 1 &   $-1.9$ & Definite/Positive \\
CB   & Model 1 &     $-5.9$ & Very Strong\\
CBJ   & Model 1 &    $-6.2$ & Very Strong\\
CJR   & Model 1 &    $-3.2$ & Strong\\
CBJRC  & Model 1 &  $-2.7$ & Definite/Positive\\
\hline

CMB & Model 2a  &  $-3.1$ & Strong \\
CB & Model 2a  &    $-5.1$ & Very Strong \\
CBJ & Model 2a  &     $-4.9$ & Strong\\
CJR & Model 2a  &     $-1.6$ & Definite/Positive \\
CBJRC & Model 2a  & $-1.1$ & Definite/Positive \\
\hline 

CMB   & Model 2b &    $-2.6$ & Definite/Positive \\
CB  & Model 2b &      $-7.0$ & Very Strong\\
CBJ & Model 2b &      $-6.1$ & Very Strong \\
CJR & Model 2b &     $-1.8$ & Definite/Positive \\
CBJRC & Model 2b &    $-1.9$ & Definite/Positive\\
\hline 

CMB & Model 3a &     $-3.3$ & Strong \\
CB & Model 3a &      $-4.0$ & Strong \\
CBJ & Model 3a &      $-5.2$ & Very Strong\\
CJR & Model 3a &       $-3.1$ & Strong\\
CBJRC & Model 3a &     $-2.3$ & Strong\\

\hline 

CMB & Model 3b &      $-2.5$ & Definite/Positive\\
CB & Model 3b &        $-3.9$ & Strong \\
CBJ & Model 3b &       $-5.4$ & Very Strong \\
CJR & Model 3b &       $-1.7$ & Definite/Positive\\
CBJRC & Model 3b &     $-1.5$ & Definite/Positive\\

\hline\hline  

\end{tabular}                                                                                                                   
\caption{The values of $\ln B_{ij}$, the logarithm of the Bayes factor for different scalar field models in this work with respect to the base $\Lambda$CDM model for different observational data sets as well as the corresponding strength of evidence for $\Lambda$CDM quantified based on the modified Jeffreys scale (see Tab.~\ref{tab:jeffreys}). From the Bayesian point of view, the negative values of $\ln B_{ij}$ indicate that the $\Lambda$CDM model is certainly preferred in respect to the scalar field models.  }\label{tab:bayesian}                                                                                                   
\end{table}
\end{center}                                                                                                                    
\endgroup 

The computation of the Bayesian evidence is now easy since one can directly use the MCMC chains that are already found to extract the parameters space for different observational data sets. 
For a detailed explanation and implementation of the Bayesian evidence for any cosmological model we refer to the original works \cite{Heavens:2017hkr,Heavens:2017afc}. Here we use the code \texttt{MCEvidence}\footnote{This code is freely available at \href{https://github.com/yabebalFantaye/MCEvidence}{github.com/yabebalFantaye/MCEvidence}.} for the computation of the Bayesian evidence of the models.                                                                                                       
In Table \ref{tab:bayesian}, we present the $\ln B_{ij}$ values calculated for all the scalar field models with respect to the base cosmological model $\Lambda$CDM obtained for different observational data sets employed in this work. The negative values of  $\ln B_{ij}$ for any scalar field model obtained for any observational data set actually indicate that the $\Lambda$CDM model is preferred over the scalar field models. From Table \ref{tab:bayesian} one can see that $\Lambda$CDM model is clearly favored over all the scalar field models. In particular, for Model 3a, this becomes much clear while one may note that, depending on different data sets, a slight changes appear in the conclusion, but however, overall the $\Lambda$CDM model is favored with a preference ranging from definite/positive to very strong.

\begin{figure}
\includegraphics[width=0.46\textwidth]{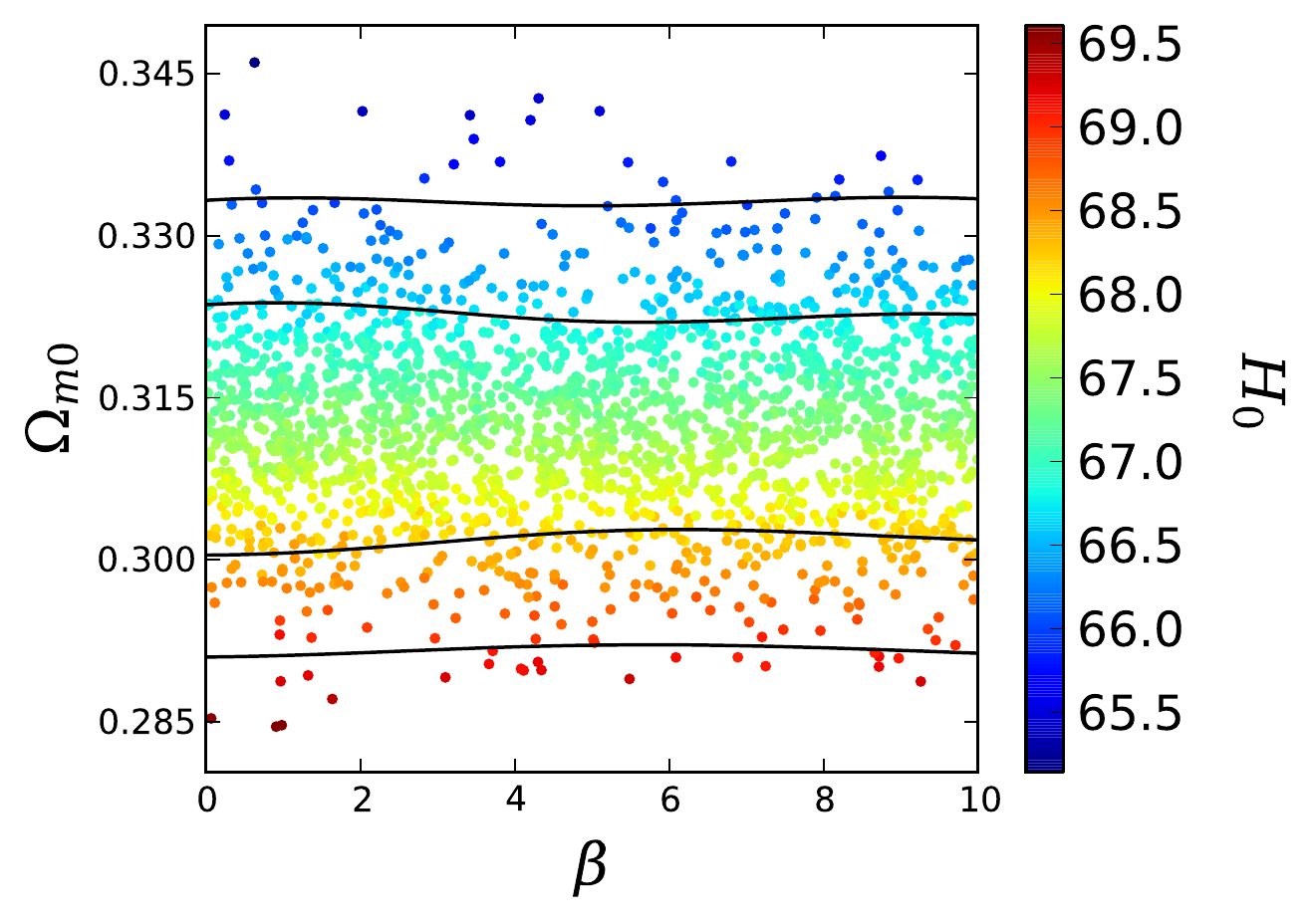}\\
\includegraphics[width=0.46\textwidth]{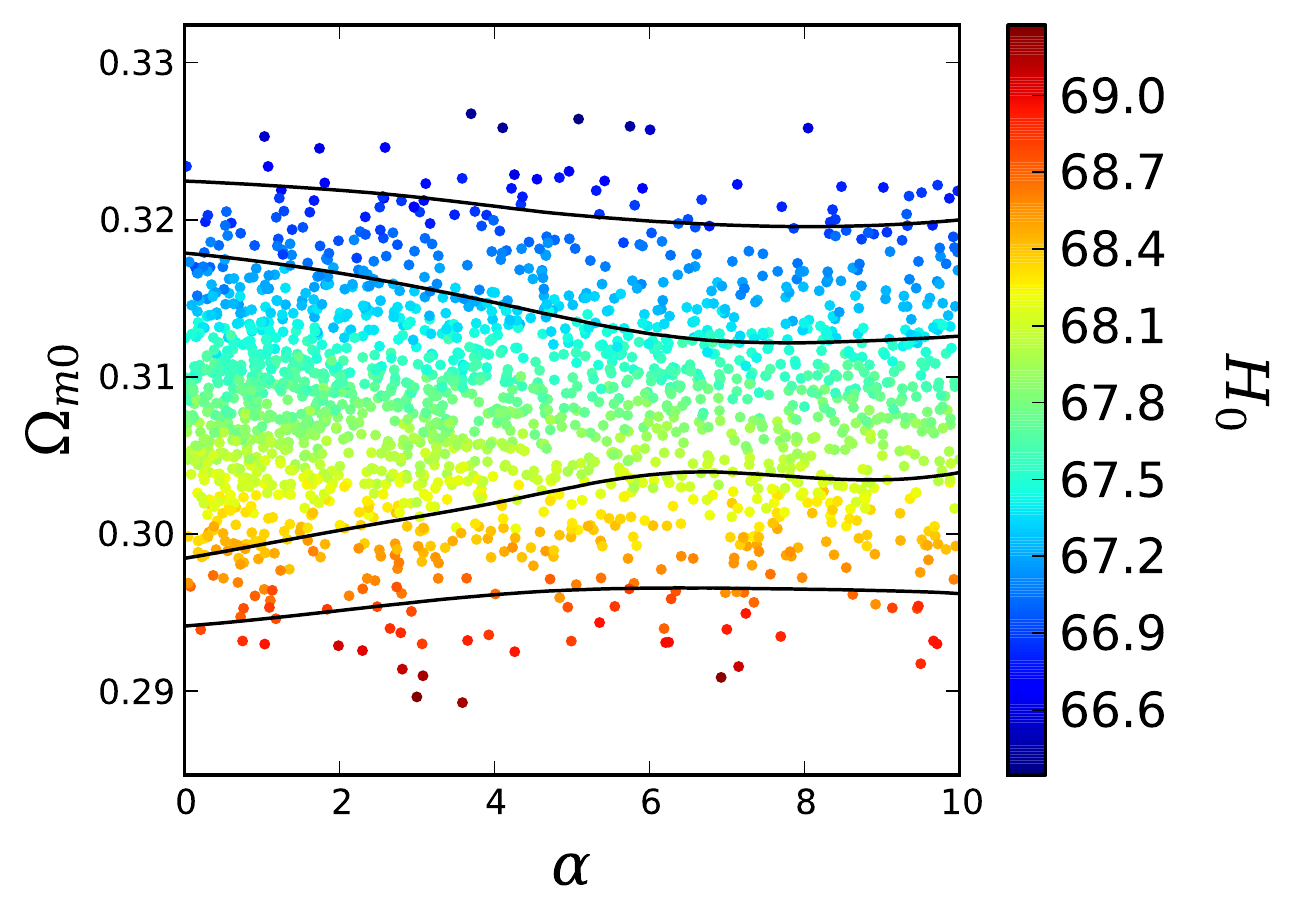}
\includegraphics[width=0.46\textwidth]{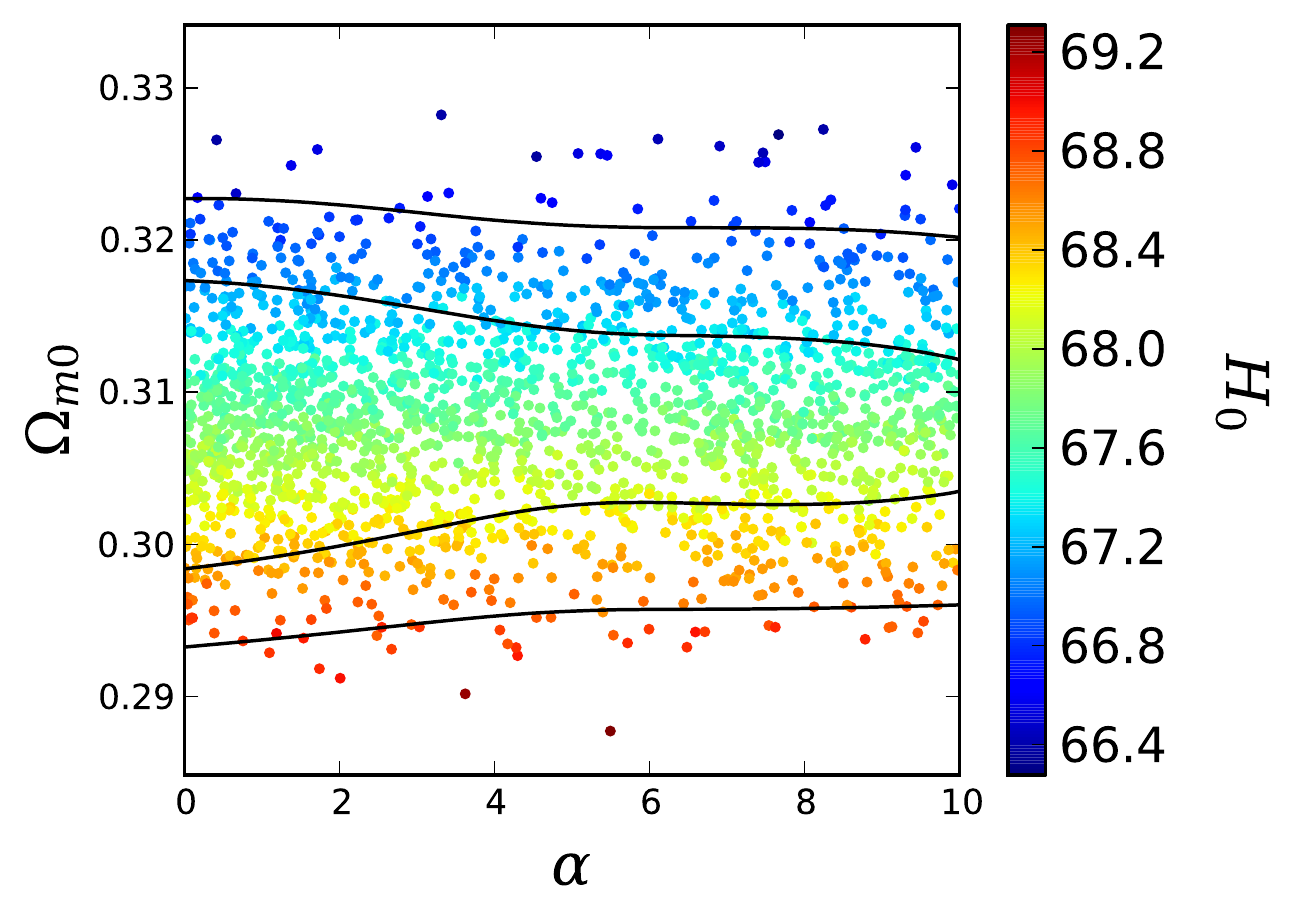}\\
\includegraphics[width=0.46\textwidth]{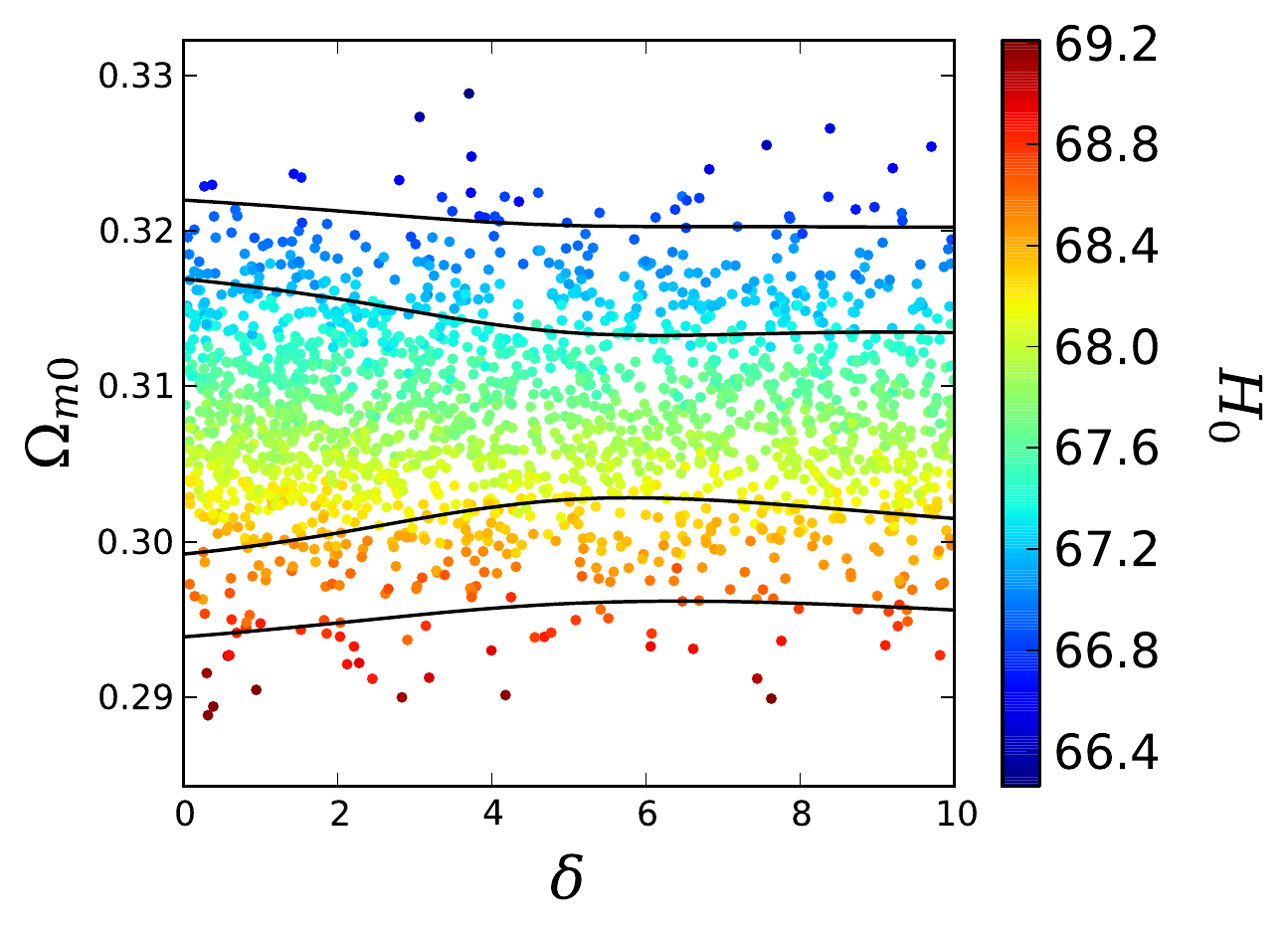}
\includegraphics[width=0.46\textwidth]{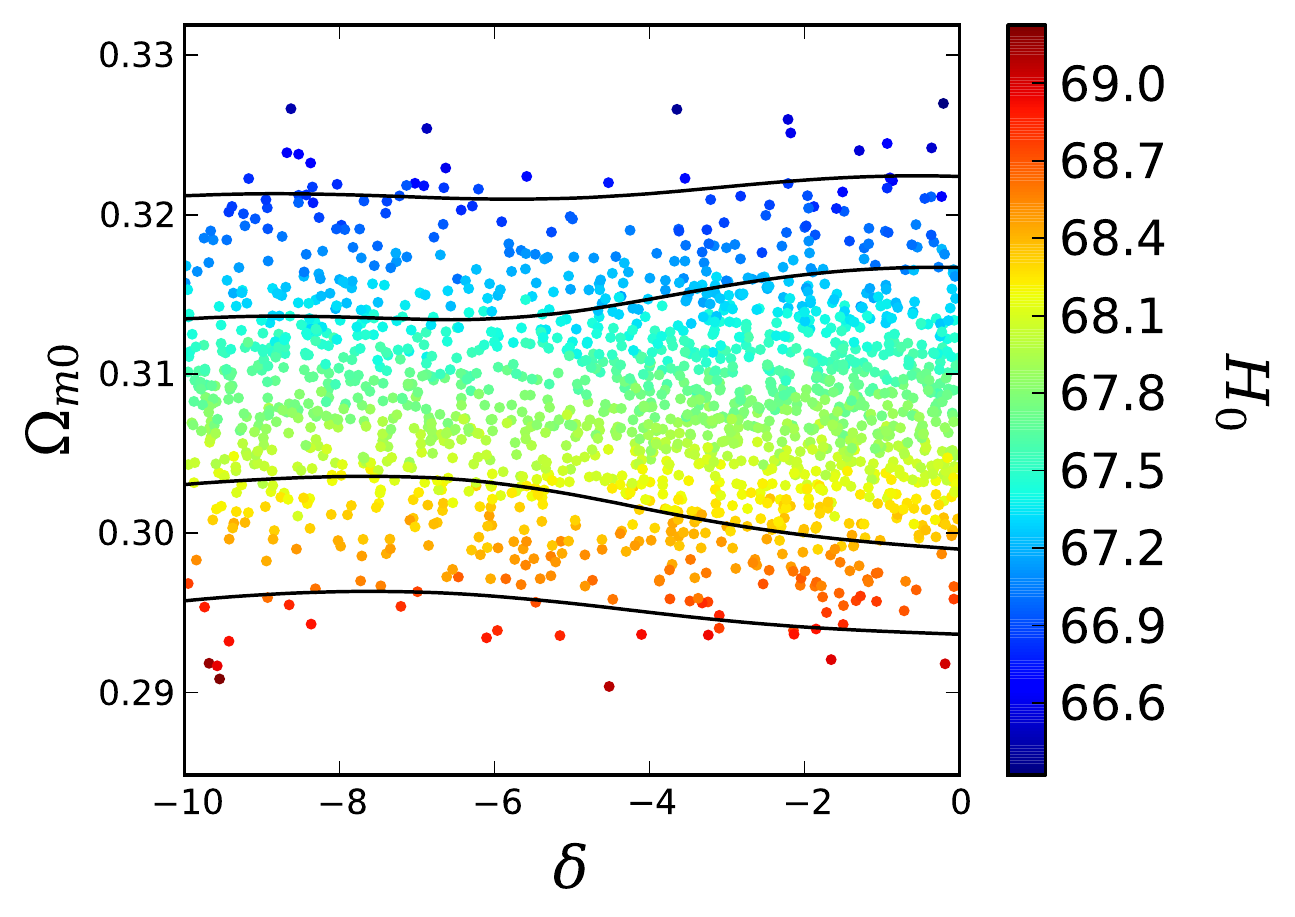}
\caption{MCMC samples in the ($\tilde{x}$, $\Omega_{m0}$) plane where $\tilde{x} = \beta$ (for Model 1), $\alpha$ (for Model 2a and Model 2b) and $\delta$ (for Model 3a and Model 3b) colored by the values of $H_0$ for the combined analysis CMB+BAO+JLA+RSD+CC. The upper panel stands for Model 1; the middle panels correspond to Model 2 (Model 2a in the left side and Model 2b in the right side); the lower panels are for Model 3 (Model 3a in the left side and Model 3b in the right side).}
\label{fig-scatter-plots}
\end{figure}
\begin{figure}
\includegraphics[width=0.46\textwidth]{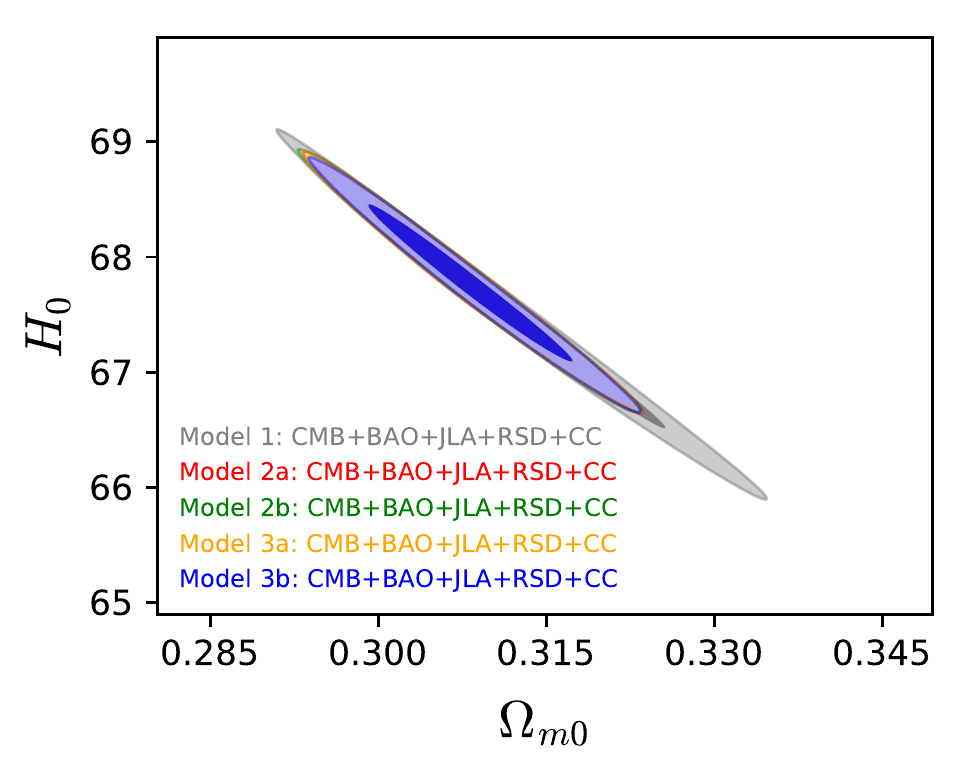}
\includegraphics[width=0.46\textwidth]{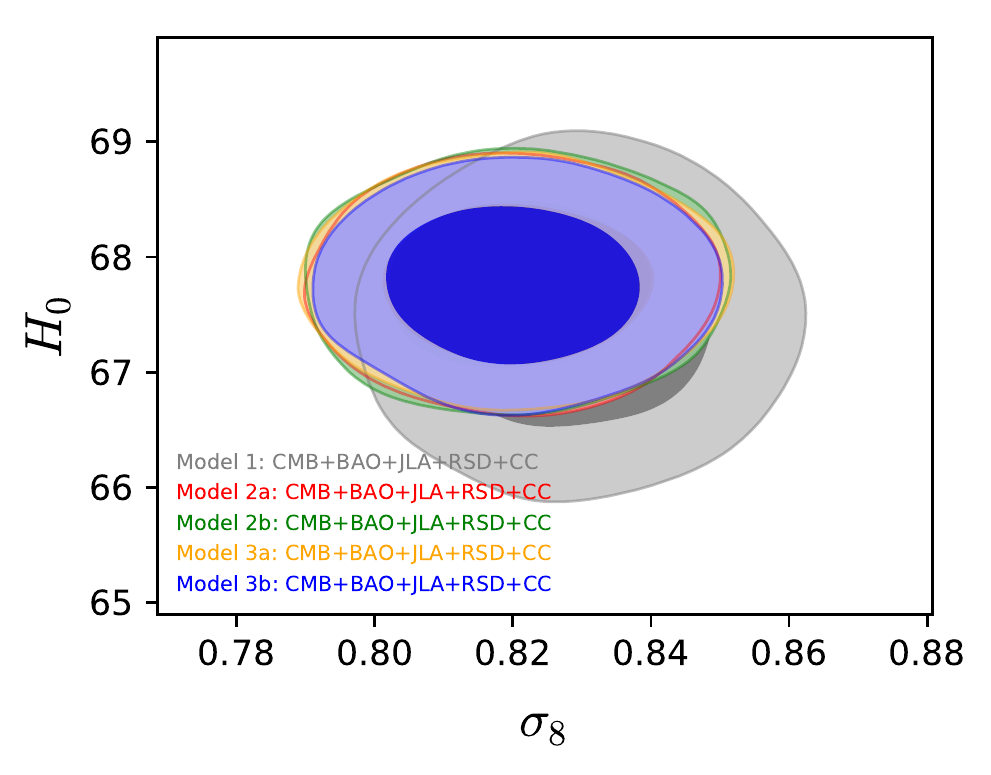}
\caption{The left panel shows a strong negative correlation between the parameters $(\Omega_{m0}, H_0)$ for the scalar field models and the right panel shows that the parameters ($\sigma_8, H_0$) for all the scalar field models are uncorrelated. We have shown both the results for the combined analysis CMB+BAO+JLA+RSD+CC, however, this result is valid for the remaining datasets. }
\label{fig-om-H0-s8}
\end{figure}

\subsection{A comparative study of the {\bf quintessence} scalar field models}
\label{sec-comparisons}

In the previous subsections, we have presented the observational constraints on the scalar field models and also performed their Bayesian evidences from which it has been found that $\Lambda$CDM is always preferred over the present scalar field models, at least according to the observational data employed in this work. In this section, we aim to offer an overall comparison between the present scalar field models (\ref{model1}), (\ref{potential}) and (\ref{model3}) focusing on some generic properties of the models independent of the observational datasets. In all scalar field models, we find that the parameter which quantifies the model from the constant potential, that means, $\beta $ for Model 1  of eqn. (\ref{model1}), $\alpha $ for Model 2 of eqn. (\ref{potential}) and Model 3 of eqn. (\ref{model3}) are very weakly constrained and within 68\% CL, the models allow the constant potential $V (\phi) = V_0$. This is not a new result in the scalar field theory. One might recall similar investigations with Peebles-Ratra potential $V (\phi) \propto \phi^{-\alpha}$ ($\alpha \geq 0$) \cite{Peebles:1987ek,Ratra:1987rm} where the parameter $\alpha$ quantifying the deviation of the model from the constant potential cannot be constrained well even with the latest cosmological datasets, see the results in \cite{Chen:2016uno}. For the exponential potential \cite{alam:2017EPJC}, a similar conclusion finding that the constant potential is allowed by the data was achieved but again the quantifying parameter was weakly constrained. Some of the models studied in this work are closely related to the  exponential potential since the first two models, Model 1, Model 2 can be expressed in terms of the exponential potential and Model 3 has similar structure to the power law type potential (although it looks similar to the power law potential \cite{Chen:2016uno} for negative values of $p$, but strictly not). However, to understand this scenario much better, in Fig. \ref{fig-scatter-plots}, we have shown the 2D contour plots between the parameters ($\tilde{x}$, $\Omega_{m0}$) where $\tilde{x} = \beta$ (for Model 1), $\alpha$ (for Model 2a and Model 2b) and $\delta$ (for Model 3a and Model 3b) colored by the $H_0$ values. The points in all the sub plots are the samples from the chains of the markov chain monte carlo (mcmc) analysis using the combined dataset CMB+BAO+JLA+RSD+CC (we only show the final analysis because the others give similar conclusions).  From this figure (Fig. \ref{fig-scatter-plots}), one can now clearly see that for  higher or lower values of $H_0$ (used different color to distinguish), the parameter $\tilde{x}$ ($= \beta, \alpha, \delta$) mentioned above, does not offer any specific changes with different (higher/lower) values of $\Omega_{m0}$. 
Apart from that, the scalar field models are very similar predicting a smooth transition from the past decelerating era to the present accelerating phase. Also, the equation of state $w_{\phi}$, of the scalar field model and the effective equation of state $w_{\rm eff}$ in eqn. (\ref{eff-eos}) of the cosmological model  are close to `$-1$', the cosmological constant limit. 
Concerning the statistical analysis, we find that there is a strong negative correlation between the parameters $\Omega_{m0}$ and $H_0$ (see the left panel of Fig. \ref{fig-om-H0-s8}) while there is no such correlation found for the parameters $(\sigma_8, H_0)$. Both the results are independent of all the datasets considered. One may recall that the correlation between $\Omega_{m0}$ and $H_0$ exists for some other well known potentials \cite{Chen:2016uno,alam:2017EPJC} and this is just a consequence of the geometric degeneracy. While the absence of correlation between $\sigma_8$ and $H_0$ for all three quintessence scalar field models is certainly an interesting issue since the correlations between these two parameters are often present in other cosmological models explored recently \cite{Yang:2017alx, Pan:2017zoh,Yang:2018euj, Yang:2018xlt,Yang:2018qmz,Yang:2018uae,Yang:2019vni}. 
As already explained in detail, $H_0$ and $\sigma_8$ are respectively more background and perturbation quantities while $\sigma_8$ is a derived parameter which is related to $A_s$ as  $\sigma_8 \propto A_s^2$. Thus, based on their individual character, it is expected that they should govern different physics irrespective of the simplicity of the model. The parameter $H_0$ governs the distance to last-scattering and thus $\Theta_s$. So,  changing $H_0$ changes the position of all the peaks in the CMB spectrum, especially the first one. On the other hand, changing $\sigma_8$ changes the amplitude of the CMB peaks. So, in principle we do not expect a strong correlation between $H_0$ and $\sigma_8$, rather the only effect is an indirect correlation between $H_0$ and $\Omega_{m0}$. Since the absence of the correlation between $H_0$ and $\sigma_8$ is valid for all three different quintessence scalar field models, thus, 
we certainly believe that  
the investigations toward this direction should be continued in order to get a more transparent picture of the quintessence scalar field models.  In fact, it is important to check whether the presence or absence of the correlations between the above three parameters are independent of the quintessence scalar field potentials or not.

\section{Summary and Conclusions}
\label{sec-conclu}

Scalar field models are well known due to their diversities in explaining various phases of the universe's evolution. They have been found to explain both early- and late-accelerating phases of the universe in a satisfactory way, see  \cite{Guth:1980zm,Linde:1981mu,Peebles:1987ek, Ratra:1987rm,Barrow:1993ah,Barrow:1993hn,Liddle:1998xm,Peebles:1998qn,Sahni:1999qe,Rubano:2001xi}. The scalar field models are also very appealing to provide with a unified picture connecting the early- and late accelerating phases, known as quintessential inflationary models, see for instance  \cite{Peebles:1998qn,deHaro:2016hpl,deHaro:2016hsh,deHaro:2016cdm,deHaro:2016ftq,deHaro:2017nui,Haro:2018jtb}. So, without any further doubt, the scalar field models might be thought to furnish a viable description for the universe evolution, and as a consequence, this triggered the investigations in this direction with an output of numerous scalar field models \cite{Copeland:2006wr}. In the present work we focus on a specific scalar field model, namely, the quintessence scalar field model. Since the potential plays the key role in determining the nature of the model, thus, it is natural to ask the observational viabilities of the models. Thus, remembering this issue, in the present work we consider a varieties of quintessence scalar field models in order to test their observational viabilities and to provide stringent constraints using the latest {\bf cosmological}  data from various sources, namely, the observations from cosmic microwave background (CMB), baryon acoustic oscillations (BAO), redshift space distortions (RSD), joint light curve analysis (JLA) from Supernovae Type Ia and the Hubble parameter measurements from the cosmic chronometers (CC). The analyses of the scalar field models have been performed using markov chain monte carlo package \texttt{cosmomc} \cite{Lewis:2002ah}, a fast converging algorithm equipped with a convergence criteria by Gelman-Rubin \cite{Gelman-Rubin}. The analyses of the results for different scalar field models, and using several observational datasets, have been presented in Table \ref{tab:results-model1} (Model 1), Table \ref{tab:constraints} (Model 2a), Table \ref{tab:constraints2} (Model 2b)
Table \ref{tab:results-model3A} (Model 3a), Table \ref{tab:results-model3B} (Model 3b) with the corresponding 1D marginalized posterior distributions of some key parameters of the models as well as the 2D contour plots between several combinations of the model parameters in Fig. \ref{fig-model1} (Model 1), Fig. \ref{fig-model2a} (Model 2a), Fig. \ref{fig-model2b} (Model 2b), Fig. \ref{fig-model3a} (Model 3a) and Fig. \ref{fig-model3b} (Model 3b).  

Our analyses show that the free parameters, $\beta$, $\alpha$ and $\delta$ of the scalar field potentials, quantifying the deviations of the models from the constant potential remain unconstrained for all the observational datasets we have used in this work. We further find that the constraints on various model parameters obtained from CMB alone dataset and from CMB+ext datasets (where `ext' means the external dataset such as BAO, JLA, RSD, CC) are almost similar. This means that the quintessence scalar field models are pretty much determined by the CMB alone dataset.  It is interesting to note that in all scalar field models, a strong negative correlation between the parameters ($H_0$, $\Omega_{m0}$) exists (the consequence of the geometric degeneracy), whilst on the other hand, the parameters ($H_0$, $\sigma_8$) are not correlated. The absence of correlation between $H_0$ and $\sigma_8$ is expected. The parameter, $H_0$, governs the distance to last-scattering and hence $\Theta_s$ ($= 100 \theta_{MC}$). So, changing $H_0$ changes the position of all the peaks in the CMB spectrum, especially the first one. Whereas changing $\sigma_8$ alters the amplitude of the CMB peaks. The effects are distinct and thus can be well distinguished between each other. So, usually we do not expect any correlation between $H_0$ and $\sigma_8$ for the scalar field models.

However, in spite of that, perhaps it will be important to  continue further the investigations along the similar lines in order to see whether the presence of correlations between ($H_0$, $\Omega_{m0}$) and the absence of correlations between ($H_0$, $\sigma_8$) in the scalar field models are generic or not. 
The generic nature of any parameter is very important to understand the nature of the models.    
Finally, the models have been further analyzed using the Bayesian analysis with respect to the base $\Lambda$CDM model for all the cosmological data sets employed in the work.  We compute the observational support of the scalar field models with respect to the the $\Lambda$CDM model (summarized in Table \ref{tab:bayesian}). The Bayesian evidence analyses report that overall the $\Lambda$CDM model is favored with a preference ranging from definite/positive to very strong.

\section*{ACKNOWLEDGMENTS}
The authors are grateful to  the referee for some essential comments that certainly improved the quality of the manuscript. 
W. Yang is supported by the National Natural Science Foundation of China under Grants No. 11705079, No. 11647153. S. Pan acknowledges the support through the Faculty Research and Professional Development Fund (FRPDF) Scheme of Presidency University, Kolkata. 
A. Wang acknowledges the financial support provided by the National Natural Science Foundation of China under Grant No. 11675145. The authors also thank  S. Vagnozzi, R. C. Nunes and E. Di Valentino for several fruitful discussions. 

\section*{Appendix A}
\label{appendix}
We use following dimensionless variables \cite{alam:2018IJMPD}
\begin{equation}
Y_1={\phi \over M_p},\quad Y_2={\dot{\phi} \over M_{p} H_{0}},\quad {\cal V}={ V(Y_1) \over M_p^2 H_0^2}.
\end{equation}
where dot ($.$) denotes  the derivative with respect to cosmic time and $H_0$ is the Hubble parameter at present epoch. We can cast Hubble and Klein-Gordon Eqs. as a system of the first-order equations 
\begin{equation}
Y_1'=\frac{ Y_2}{h(Y_1,Y_2)} \,
\label{evol1d}
\end{equation}
\begin{equation}
Y_2'= -3Y_2-{1 \over h(Y_1, Y_2)}\Big[{d {\cal V}(Y_1) \over dY_1} \Big]
\label{evol2d}
\end{equation}
The prime ($'$) represents  the derivative with respect to $\ln(a)$, and the function
$ h(Y_1, Y_2)=H/H_0$ is given by,
\begin{equation}
 h(Y_1, Y_2)=\sqrt{\left[{Y_2^2 \over 6}+ {{\cal V}(Y_1) \over 3} +{\Omega_{m0} e^{-3a}} +{\Omega_{r0} e^{-4a}}  \right]} \label{hubble}
\end{equation}
Here, $\Omega_{r0}$ and $\Omega_{m0}$ are the energy density parameters of radiation and matter, respectively at the present epoch. We solve the evolution equations (\ref{evol1d}) and (\ref{evol2d}) numerically, and the results are shown in Figures \ref{fig:1} $-$ \ref{fig:3b} for different potentials.

Above Eqs. are first order differential equations, and to solve them we need two initial conditions of $Y_1$ and $Y_2$. We choose initial values of $Y_1$ and $Y_2$ such that at the present epoch $\Omega_{m0} \simeq 0.3$, $\Omega_{r0} \simeq 10^{-4}$, $\Omega_{\phi_0} \simeq 0.7$ and $h(z)=\frac{H(z)}{H_0}\vert_{ z=0}=1$, where $H_0$ is the Hubble constant at present epoch and  $z$ is redshift.

\section*{Appendix B}
\label{sec-app2}

In this section we present the first order perturbations equations in the synchronous gauge. We consider the perturbed metric in the above gauge that takes the form \cite{Mukhanov,Ma:1995ey, Malik:2008im}
\begin{eqnarray}
\label{perturbed-metric}
ds^2 = a^2(\eta) \left [-d\eta^2 + (\delta_{ij}+h_{ij}) dx^idx^j  \right],
\end{eqnarray}
in which $\eta$ refers to the conformal time; $\delta_{ij}$,  $h_{ij}$ are the unperturbed and the perturbed metric tensors, respectively. Now, the conservation equations of the energy and momentum for the $i$-th component of the fluid for a mode with 
wavenumber ${k}$ are recasted into:

\begin{eqnarray}
&&\delta'_{i}  = - (1+ w_{i})\, \left(\theta_{i}+ \frac{h'}{2}\right) - 
3\mathcal{H}\left(\frac{\delta p_i}{\delta \rho_i} - w_{i} \right)\delta_i - 9 
\mathcal{H}^2\left(\frac{\delta p_i}{\delta \rho_i} - c^2_{a,i} \right) (1+w_i) 
\frac{\theta_i}
{{k}^2}, \label{per1} \\
&&\theta'_{i}  = - \mathcal{H} \left(1- 3 \frac{\delta p_i}{\delta 
\rho_i}\right)\theta_{i} 
+ \frac{\delta p_i/\delta \rho_i}{1+w_{i}}\, {k}^2\, \delta_{i} 
-{k}^2\sigma_i,\label{per2}
\end{eqnarray}
where the symbols used in the above equations have the following meanings: (i) The prime associated to each quantity refers  the derivative with respect to conformal time,  (ii) $\mathcal{H}= 
a^{\prime}/a$ is the conformal Hubble parameter, (iii) the quantity $\delta_i = \delta \rho_i/\rho_i$ refers to the density perturbation of the $i$-th fluid,  (iv) $\theta_{i}\equiv i k^{j} v_{j}$ is the divergence of the $i$-th fluid 
velocity, (v) $h = h^{j}_{j}$ is the trace of the metric perturbations $h_{ij}$, (vi) 
$\sigma_i$ denotes the anisotropic stress associated with the $i$-th fluid, (vii) $w_i$ is the equation of state of the $i$-th fluid, so for scalar field model, $w_{\phi} = p_{\phi}/\rho_{\phi}$.  
Finally, $c_{a,i}^2 = \dot{p}_i/\dot{\rho}_i$, is the adiabatic speed of sound of the $i$-th 
fluid with $ c^2_{a,i} =  w_i - \frac{w_i^{\prime}}{3\mathcal{H}(1+w_i)}$. The quantity $c_{a,i}^2$ should be taken to be non-negative in order to avoid any kind of instabilities.  


\end{document}